\documentclass[12pt]{article}

\usepackage{verbatim}
\usepackage{cite}
\usepackage[utf8]{inputenc}
\usepackage{amsmath,pifont,bbding}
\usepackage{epsfig}
\usepackage{amsfonts}
\usepackage{amssymb}
\usepackage{multirow}
\usepackage{bbm}
\usepackage{array}
\usepackage{enumerate}
\usepackage{subfigure}
\usepackage{slashed}
\usepackage{fouridx}
\usepackage{tensor}
\usepackage{todonotes}
\usepackage{hhline}
\usepackage{booktabs}
\usepackage{tabulary}

\usepackage[bottom]{footmisc}

\definecolor{cobalt}{RGB}{44, 98, 120}
\definecolor{celadon}{rgb}{0.67, 0.88, 0.69}
\definecolor{dm}{cmyk}{.20, 0, .30, 0}
\definecolor{burgundy}{rgb}{0.5, 0.0, 0.13}
\definecolor{plotBlue}{RGB}{94, 130, 181}
\definecolor{cornellRed}{HTML}{B31B1B}
\definecolor{cornellBlue}{HTML}{0068AC}
\definecolor{cornellGreen}{HTML}{6EB43F}

\definecolor{crimsonred}{RGB}{153,0,0}		% Neutral red, good for dark or light bg
\definecolor{darkcharcoal}{RGB}{25,25,25}		% Darker gray
\definecolor{charcoal}{RGB}{51,51,51}		% Darker gray
\definecolor{ash}{RGB}{100,100,100}			% medium gray
\definecolor{paleblue}{RGB}{0,102,102}		% More of an `ocean' color
\definecolor{turtlegreen}{RGB}{51,153,0}	% A more neutral green
\definecolor{paleale}{RGB}{204,204,102}		% Only for dark BG
\definecolor{lager}{RGB}{140,110,10}		% Use instead of pale ale for white BG
\definecolor{regal}{RGB}{90,0,120}			% A more neutral purple
\definecolor{jeans}{RGB}{20,30,150}			% A more neutral blue

\RequirePackage[colorlinks=true
,urlcolor=cobalt
,anchorcolor=cobalt
,citecolor=cobalt
,filecolor=cobalt
,linkcolor=cobalt
,menucolor=blue
,pagecolor=blue
,linktocpage=true
,pdfproducer=medialab
]{hyperref}

\usepackage{color}

\usepackage{subfigure}

\usepackage[labelfont=bf]{caption}

\newcommand{\be}{\begin{equation}}
\newcommand{\ee}{\end{equation}}
\newcommand{\bee}{\begin{equation*}}
\newcommand{\eee}{\end{equation*}}
\newcommand{\bea}{\begin{eqnarray}}
\newcommand{\eea}{\end{eqnarray}}
\newcommand{\bean}{\begin{eqnarray*}}
\newcommand{\eean}{\end{eqnarray*}}

\newcommand{\ud}{\mathrm{d}}
\newcommand{\lab}[1]{{\mathrm{#1}}}
\newcommand{\vol}{\mathrm{vol}}
\newcommand{\hodge}{\mathord{*}}

\newcommand{\units}[1]{\,\,{\mathrm{#1}}}

\newcommand{\pertTensor}[3]{{\fourIdx{}{}{{\scriptscriptstyle(#2)}}{#3}{\smash{#1}}}}

\newif\iffastcompile
\fastcompilefalse

\iffastcompile
\newcommand{\js}[1]{}
\newcommand{\lm}[1]{}
\newcommand{\aw}[1]{}
\newcommand{\ps}[1]{}
\else
\newcommand{\js}[1]{ \todo[color=cobalt!30,size=\scriptsize, bordercolor=cobalt!30]{JS: #1} }
\newcommand{\lm}[1]{ \todo[color=dm!90, size=\scriptsize, bordercolor=dm!90]{LM: #1} }
\newcommand{\aw}[1]{ \todo[color=red!30, size=\scriptsize, bordercolor=red!30]{AW: #1} }
\newcommand{\ps}[1]{ \todo[color=orange!30, size=\scriptsize, bordercolor=orange!30]{PS: #1} }
\fi
\newcommand{\minus}{{\scalebox {0.75}[1.0]{$-$}}}
\DeclareRobustCommand\circled[1]{\tikz[baseline=(char.base)]{
            \node[shape=circle,draw,inner sep=0.75pt, scale=0.95] (char) {#1};}}

\newcommand{\labbot}{\cup}
\newcommand{\labtop}{\cap}
\newcommand{\n}{N}
\newcommand{\N}{N}
\newcommand{\cbr}{\gamma_\lab{br}}
\newcommand{\cbg}{\gamma_\lab{bg}}

\begin{document}

\setcounter{page}{0}
\thispagestyle{empty}

\begin{flushright}
DESY-16-159\\
MITP/16-108
\end{flushright}

\vskip 8pt

\begin{center}
{\bf \LARGE {
Runaway Relaxion Monodromy
}}
\end{center}

\vskip 12pt

\begin{center}
{{\fontsize{14}{30}\selectfont  {Liam McAllister,$^{a}$  Pedro Schwaller,$^{b,c}$ Geraldine Servant,$^{b,d}$\\ John Stout,$^{a}$ and Alexander Westphal$^{b}$ }}}
 \end{center}
\vskip 16pt

\begin{center}
\centerline{$^{a}${\it Department of Physics, Cornell University, Ithaca, New York, 14853, USA}}
\centerline{$^{b}${\it DESY, Notkestrasse 85, D-22607 Hamburg, Germany}}
\centerline{$^{c}${\it PRISMA Cluster of Excellence, Institut f\"ur Physik,
Johannes Gutenberg-Universit\"at, Mainz, Germany}}
\centerline{$^{d}${\it II. Institute of Theoretical Physics, Univ. Hamburg, D-22761 Hamburg, Germany}}

\vskip .3cm
\centerline{\tt mcallister@cornell.edu, pedro.schwaller@uni-mainz.de, geraldine.servant@desy.de,}
\centerline{\tt  jes554@cornell.edu, alexander.westphal@desy.de}
\end{center}

\vskip 10pt

\begin{abstract}
\vskip 3pt
\noindent
\fontsize{12}{14}\selectfont We examine the
relaxion mechanism in string theory.
An essential feature
is that an axion winds over $\n \gg 1$ fundamental periods.
In
string theory realizations via axion monodromy,
this winding number corresponds to a physical charge carried by branes or fluxes.
We show that this monodromy charge backreacts on the compact space, ruining the structure of the relaxion action.
In particular, the barriers generated by strong gauge dynamics have height $\propto e^{-\n}$,
so the relaxion does not stop when the Higgs acquires a vev.
Backreaction of monodromy charge
can
therefore spoil the relaxion mechanism.
We comment on the limitations of technical naturalness arguments in this context.
\end{abstract}

\vspace*{\fill}
\today
\newpage

\tableofcontents

\vskip 13pt

\newpage

\section{Introduction}  \label{sec:intro}
Why is the Higgs mass so small?
Graham, Kaplan, and Rajendran (GKR) have  proposed a novel solution to the electroweak hierarchy problem, the \emph{relaxion mechanism}, in which the evolution of an
axion field $\phi$ drives the Higgs mass $m_h$ to relax dynamically to a value much smaller than the cutoff, $|m_h^2| \ll M^2$ \cite{Graham:2015cka}.
Achieving a large hierarchy in this way requires very small dimensionless couplings, as well as field excursions $\Delta\phi \gg M$, but GKR argued that the requisite couplings are technically natural.

In this work, we study the impact of  ultraviolet completion
on the relaxion mechanism.
The large field excursions required by the
mechanism, while technically natural in effective field theory, turn out to be \emph{source terms} in string theory!   Winding an axion $\phi$
over $\n \gg 1$ fundamental periods
leads to the accumulation of $N$ units of monodromy charge, providing a
large source term in ten dimensions.  This changes the shape of the compactification and alters the couplings of the effective theory,
eliminating the barrier that is needed to stop the relaxion once the Higgs acquires a vev.

The root of the problem is that new states linked to the monodromy charge, which are too massive in the initial configuration to be visible, are eventually drawn below the cutoff $M$.  These  new light states induce changes in the couplings of the effective theory.
In particular, the gauge coupling $g_\lab{YM}$ of the gauge theory that generates the stopping potential receives a correction $\delta g_\lab{YM}^{-2} \sim \n$.
This leads to an exponential suppression  of the stopping potential, with barrier heights $\sim e^{-\n}$, and therefore to a runaway relaxion.
This problem persists even in the limit in which the relaxion shift symmetry appears to be restored.

Although we work in string theory, and quantum gravity completion is the central question, our results do not hinge on super-Planckian displacements $\Delta\phi \gg M_{\rm{pl}}$, which are famously challenging in quantum gravity.  The problems that we expose occur even for  $\Delta\phi \ll M_{\rm{pl}}$.
The core issue is indeed one of large displacements, but here large means compared to the natural scale (or periodicity) of the effective theory.  When $\phi$ is an axion with decay constant $f$, the backreaction of monodromy charge is significant for $\Delta\phi \gg f$.

Our analysis does not amount
to a complaint that the effective theories given in \cite{Graham:2015cka} contain small dimensionless parameters.
Constructing a solution
of string theory that yields an effective field theory containing small numbers plausibly requires fine-tuning, e.g.~of the discrete data of a compactification.
Quantifying this obvious issue is not our aim.  The backreaction phenomenon that
we identify is a much more severe problem: even granting fine-tuned data that gives rise to an apparently-suitable relaxion Lagrangian in the probe approximation that omits the monodromy charge as a source in ten dimensions, the full Lagrangian beyond the probe approximation is
not of the form given in \cite{Graham:2015cka}, and does not allow for relaxation of a hierarchy.

Our goal is to identify the challenges that confront the relaxion mechanism in string theory.
Though we analyze a specific realization in type IIB string theory, we find a set of surprising, plausibly general, qualitative lessons about the nature of hierarchies and technical naturalness in low energy effective field theories descending from string theory.

The remainder of \S\ref{sec:intro}
is a microcosm of the paper.  We begin with a review of the relaxion mechanism and then provide an overview of our
results,
leaving detailed analysis for the main text.  The casual reader need only read \S\ref{sec:intro}.
 		
\subsubsection*{Overview of the Relaxion}  \label{ss:rintro}

	The simplest model of electroweak scale relaxation involves adding to the Standard Model a single axion $\phi$, the relaxion, with the potential\footnote{We follow the same notation as \cite{Graham:2015cka}, except that we take the coupling $g$ to be dimensionless, $g_\lab{GKR} = g \times M$, and shift the origin of the relaxion $\phi$ field space.}
	\begin{equation}
		V(\phi, h) = \underset{\circled{A}}{\Bigl(M^2 - g M \left(\phi_\lab{init} - \phi\right)\Bigr) |h|^2} + \underset{\circled{B}}{\vphantom{\Bigl(M^2 - g M \left(\phi_\lab{init} - \phi\right)\Bigr) |h|^2}g M^3 \phi} + \underset{\circled{C}}{\vphantom{\Bigl(M^2 - g M \left(\phi_\lab{init} - \phi\right)\Bigr) |h|^2}V_\lab{stop}(\phi, v)}
		.\label{eq:relaxpotential}
	\end{equation}
Here $h$ is the Higgs field and $v$ is its vacuum expectation value, $v^2 \equiv \langle |h|^2 \rangle$, $M$ is the cutoff of the effective field theory, and  $g$ is a dimensionless parameter that controls the explicit (albeit
weak) complete breaking of the relaxion's perturbatively exact continuous shift symmetry $\phi \mapsto \phi + \lab{const}$.  The coupling \circled{A}
promotes the Higgs mass $m_h^2$ to a dynamical variable, so that evolution of $\phi$ scans over a range of Higgs masses, while \circled{B} is a potential that forces $\phi$ to smaller values, $\phi_\lab{final} \ll \phi_\lab{init}$. Finally, \circled{C} is a non-perturbatively generated, oscillatory ``stopping potential'' $V_\lab{stop}(\phi, v) = V_\lab{stop}(\phi+ f, v)$, whose height grows with the Higgs vev $v$.
	For now, we take this potential to be
	\begin{equation}
		V_\lab{stop}(\phi, v) = \Lambda_c^3 \,v \cos \left(\frac{2 \pi \phi}{f}\right)
		\label{eq:Vstop}
	\end{equation}
	with $\Lambda_c$ the confinement scale of a gauge theory $G$ to which $\phi$ has an axionic coupling, though we will consider more general potentials in \S\ref{reviewrelaxion}.
	This potential is generated
by strong gauge dynamics
 and disappears when the theory is in a phase with unbroken chiral symmetry, i.e. in a phase with massless quarks.  Thus, the stopping potential vanishes unless the Higgs has developed a vev.
	
	\begin{figure}[]
		\centering
		\includegraphics[width=0.85\textwidth]{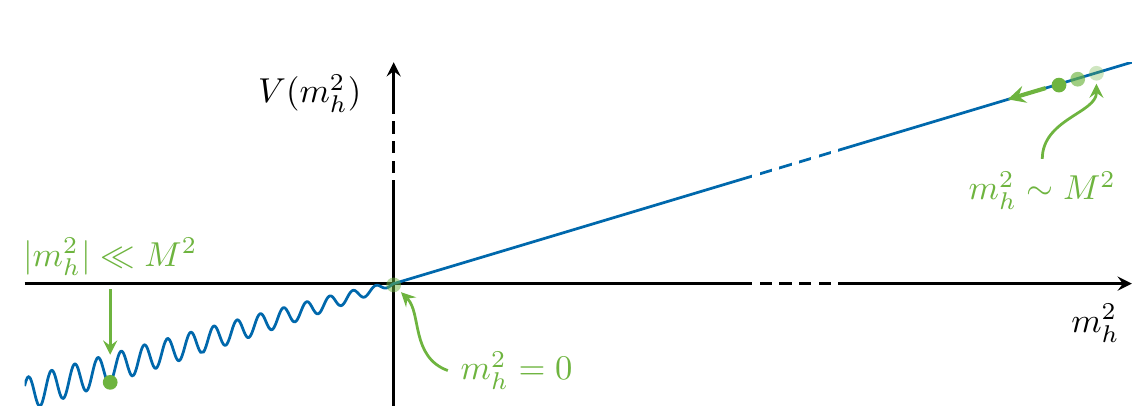}
		\caption{Schematic plot of the relaxion potential (\ref{eq:relaxpotential}). \label{fig:potplot}}
	\end{figure}
	
	The mechanism is illustrated in Fig.~\ref{fig:potplot}. The relaxion starts at a large value $\phi_\lab{init}$, where $m_\lab{h}^2 \sim M^2$, and begins to slowly roll down the linear potential \circled{B}. For generic initial conditions, the relaxion will roll a distance
	\begin{equation}
	\Delta \phi \sim M/g
	\label{eq:fieldexcursion}
	\end{equation}
	in field space before the Higgs becomes massless, $\circled{A} = m_h^2 = 0$. The Higgs then develops a vev and the stopping potential is generated. The relaxion continues to roll, halting once the stopping potential grows strong enough to counterbalance the linear potential---roughly when
	\begin{equation}
		\frac{v}{M} \sim g\left(\frac{M}{\Lambda_c}\right)^3\!\frac{f}{M} .
		\label{eq:hierarchy}
	\end{equation}
The hierarchy between the Higgs vev and the cutoff of the theory is thus controlled by the shift-symmetry breaking parameter $g$.
In effective field theory, it is technically natural for $g$ to be arbitrarily small.
However, we will see that there are obstacles to such a structure in string theory.

\subsubsection*{Requirements for Relaxation}
We now summarize the necessary ingredients for a successful relaxation of the electroweak scale.
	\begin{enumerate}
		\item The Higgs mass must be made
dynamical by  introducing an axion\footnote{As is clear from the name, it is important the relaxion $\phi$ be an axion: the axionic shift symmetry protects the potential against undesirable corrections.
One could envision a more general relaxation scenario involving a field $\phi$ that is not an axion, but it would then be necessary to explain how the structures in (\ref{eq:relaxpotential}) could be technically natural.} field $\phi$ with a coupling to the Higgs of the form
		\begin{equation}
			\mathcal{L}_h \supset \mathcal{G}(\phi) |h|^2
		\end{equation}
		where $\mathcal{G}(\phi)$ is some polynomial in $\phi$. Evolution in $\phi$ scans over Higgs masses.
		
		\item The dynamics of $\phi$ must be attractive, with the
late-time (when $m_h^2 \sim 0$)
 behavior of $\phi$ being independent of the initial conditions.

		\item $\phi$ must stop when the Higgs mass is approximately its observed, unnatural value.
	\end{enumerate}

For the evolution of the relaxion to be both attractive and dominated by classical dynamics, some friction is necessary.  Therefore, the relaxion scenario has been assumed to take place during inflation (for an alternative source of friction from particle production see \cite{Hook:2016mqo}). In this paper we will not discuss the underlying model of inflation (e.g. see~\cite{Patil:2015oxa,DiChiara:2015euo,Kobayashi:2016bue}
), nor its possible realization in string theory; these issues are the subject of an extensive literature (see for example \cite{Baumann:2014nda}).  We assume inflation to be operative, and  concentrate instead on the relaxion potential and examine how it may arise in string theory constructions.

Typically,  the stopping potential is generated by non-perturbative effects and is $f$-periodic. This ensures that only \circled{A} and \circled{B} explicitly break the discrete shift symmetry $\phi \mapsto \phi + f$, and protects against
possibly disastrous corrections.  The height of
the stopping potential must depend on the Higgs vev.
Furthermore, we require the minima of (\ref{eq:relaxpotential}) to scan through Higgs masses finely enough so that a small overshoot does not dramatically increase the final electroweak scale; since the stopping potential minima are spaced roughly $\Delta \phi \sim f$ apart, this translates into the requirement that $\mathcal{G}'(\phi) f \ll v^2$.

While this appears to be a beautiful solution to the Higgs hierarchy problem, there is some cause for concern: $g$ must be an exceptionally small number in order to generate
 a sizable hierarchy.  The simplest model of \cite{Graham:2015cka} requires $g \sim 10^{-28}$; see \S\ref{reviewrelaxion} for the requirements in variants of the model.  It is reasonable to ask whether the associated large number $1/g$ infects any other terms in the effective action.

Note that although we used the same $g$ in ${\circled{A}}$ and ${\circled{B}}$, these two terms could in principle be different.
Let us temporarily distinguish them and denote  the coupling in ${\circled{A}}$ as $g_h$. If $g \ll g_h$
at tree level,
Higgs loops will
drive the coupling in ${\circled{B}}$ to be of order
$g_h$ so that the two couplings in
${\circled{A}}$ and ${\circled{B}}$ are not very different. If, on the other hand, we take $g \gg g_h$ at tree level, this hierarchy is stable but the required field excursion in (\ref{eq:fieldexcursion}) increases
to $M/g_h \gg M/g$. So, models with $g_h \sim g$ undergo the smallest field excursion, and for this reason we only consider one $g$ coupling in Eq. (\ref{eq:relaxpotential}).

Although all the phenomena that we will uncover in this work can be encoded in an effective field theory, appropriately extended to include the effects of states that enter the spectrum as the relaxion makes its long excursion, these effects are not easily seen without the perspective of an ultraviolet theory.
This is to say that the technical naturalness reasoning of \cite{Graham:2015cka} amounts to a set of premises about the field content and interactions of an effective theory, together with conclusions that indeed follow from those premises.  In this work we question these
premises, asking whether string theory imposes restrictions or refinements on the possible effective theories. We first critically examine technical naturalness arguments in this context and then turn to a string theory embedding of the relaxion.

\subsubsection*{Technical Naturalness and Large Displacements}

Technical naturalness is often used as a panacea in model building: one begins with a symmetry that protects against potentially disastrous quantum corrections and then weakly breaks it, confident that all corrections induced by this breaking are necessarily small.
If $g$ is a dimensionless parameter measuring the weak symmetry breaking, and the symmetry is restored for $g \to 0$, corrections in the effective theory are proportional to positive powers of $g$, and so are well-controlled for $g\ll 1$.
This logic must be used with
care in the presence of field excursions $\Delta\phi$ that are large compared to the
effective theory's cutoff $M$.
The essential problem is that $\Delta\phi/M$ provides a new large parameter and
corrections can depend both on $g$ and on $\Delta\phi/M$.

As a toy example, consider a four-dimensional effective theory for a scalar field $\phi$ with Lagrangian
\begin{equation} \label{Lexpd}
{\cal L} = -\frac{1}{2}(\partial\phi)^2 -  M^4 \sum_{i=1}^{\infty} c_i\,\left(\frac{\phi}{M}\right)^{d_i} g^{e_i}\,,
\end{equation} where $M$ is a physical ultraviolet cutoff (the scale of some new physics), $g \ll 1$ is a dimensionless parameter, the $c_i$ are dimensionless Wilson coefficients, and the $d_i$ and $e_i$ are non-negative numbers. As long as
\begin{equation}\label{sufficesfortn}
e_i \neq 0  \qquad \forall i,
\end{equation} all quantum corrections are proportional to powers of $g$, and the continuous shift symmetry $\phi \mapsto \phi + \lab{const.}$ is restored in the limit $g \to 0$.
However, we stress that (\ref{sufficesfortn}) must be checked for every term in (\ref{Lexpd}), as any $e_i = 0$ term, no matter how irrelevant, could potentially provide disastrous corrections.

At large displacements $\phi \gg M$, the condition (\ref{sufficesfortn}) is far from sufficient to ensure that quantum corrections are under control at small but finite $g$.
The theory contains a new large parameter, $\phi/M$, and corrections proportional to $g^{e_i} (\phi/M)^{d_i}$ are not necessarily small for $g \ll 1$ and $\phi/M \gg 1$. Ensuring that the
corrections to the classical equations of motion
are small requires knowledge about the \emph{entire sequence} $\{c_i, d_i, e_i\}$, and so the full Lagrangian (\ref{Lexpd}).

In systems allowing axion monodromy, there is an additional subtlety: the limit $g \to 0$ is not smooth,\footnote{This observation led the authors of \cite{Gupta:2015uea} to argue that the $g \to 0$ limit is not technically natural.} 
because the field space \emph{discontinuously} changes from a helix (for $g\neq 0$) to a circle (for $g = 0$).  Standard technical naturalness arguments that rely on the $g \to 0$ limit can therefore become problematic.

Now suppose one obtains an effective theory from the top down, beginning in a vacuum of quantum gravity and integrating out Planck-scale degrees of freedom, for example by performing dimensional reduction in a string compactification with stabilized moduli.  Then the low-energy theory in four dimensions could still take the form (\ref{Lexpd}), but with two important caveats.  First, the exponents $d_i$, $e_i$ are dictated by the vacuum configuration of the underlying theory, and the condition (\ref{sufficesfortn}) must be established rather than assumed.  Second, in configurations with $\le 4$ supercharges in four dimensions, in practice one never obtains complete information about the infinite sum in (\ref{Lexpd}): some terms can be computed in different approximations, but other terms remain incalculable, although they are in principle determined by the underlying vacuum.

Because we do not have the ability to compute every term in (\ref{Lexpd}) in {\it{any}} halfway-realistic solution of string theory,
it is difficult
to prove
that (\ref{sufficesfortn}) is possible in quantum gravity.  As a result, there is a disjunction between bottom-up reasoning based on technical naturalness, and top-down reasoning based on obtaining effective theories from quantum gravity: the former strictly requires the condition (\ref{sufficesfortn}), which appears not to be provable in quantum gravity.

In our view, the difficulty in establishing (\ref{sufficesfortn}) in any particular solution of string theory is not just that the computation is challenging;
it is that plausible general reasoning about black hole thermodynamics in quantum gravity suggests that
(\ref{sufficesfortn}) is in fact false.   Exact continuous global internal symmetries are thought by many to be impossible in quantum gravity and have not appeared in string theory to date. We therefore expect quantum gravity to dramatically affect the $g\to 0$ limit.  Although our results will turn out to be compatible with this  general expectation, we do \emph{not} rely on bottom-up reasoning about quantum gravity at any point in our analysis.  In particular, we do not assume any form of the Weak Gravity Conjecture (WGC).\footnote{For work applying the WGC to the relaxion, see e.g.~\cite{Ibanez:2015fcv,Hebecker:2015zss}.}

We will argue that axion monodromy in string theory is very generally characterized by the existence of one or more terms in the effective action (\ref{Lexpd}) with $e_i=0$, and the theory is poorly-controlled in the limit $g\to 0$, $\phi/M \to \infty$.
The physical origin of these problematic terms is \emph{backreaction by monodromy charge}, as we now explain.

\subsubsection*{New States from Monodromy}

In a viable relaxion theory, we must find that every shift symmetry breaking term in the relaxion Lagrangian is proportional to a power of $g$, the parameter that controls the weak breaking
in (\ref{eq:relaxpotential}).
However, as we will explain qualitatively now and quantitatively in \S\ref{analysis},
the monodromy charge
\begin{equation}
\n \equiv  \frac{\Delta \phi}{f}\,,
\end{equation}
leads to corrections that are \emph{not} dressed by powers of $g$, so that (\ref{sufficesfortn}) does not hold.
We will begin with an example and then draw more general lessons.

Suppose (cf. the detailed discussion in \S\ref{sec:genrelaxmon}) that the relaxion is associated to a two-cycle wrapped by an NS5-brane.
Further, suppose that the stopping potential arises from the dynamics of a strongly-coupled non-Abelian gauge theory, with group $G$, living on
a stack of D7-branes wrapping a four-cycle $\Sigma_4$.
The height of the stopping potential depends on the
coupling $g_{\lab{YM}}$ of this D7-brane gauge theory:
\begin{equation}
|V_\lab{stop}| \propto \Lambda^3_c \propto \exp\left(-\frac{8\pi^2}{g_\lab{YM}^2 \,c_G}\right)\,.
\end{equation}
Here the constant $c_G$ is determined by the type of non-perturbative effects that generate $V_\lab{stop}$, and may be set to unity for our purposes.
The gauge coupling function of $G$ is proportional to the warped volume of $\Sigma_4$, cf.~(\ref{eq:d7g}):
\begin{equation} \label{onegsq}
		\frac{1}{g_\lab{YM}^2} = \frac{\vol_\lab{W}(\Sigma_4)}{2 \pi \ell_s^4}\,.
\end{equation}

When the system is wound up over $\n$ cycles, $\n$ units of monodromy charge---which in this scenario is D3-brane charge---accumulate on the NS5-brane.
This charge is a source in the ten-dimensional Einstein equations, and so leads to changes in the metric of the internal space and the warp factor. The backreaction thus alters the warped volume $\vol_\lab{W}(\Sigma_4)$.
Then, through (\ref{onegsq}),
the gauge coupling function---and hence the height of the barriers---depend on $\n$.
In \S\ref{sec:D7backreaction}, we will show that
\begin{equation} \label{dgon}
\delta\left(\frac{8\pi^2}{g_\lab{YM}^2}\right) \sim N,
\end{equation}
\emph{without} dependence on $g$.

The correction (\ref{dgon}) can be understood in a dual description as resulting from new light states associated to the source of monodromy.
The one-loop $\overline{\lab{MS}}$ $\beta$-function in a Yang-Mills theory with $n_\lab{F}$ fermions, $n_\lab{S}$ complex scalars, and coupling constant $g_\lab{YM}$ can be written
\begin{equation}\label{eq:1loopbeta}
		\frac{\ud}{\ud \log \mu}\left(\frac{8 \pi^2}{g_\lab{YM}^2}\right) = \frac{11}{3} T(\lab{Ad}) - \frac{2}{3} \sum_{i = 1}^{n_\lab{F}} T(\lab{R}_i) - \frac{1}{3}\sum_{a= 1}^{n_\lab{S}} T(\lab{R}_a),
\end{equation}
where $T(\lab{R}_i)$ is the index of representation $\lab{R}_i$ and $\lab{Ad}$ denotes the adjoint representation.
The introduction of $\n$ light states will typically lead to a change
\begin{equation} \label{dgono}
\delta\left(\frac{8\pi^2}{g_\lab{YM}^2}\right) = \cbr  \n\,,
\end{equation}
with $\cbr $ a constant independent of $\n$.

Where do these new light states come from?
The $N$
units of D3-brane charge in the NS5-brane can be viewed as resulting from $\N$ actual D3-branes (up to a binding energy that does not affect our argument).
So there are $\N$
new states in the theory,
corresponding to strings stretching from the D7-brane stack, where the gauge theory lives, to the D3-branes.
These states transform in the fundamental of $G$, and so may be described as $\N$ species of quarks from the viewpoint of $G$.  Including these species in loops leads to (\ref{dgono}).

The lesson is that ${\cal{O}}(\N)$ new states associated with the source of monodromy---in our examples, fundamental strings stretching from the source of monodromy to the gauge theory D-branes---can give large loop corrections.
These states could easily be missed in field theory, but in a string theory configuration with two D-brane gauge theories $G_1$, $G_2$, the presence of bifundamentals is hard to avoid.  The only question is whether the bifundamentals are so massive that they are physically unimportant.  In our setting,
we will find (cf. Appendix \ref{d7pf}) that arbitrarily short---and hence, light---bifundamental strings are present.

The fact that for each unit of monodromy charge
there is a new state coming down in mass that contributes to the gauge coupling of the effective theory---even though this state was far above the cutoff in the vacuum at
zero winding---is a consequence of the structure of the ultraviolet completion.
The new states described above arise from stretched strings, and so obviously have their origin in string theory \emph{per se}, but there are also new states that arise simply from the presence of extra dimensions: these are Kaluza-Klein (KK) states made light by monodromy.
Thus, our considerations can be extended to extra-dimensional ``partial'' ultraviolet completions of four-dimensional field theories, without invoking string theory.

Perhaps the simplest illustration of this phenomenon is the model of \cite{Furuuchi:2015foh} (see also~\cite{Kaloper:2016fbr}), which describes axion monodromy arising from a Stueckelberg massive $\lab{U}(1)$ gauge field coupled to a massless charged scalar field in a five-dimensional spacetime with the extra dimension compactified on a circle,
\be
S_{\lab{5D}}=\int \!\ud^4 x \int_{\lab{S}^1} \!\lab{d}y\, \sqrt{-g}\left(-\frac14 F_{MN}F^{MN} - \frac{1}{2} m^2 {\cal A}_M {\cal A}^M - \left(D_M\Phi \right)^\dagger \left(D^M\Phi\right)\right)\,,
\ee
where $D_m=\partial_M-i q A_M$, $F_{MN}=\partial_{[M}A_{N]}=\partial_{[M}{\cal A}_{N]}$, and ${\cal A}_M=A_M-i e^{i\theta}\partial_M e^{-i\theta}$ denotes the Stueckelberg covariant $\lab{U}(1)$ gauge field. Now we perform a KK reduction on the circle, whose circumference we denote by $2\pi R$.
We decompose the five-dimensional fields into an infinite series of discrete Fourier (KK) modes on the circle, and focus on the KK modes of the scalar $\Phi$,
\begin{equation}
	\Phi(x^\mu, y) = \frac{1}{\sqrt{2 \pi R}} \sum_{n \in \mathbbm{Z}} \Phi_n(x^\mu) \exp\left(\frac{i n y}{R}\right)	
\end{equation}
This yields the effective four-dimensional action
\begin{equation}
	S_\lab{5D} \supset \int\!\ud^4 x\, \left(\frac{1}{2} m^2 \phi^2 + \sum_{n \in \mathbbm{Z}} \left(\frac{n}{R} - q \phi\right)^2 \left|\Phi_n\right|^2\right).
\end{equation}
Here, $\phi \sim A_5^{(0)}$ denotes the four-dimensional axion field corresponding to the five-dimensional gauge field Wilson line around the $\lab{S}^1$.
The axion $\phi$ evidently experiences monodromy, acquiring a quadratic potential.

The key observation is that the  masses of the KK modes $\Phi_n$,
\begin{equation} \label{kkm}
m_n^2=\left(\frac{n}{R}-q \phi\right)^2\,
\end{equation}
depend on the vev of the axion $\phi$. As $\phi$ scans across its field space, one KK mode after another falls below the cutoff $R^{-1}$ in mass and thus enters the spectrum of the low-energy effective theory. In particular, as $\phi$ moves over $\n$ units of its fundamental domain, $\n$ KK modes fall below the cutoff $R^{-1}$, in analogy with the string theory effect discussed above.

We should clarify that in our examples, monodromy affects  mass spectra in two very different ways.
One effect is \emph{shifting},  in which $\phi \mapsto \phi +  f$ leaves the set of masses $m$ in a sector invariant, but permutes the states associated with these masses.  For example, in (\ref{kkm}),
changing $\phi \mapsto \phi + (qR)^{-1}$ increases by one unit the Kaluza-Klein charge of the state at each mass level.

The other effect is \emph{compression}, in which a monodromy $\phi \mapsto \phi + f$ changes the mass spectrum.
Typically, as the axion winds up and stores more energy, the masses in affected sectors are reduced.  A shifting spectrum is compatible with an exact discrete shift symmetry of the theory; the number of states below a fixed cutoff does not change, but the labels of the states change.  Compression violates even a discrete shift symmetry, as the number of states below a fixed cutoff depends on $\phi$.

With this terminology, we remark that the five-dimensional example above displays only shifting, not compression.  This is a consequence of the oversimplified nature of the model.
We will show below that axion monodromy also causes compression of the mass spectrum of Kaluza-Klein excitations of an NS5-brane.  Thus, stretched string states are not the only states that experience compression, and we expect that compressed spectra can arise in purely extra-dimensional scenarios without string theory.

Why do we not provide a purely four-dimensional field theory toy model showing the effects of shifting and compression, for instance in the case of axion monodromy from a four-form field strength~\cite{Kaloper:2016fbr, KaloperSorbo1,KaloperSorbo2,ConstraintsKaloperKleban}?
The issue is that although the core mechanism of axion monodromy arising via the Stueckelberg mechanism can be described in four-dimensional field theory, the results of~\cite{Kaloper:2016fbr} make it clear that backreaction effects, including those of massive states entering the spectrum, are described by higher-derivative corrections arising from higher powers of the four-form field strength. These corrections must be determined in the ultraviolet completion of gravity, as explicitly noted in \cite{Kaloper:2016fbr} as well.  That is, the two-derivative, four-dimensional field theory Kaloper-Sorbo model of axion monodromy \cite{KaloperSorbo1} \emph{is not} a magic wand that suppresses or controls backreaction effects.

\subsubsection*{Exponential Suppression of the Stopping Potential}  \label{ss:sintro}

We have argued that backreaction by $\n$ units of monodromy charge
leads to a large correction to the gauge coupling (\ref{dgono}).
Thus,
\begin{equation} \label{wbr}
|V_\lab{stop}| \propto \exp\left(-\cbr \N\right),
\end{equation} where $\cbr$ is a number that has no parametric dependence on $\N$ or on the shift symmetry breaking parameter $g$.
When $\cbr$ is positive, the immediate and fatal consequence is the exponential suppression of the stopping potential.\footnote{One might ask whether $\cbr$ can be fine-tuned to be small. In \S\ref{sec:warpsupp} we explain what such a tuning would correspond to in terms of compactification parameters, but it is already clear that fine-tuning $\cbr$ cannot be a satisfactory solution. One would need to arrange that $\cbr \lesssim \mathcal{O}(\N^{-1})$, which by (\ref{eq:fieldexcursion}) reintroduces a tuning on the order of the hierarchy that was supposed to be explained naturally by the mechanism.}
The stopping potential, including the backreaction effect encoded in (\ref{wbr}), is far too small to halt the evolution
when the Higgs is almost massless, $|m_\lab{h}|^2 \ll M^2$.  The result is a runaway relaxion.
If instead $\cbr$ is negative, the story is more involved. But, as we will see in \S\ref{sec:D7backreaction}, the result is still exponential suppression of the stopping potential.

Now to make things worse, there are two independent requirements that necessitate placing the source of monodromy in a region with a large background D3-brane charge $N_\lab{D3}$
that obeys
$N_\lab{D3} \gg N$.
This background charge introduces an additional, larger exponential suppression of the stopping potential.

First, achieving a large hierarchy between the Higgs vev $v$ and the cutoff $M$ necessitates an extremely small $g$. This parameter controls how strongly the relaxion shift symmetry is broken and is determined by the amount of energy introduced into the configuration per unit of monodromy charge. Since the source of monodromy corresponds to a physical quantized object, the amount of energy introduced by an additional winding is, in a sense, irreducible.
However, \emph{warping} the source of monodromy reduces this quantum of energy compared to other scales in the problem.
So, an extremely small $g$---and thus a large hierarchy---may be realized by placing the source of monodromy in a strongly warped region, as in Figure \ref{fig:coreelements} on page~\pageref{fig:coreelements}.  As we will show in \S\ref{sec:fivebrane}, we may characterize this warping by the amount of effective D3-brane charge $N_\lab{D3}$ needed to create the warped throat, and
	\begin{equation} \label{gvsN}
		g \propto N_\lab{D3}^{-1}.
	\end{equation}

Moreover, we must require a large background D3-brane charge to retain computational control and to ensure stability of the ten-dimensional configuration.
The D3-brane charge induced as the relaxion is wound must be a small correction to the charge of the ambient space
for the backreaction not to overwhelm the background configuration, and we therefore must require $N_\lab{D3} \gg \N$.

Now as in the preceding section, the D3-brane charge $N_\lab{D3}$ of the ambient space has an effect akin to that of $N_\lab{D3}$ actual D3-branes, which would give rise to $N_\lab{D3}$ species of quarks in the fundamental of $G$.  Loops of these quarks yield
\begin{equation} \label{Ncorr}
\delta\left(\frac{8\pi^2}{g_\lab{YM}^2}\right) = \cbg N_\lab{D3}\,,
\end{equation} where $\cbg$ is a positive constant.

We conclude that the stopping potential is \emph{exponentially suppressed} by the warping required to achieve a weak shift symmetry breaking  $g \ll 1$  and to maintain control over the model.
Schematically, we have
\begin{equation} \label{bigN}
|V_\lab{stop}| \propto \exp\left(-\cbg N_\lab{D3}\right)\,,
\end{equation} with $N_\lab{D3} \gg \n$.  Here we remind the reader that $\N \gg 1$ is the
large number of windings required to substantially ameliorate the hierarchy problem.  The suppression (\ref{bigN}) renders the barriers utterly negligible.\footnote{This effect holds regardless of the sign of $\cbr$ in (\ref{wbr}). But, for $\cbr<0$,  (\ref{bigN}) is the only relevant exponential suppression, while for $\cbr>0$ (\ref{wbr}) amounts to an independent suppression $\sim \!e^{-N}$ which even on its own is sufficient to cause a runaway.}
Using (\ref{gvsN}), (\ref{bigN}) can be written as
\begin{equation}\label{BURN}
	|V_\lab{stop}| \propto \exp\Bigl(-\mathcal{O}\left(1/g\right)\Bigr),
\end{equation}
so we see that the generated hierarchy is no longer proportional to $g$.
Instead, suppressing the shift symmetry breaking scale
simultaneously suppresses the stopping potential barriers, leading to a relaxion runaway.

\subsubsection*{Overview of the paper}

The organization of this paper is as follows.
In \S\ref{reviewrelaxion} we briefly survey relaxion models constructed in effective  field theory,  and identify the parameter ranges that allow relaxation of a hierarchy.
In \S\ref{sec:axmon} we introduce axion monodromy in string theory, emphasizing the fact that monodromy results from a physical, quantized source.
We review the scenario of axion monodromy on NS5-branes, and then explain how the relaxion mechanism could be realized in this setting.
In \S\ref{analysis} we determine the microphysical constraints that arise in  such a realization.  An executive summary appears in \S\ref{Overview}.
We discuss generalizations in \S\ref{sec:summary}, and conclude in \S\ref{conc}.
The appendices contain more technical material.  Appendix \ref{app:dictionary} provides
background on axions in string theory.
In Appendix \ref{d7pf} we prove that the D7-branes responsible for the stopping potential must intersect the NS5-brane.
In Appendix \ref{sec:conventions} we give
the actions for D5-branes and NS5-branes in warped compactifications of type IIB string theory.
In Appendix \ref{sec:internalbr} we analyze the backreaction of D3-brane and anti-D3-brane charge and tension on the metric of the internal space.

\section{Relaxion Zoology} \label{reviewrelaxion}

We now briefly overview a selection of existing relaxion models in field theory. To present a unified synopsis of the genus \emph{Relaxion} in its various speciations, we discuss these models in a consistent notation and, since the number of windings $\n$ is severely constrained in string theory, we pay special attention to the field excursions required to generate a large hierarchy.

For generic initial displacements,
the relaxion must scan a field range $\Delta \phi \sim M/g$ 
to reach $m_h^2 = 0$. If we generalize (\ref{eq:Vstop}) and (\ref{eq:hierarchy}) by including a more generic dependence on the Higgs vev $v$ as in~\cite{Espinosa:2015eda}, schematically
\begin{equation}
	V_\lab{stop}(v, \phi) = \Lambda^4 (v) \cos \left(\frac{2 \pi \phi}{f}\right) =  \epsilon \,\Lambda_c^4 \left(\frac{v}{\Lambda_c}\right)^r \!\cos \left(\frac{ 2 \pi \phi}{f}\right)
	\label{eq:barrier}
\end{equation}
and
\begin{equation}
	\left(\vphantom{\frac{M}{\Lambda_c}}\frac{v}{M}\right)^r \sim \frac{g}{\epsilon} \frac{f}{M} \left(\frac{M}{\Lambda_c}\right)^{4-r}
	\label{eq:hierarchy2}
\end{equation}
with $\epsilon$ a constant coefficient, this excursion implies a winding charge of
\begin{align}
	\n = \frac{\Delta \phi}{f} \sim \frac{1}{\epsilon} \left(\frac{M}{\Lambda_c}\right)^4 \left(\frac{\Lambda_c}{v}\right)^{r}.
	\label{eq:masterformula}
\end{align}
For $\Lambda_c \sim v$, $\n$ scales with the fourth power of the ratio of the cutoff scale $M$ to the weak scale, further increasing if $\Lambda_c \ll v$.

\subsection{Original models}

Two explicit constructions were originally proposed in \cite{Graham:2015cka}, and the dynamics of these models were explained in \S\ref{ss:rintro}. The relaxion in the first model (GKR1) is the QCD axion
and the potential barriers are generated by strong chromodynamic forces. The potential barriers in (\ref{eq:barrier}) then scale as $\Lambda^4(v) \sim \Lambda_\lab{QCD}^3 m_u$, i.e. $r = 1$, $\Lambda_c \sim \Lambda_\lab{QCD}$, and $\epsilon$ is the up-quark Yukawa coupling $y_u$.
The main drawback of this model is that it destroys the solution to the strong CP problem.
The PQ solution may be restored, as discussed in \cite{Graham:2015cka}, by introducing additional dynamics at the end of inflation, which removes the slope of the relaxion potential at the end of inflation. However, in this case the cutoff scale cannot be pushed  higher than
$M \approx 30~{\rm{TeV}}$.  The hierarchy (\ref{eq:masterformula}) is then multiplied by a factor of the QCD angle $\theta_\lab{QCD}$. In either case, the number of windings obeys $N \geq (M/\Lambda_\lab{QCD})^4$.

Because the generated hierarchy (\ref{eq:masterformula}) grows with the confinement scale $\Lambda_c$, the second model (GKR2) introduces a new strongly-interacting gauge sector $G$, whose axion is the relaxion. The PQ solution to the strong CP problem is then untouched and $\Lambda_c$ can be much larger than $\Lambda_\lab{QCD}$. However, this does not allow one to make the barrier arbitrarily high. New electroweak scale fermions couple the Higgs sector to this new sector and the barrier height depends quadratically ($r = 2$) on the Higgs vev. But, a constant term ($r = 0$) will also be generated by quantum corrections so the barriers can schematically be written as \cite{Espinosa:2015eda}
\begin{equation}
	\Lambda^4(v) \sim \epsilon\,\Lambda_c^4 \left(1 + \left(\frac{v}{\Lambda_c}\right)^{\!2}\,\right).
\end{equation}
So, the barrier will not depend strongly enough on the Higgs vev $v$ for the relaxion mechanism to work unless $\Lambda_c \lesssim v$. Still, GKR2 can generate a much larger hierarchy $M \lesssim 10^8 \units{GeV}$ than GKR1, with a similar parametric scaling of the number of windings $N\geq (M/v)^4$. Unfortunately, GKR2 requires that new electroweak scale fermions be put in by hand, and this coincidence of scales must be explained.

\subsection{CHAIN}

A solution to this coincidence problem was suggested in \cite{Espinosa:2015eda}, and involves taking the barrier height to depend on an extra scalar field. The relaxion mechanism is then able to explain the near-criticality of the Higgs without a coincidence of scales. Instead, there is only one scale in the problem, the cutoff $M$, which also sets the barrier height $\Lambda_c \sim M$.
The extra scalar $\sigma$, which need not be an axion, controls the height of the stopping potential,
 \be
 \Lambda^4(h,\phi,\sigma) =  \epsilon M^4 \left(\beta + c_{\phi} \frac{g \phi}{M}- c_{\sigma} \frac{g_{\sigma} \sigma}{M} + \frac{h^2}{M^2}\right).
 \label{eq:chainbarrier}
 \ee
The initial conditions are very different from both GKR1 and GKR2. At first, the barriers are large and the relaxion is stuck in one of its minima.
As the second field $\sigma$ evolves, its vev will eventually cancel this barrier and allow the relaxion to roll.
In contrast with GKR2, there are no constraints on the decay constant $f$ from reheating.

Given that now the barriers are allowed to be high, $\Lambda_c \gg v$, one might hope that the required number of windings for a given cutoff scale is substantially reduced.  A more careful analysis, however, reveals that this is not the case. Instead, because classical evolution must dominate over quantum fluctuations, we require that $\epsilon \lesssim v^2/M^2$, while 
imposing that the Higgs barrier in (\ref{eq:chainbarrier}) is solely responsible for stopping $\phi$ 
requires that $v^2 \sim g M f/\epsilon$. Together, these imply that
\begin{align}
	\n \sim \frac{M}{g f} \gtrsim \left( \frac{M}{v}\right)^4,
\end{align}
and so the CHAIN model also requires a large number of windings to
resolve the hierarchy problem.

\begin{figure}
\centering
\includegraphics[width=0.62\textwidth]{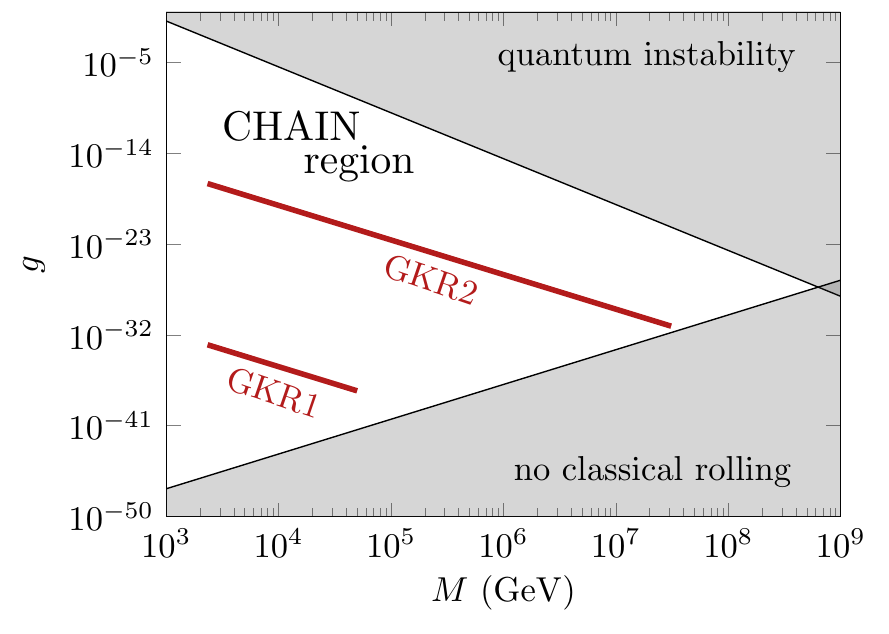}
\caption{
Schematic parameter space in the three main non-supersymmetric relaxion models.
See \cite{Espinosa:2015eda}  for the derivation of the constraints on the parameter space.}
\label{fig:parameterspace}
\end{figure}

A comparison of the three models is shown in Fig.~\ref{fig:parameterspace} and Tab.~\ref{tab:comparison}.

\begin{table}[t]
\begin{center}
\begin{tabular}{c c c c }
\toprule
 &GKR 1&GKR 2& CHAIN with $f\sim M$ \\
\cmidrule{2-4}
$f$ &$f_{PQ}\sim 10^{10}-10^{12} \units{GeV}$ & $\gtrsim M_\lab{GUT} \sim 10^{16} \units{GeV}$  & $\gtrsim M$\\
\cmidrule{2-4}
$g$  & $ \big(\frac{\Lambda_\lab{QCD}}{M}\big)^4  \frac{M}{f_\lab{PQ}} \theta_\lab{QCD} \lesssim 10^{-36}$&  $\big(\frac{\Lambda_\lab{EW}}{M}\big)^4 \frac{M}{M_\lab{GUT}} \sim 10^{-30}$--$10^{-20}$  &$\lesssim {v^4}/{M^4}\sim 10^{-26}$--$10^{-6}$ \\
\cmidrule{2-4}
$M_\lab{max} $&$3\times 10^{3} \units{GeV}$&$10^{8}\units{GeV}$ &$10^{9} \units{GeV}$\\
\cmidrule{2-4}
$m_{\phi}$ &$\frac{\Lambda^2_\lab{QCD}}{f_\lab{PQ}}\lesssim 10^{-11} \units{GeV}$ &$\frac{\Lambda^2_\lab{EW}}{M_\lab{GUT}}\lesssim 10^{-12}\units{GeV}$&$\sqrt{{g M^4}/{v^2}}\lesssim v$\\
\cmidrule{2-4}
$\Delta \phi/f $&$\theta_\lab{QCD}^{-1}  \big(\frac{M}{\Lambda_\lab{QCD}}\big)^4  \gtrsim 10^{30}$&$({M}/{\Lambda_\lab{EW}})^4\sim 10^8-10^{24}$&$g^{-1}\sim 10^6-10^{26}$\\
\bottomrule
\end{tabular}
\end{center}
\caption{ \small  Summary of parameter values in the three non-supersymmetric relaxion models discussed in \S\ref{reviewrelaxion}.}
\label{tab:comparison}
\end{table}

\subsection{Supersymmetric models}

Inflation limits the achievable cutoff scale to $M\sim 10^9$~GeV.  The energy stored in the relaxion must not dominate over the energy driving inflation,
\begin{align}
	M^4 < H^2 M_{\rm pl}^2\,,
\end{align}
where $H$ is the Hubble rate during inflation, and
barriers cannot form unless $H < \Lambda_c$. This
immediately implies a bound on the cutoff $M \lesssim \sqrt{v M_{\rm pl}} \sim 10^9$~GeV for GKR2. While this argument does not directly apply to the CHAIN model, there one finds the same bound $M \lesssim 10^9$~GeV.
Since we must also explain the remaining hierarchy between the cutoff $M$ and the Planck scale $M_\lab{pl}$, a natural candidate solution is that supersymmetry is restored above $M$ and the relaxion is embedded within a supersymmetric model.

A supersymmetric version of GKR1 was presented in~\cite{Batell:2015fma}, on which the following discussion is based. The relaxion becomes part of a chiral superfield $S$:
\begin{align}
	S = \frac{s + i a} {\sqrt{2}} + \sqrt{2} \theta \tilde{a} + \theta^2 F + \dots\,,
\end{align}
which
contains the (dimensionless) relaxion $a=\phi/f$, a srelaxion field $s$, and the relaxino $\tilde{a}$.
The Peccei-Quinn symmetry acts as $S \mapsto S+i \alpha$.
The linear term \circled{B} in the relaxion potential (\ref{eq:relaxpotential}) descends from the superpotential term
\begin{align}
	W \supset \frac{1}{2} m f^2 S^2\,. \label{eqn:susy_relaxion_mass}
\end{align}
Small $m \ll f$ is technically natural since $m$ breaks the PQ symmetry, which is non-linearly realized via the term
\begin{align}
	W \supset \mu_0 \,e^{-q S} H_u H_d.
\end{align}
Apart from $S$, the model contains only SM particles and their superpartners (including the usual second Higgs doublet).
The effective potential for $s$ and $a$ is then
\begin{align}
	V = \frac{1}{2} m^2 f^2 \left(s^2 + a^2 \right) \kappa(s)\,,
\end{align}
with $\kappa(s)$ a function of $s$. As in all relaxion models, $a$ starts out at a field value far away from its minimum at $a=0$,
and so breaks supersymmetry, with $F \propto ma$.
As $a$ evolves towards its minimum, it scans the SUSY breaking scale, and therefore the soft masses of the gauginos and the scalar superpartners.
In particular, the determinant of the Higgs mass matrix was shown to scale as $a^4$ for $a \gg \mu_0/m$,  far away from any electroweak symmetry breaking minima. As $a$ approaches the critical value $a_* = \mu_0/m$, electroweak symmetry is broken, the Higgs(es) obtain a vacuum expectation value,
and barriers appear that halt the evolution of $a$.

For a suitable choice of parameters, the model explains the hierarchy between the electroweak symmetry breaking scale $v$ and the mass scale of the superpartners $\mu_0 \gg v$, thus solving the supersymmetric little hierarchy problem. According to~\cite{Batell:2015fma}, the number of windings scales as
\begin{align}
	\n \sim \Delta a \sim \frac{f^2 \mu_0^2}{\Lambda_\lab{QCD}^4}\,,
\end{align}
where $f \sim 10^9$--$10^{12}\units{GeV}$ is the QCD axion decay constant and $\mu_0$ plays the role of the UV cutoff. For $\mu_0 =10^5\units{GeV}$, a field excursion of $\Delta a \sim 10^{30}$ is required.
Without further modifications, this
model is phenomenologically unacceptable since it predicts  $\theta_\lab{QCD} \sim \mathcal{O}(1)$. A variation with a non-QCD axion similar to GKR2 is briefly discussed in~\cite{Batell:2015fma}.
In this case, a larger range of decay constants
is allowed---for $f=\mu_0$ and $\Lambda \sim v$ one obtains the same scaling as in GKR2, namely $\n \sim \mu_0^4/v^4$.

A supersymmetrization of the CHAIN model was proposed in~\cite{Evans:2016htp}. The philosophy is similar to the above discussion---now both the relaxion and the additional scalar $\sigma$ are promoted to chiral superfields. The barriers for the relaxion are generated by a new $\lab{SU}(N_g)$ gauge theory, with confinement scale $\Lambda_{g}$, which communicates with the Higgs sector via a set of vector-like leptons.
The required field excursion in this model is
\begin{align}
	\n \equiv \frac{\Delta \phi}{f} \gtrsim \frac{m_{\rm SUSY}}{|m_{\rm S}|}\,,
\end{align}
where $m_{\rm SUSY} \sim \mu_0$ is the supersymmetry-breaking scale, and $m_{\rm S}$ is the relaxion mass coming from a term similar to~(\ref{eqn:susy_relaxion_mass}). For $\Lambda_{g} \sim f \sim m_{\rm SUSY}$, a supersymmetry-breaking scale~${\sim 10^9\units{GeV}}$ may be
generated through a field excursion of $\n \sim 10^{27}$. So while the field excursion seems to grow more moderate as a function of the cutoff, it is still as large as in the earlier models for the largest possible cutoff (e.g. $M^4/v^4$ is of order $10^{26}$). Instead, even the best case scenarios with $m_{\rm SUSY}\sim 10^4$~GeV require a large number of windings $\n\gtrsim 10^8$.

\subsection{Summary}
The models presented above do not represent a complete classification of genus \emph{Relaxion}. In particular, we have omitted models that either rely on the alignment of multiple axions or use friction from particle production to halt the evolution of $\phi$.  We discuss both of these further in \S\ref{sec:summary}. However, the models that we examine represent a large cross-section of  \emph{Relaxion} and share a common trait: the required field excursion scales parametrically with the hierarchy generated, and so the associated number of windings around the relaxion field space $\N \equiv \Delta \phi/f$ is enormous. In what follows, we will argue that $\N$ is a \emph{physical charge} in string theory, which backreacts on the ten-dimensional configuration and tragically destroys the structures in (\ref{eq:relaxpotential}), allowing for a runaway relaxion.

\begin{table}[t]
\begin{center}
\begin{tabulary}{\textwidth}{C C}
	\toprule
		\textbf{Relaxion Quantity} & \textbf{String Theory Origin}\\ \midrule
		
		Axion $\phi$ & NS-NS or R-R $p$-form gauge field, dimensionally \newline reduced along non-trivial $p$-cycle, \S\ref{sec:fivebrane} \\ \midrule
		
		Discrete shift symmetry \newline $\phi \mapsto \phi + f$  & Ten-dimensional NS-NS or R-R gauge symmetry,
		exact in absence of brane or flux, \S\ref{sec:fivebrane} \\ \midrule
Physical source of monodromy explicitly breaks $\phi \mapsto \phi +  f$& Wrapped brane or flux along axion $p$-cycle, \S\ref{sec:fivebrane}\\ \midrule
 Shift symmetry-breaking  scale $g M^3 f$ &  Warped brane tension, \S\ref{sec:fivebrane}\\
 \midrule
 Winding number  $\n \equiv \Delta \phi/f$& Quantized monodromy charge,  \S\ref{sec:fivebrane}\\
\midrule
Axion decay constant $f$ & Set by internal six-dimensional geometry, \S\ref{app:dictionary} \\
\midrule
Stopping potential barrier \newline height $\Lambda(v)$   & Set by warped volume of a four-cycle,  \S\ref{sec:genrelaxmon} \\
\bottomrule
\end{tabulary}
\end{center}
\caption{ \small
A quick string theory-relaxion dictionary.
This is a summary table, with more extended explanations given throughout the paper and in the appendices.}
\label{tab:dictionary}
\end{table}

\section{Relaxion Monodromy}\label{sec:axmon}
		\subsection{Axion monodromy in string theory}

Axions are commonplace in string compactifications,\footnote{A detailed treatment of the material that follows can be found in  \cite{Baumann:2014nda}, \S5.4.2.  An overview is given in
Appendix \ref{app:dictionary}, and Table~\ref{tab:dictionary} gives a simple dictionary.} and arise when a $p$-form gauge potential---either the NS--NS two-form $B_2$ or an R--R $p$-form $C_p$---is dimensionally reduced along a non-trivial cycle $\Sigma_p$ in the compactification manifold $X_6$. The ten-dimensional supergravity action is invariant under the gauge symmetry $B_2 \mapsto B_2 + \ud \Lambda_1$ and $C_p \mapsto C_p + \ud \Lambda_{p-1}$ which, upon reduction to four dimensions, ensures that the axion enjoys a perturbatively exact shift symmetry. For an axion $a$ associated with a non-trivial cycle $\Sigma_p$, the shift symmetry $a  \mapsto a + \text{const}.$ may be broken to a discrete shift symmetry by non-perturbative effects, or completely broken by a brane wrapping $\Sigma_p$. In the latter case, the explicit breaking is proportional to the brane's tension.
For example, if one wraps an NS5-brane along a two-cycle $\Sigma_2$ in the compactification manifold,
the axion field $c$, defined by
\begin{equation}
c \equiv \frac{1}{\ell_s^2}\int_{\Sigma_2} C_2\,,
\end{equation}
experiences monodromy.  The four-dimensional action for the dimensionless field $c$ takes the form \cite{McAllister:2008hb}
\begin{equation}
{\cal{L}} = -\frac{1}{2}f^2 (\partial c)^2 - \varepsilon \mu_0^3 f c\,,
\label{eq:simpleL1}
\end{equation}
where $f$ is the axion decay constant,\footnote{The axion decay constant depends on the topology and geometry of the six-dimensional compact manifold $X_6$: see Appendix \ref{app:dictionary}.} $\mu_0$ is a parameter with dimensions of mass, determined by the geometry of $X_6$, and $\varepsilon$ parameterizes the warp factor at the location of the NS5-brane.  (In terms of the warped line element (\ref{eq:warpedelt}), we have $\varepsilon=e^{4A_\labbot}$.)
We will explain the potential (\ref{eq:simpleL1}) in more detail in \S\ref{sec:fivebrane}.

Defining the canonically normalized axion $\phi \equiv f c$, we have
\begin{equation}
{\cal{L}} = -\frac{1}{2}(\partial\phi)^2 - \varepsilon \mu_0^3 \phi\,.
\label{eq:simpleL}
\end{equation}
Comparing to the relaxion potential (\ref{eq:relaxpotential}), we have the correspondence
\be
g M^3 = \varepsilon \mu_0^3\,.
\ee
So $\varepsilon \ll 1$ corresponds  to $g \ll 1$ in the relaxion model.
Since the breaking of the shift symmetry $\phi \mapsto \phi + \lab{const.}$ is proportional to the warp factor at the location of the fivebrane,
strong warping could lead to weak breaking of the symmetry, and hence to the small values of $g$ required for a relaxion model.

The potential (\ref{eq:simpleL}) has the desirable property that the entire potential is proportional to the warp factor, so it appears completely natural to make this potential small.  However, a central observation of this paper is that achieving small $g$ through warping, without unintended consequences elsewhere in the action, is challenging.

\begin{figure}
\centering
\includegraphics[width=1\textwidth]{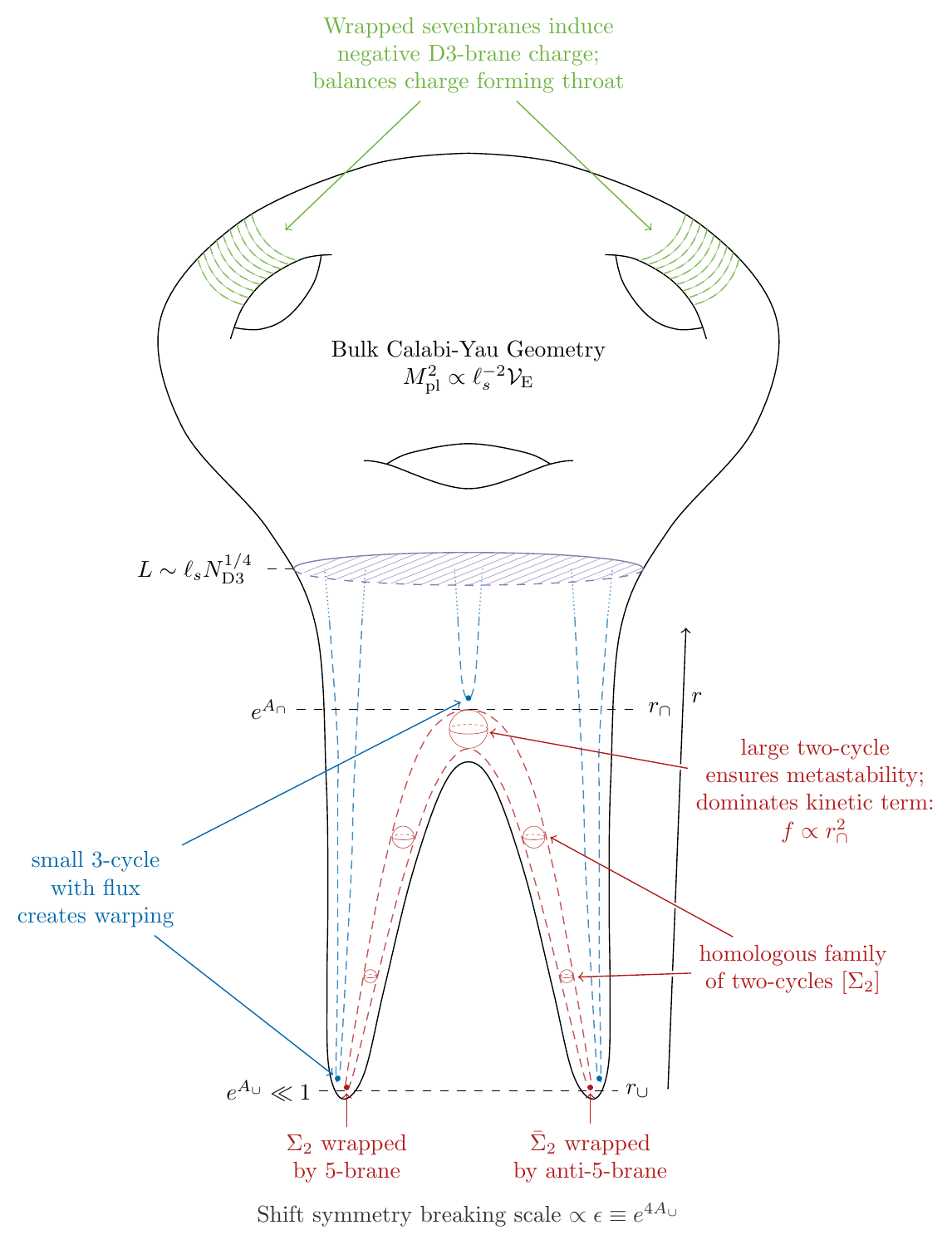}
\caption{Minimal bifurcated warped throat setup for relaxion monodromy with 5-branes.
}
\label{fig:fdoublethroat}
\end{figure}

\subsubsection*{Requirements for Axion Monodromy}\label{sec:obstacles}

Let us first summarize the core ingredients mentioned above.  For a model of axion monodromy in string theory, one requires:

\begin{enumerate}
\item[1.]  An axion field descending from a $p$-form, and a {\it{source of monodromy}}: a brane, flux, or other physical ingredient that causes the configuration space to be a multi-cover of the axion circle, rather than just a single circle.
\item[2.]  To have a plausible mechanism for making the breaking of the shift symmetry {\it{weak}}, the source of monodromy should be in a {\it{warped region}}.
\item[3.]  Most of the issues that arise as possible obstacles become visible only in vacua with stabilized moduli: if one ignores the moduli sector, many problems disappear.
But, of course, moduli stabilization is needed for a cosmological model.  So the axion and the source of monodromy must be situated in a {\it{vacuum with stabilized moduli}}.
\item[4.]  Since the compactification must have finite volume in order to lead to a finite four-dimensional Newton constant,
Gauss's law imposes strict constraints on the charges in the compact space $X_6$, and so we must {\it{satisfy all tadpole conditions}}.
\end{enumerate}

There are many mechanisms in the literature that achieve (1), for instance \cite{Silverstein:2008sg}. But there is only one model currently available that achieves (1)-(3) \cite{McAllister:2008hb,Flauger:2009ab}: this is a model with an NS5-brane/anti-NS5-brane pair in a warped throat region of a type IIB flux compactification whose complex structure moduli are stabilized by fluxes, and whose K\"ahler moduli are stabilized by nonperturbative effects and possibly also by perturbative effects.  We will call this model, whose detailed properties we will review in \S\ref{sec:fivebrane}, the {\it{NS5-brane model}}.

The central physics of the NS5-brane model is that transporting the dimensionless axion over a period induces one unit of D3-brane charge on the NS5-brane, and one unit of anti-D3-brane charge on the anti-NS5-brane.  That is, ``winding up'' the axion by one cycle develops a D3-brane dipole in the compact space; the axis of the dipole is the line from the NS5-brane to the anti-NS5-brane.  The entire dipole is in the infrared region of the warped throat where the fivebrane pair lives. See Figure \ref{fig:fdoublethroat}.

The key point is that the Lagrangian (\ref{eq:simpleL}) arising from the NS5-brane DBI action, which is intended to be the relaxion Lagrangian, holds in the so-called \emph{probe approximation}.  That is, the potential in (\ref{eq:simpleL}) follows from including the tension of the D3-branes and anti-D3-branes as a contribution to the four-dimensional vacuum energy, i.e.~as a source in the four-dimensional Einstein equations, but {\emph{not}} including this tension as a source in the ten-dimensional Einstein equations.  The effects of a particular source on the ten-dimensional field configuration are termed the \emph{backreaction} of that source, and so the probe approximation consists of neglecting the backreaction of D3-branes and anti-D3-branes.\footnote{For brevity we will often speak of ``D3-branes,'' ``D3-brane backreaction,'' etc., with the understanding that both D3-branes and anti-D3-branes are included.}

An immediate question is whether applying the probe approximation is consistent; in other words, can the backreaction of D3-branes be neglected?  In the context of axion monodromy inflation in string theory, this question has been addressed, with the outcome that backreaction can be suppressed to some degree, by a variety of mechanisms, but nevertheless remains as a leading constraint on model-building \cite{Flauger:2009ab}.  However, the sources of backreaction are the D3-brane charge and tension, both proportional to the number of windings $\n$ of the axion.  In the present context of relaxion monodromy, $\n$ needs to be extremely large, and so the problem of backreaction is much more severe than in the corresponding inflationary models.  The constraints examined in \cite{Flauger:2009ab} must therefore be revisited under this more severe test.

In this work, we will carefully examine the consequences of D3-brane backreaction for the NS5-brane model of relaxion monodromy in string theory.
The first step is to explain how to compute the backreaction in this scenario.

\subsection{Fivebrane axion monodromy}\label{sec:fivebrane}

Our analysis will rely on detailed properties of the action for NS5-branes wrapping curves in a warped region of a type IIB flux compactification, so we now give some essential background.  We will begin by discussing D5-branes, to facilitate comparison with the string theory literature, even though our eventual interest will be NS5-branes.

The action of a D5-brane
is the sum of a Dirac-Born-Infeld term related to the worldvolume $\mathcal{W}$ of the brane,
\begin{equation} \label{dbi}
S_{\rm{DBI}}=-g_s T_5 \int_{\mathcal{W}} \!\ud^{6}\sigma \,e^{-\Phi}\sqrt{-{\rm{det}}(G_{ab}+{\cal{F}}_{ab})}\,,
\end{equation}
and a Chern-Simons term encoding the coupling of the D5-brane to
the Ramond-Ramond $p$-form potentials $C_0$, $C_2$, $C_4$, and $C_6$,
\begin{equation} \label{cs}
S_{\rm{CS}} =  \mu_5 \int_{\mathcal{W}} \sum_{p} C_p \wedge e^{{\cal{F}}} \,,
\end{equation}
with ${\cal{F}}=B + 2 \pi \alpha' F$.
Here $g_s$ is the string coupling, $T_5$ is the D5-brane tension, $G_{ab}$ is the metric induced on the D5-brane, $\mu_5$ is the D5-brane charge, and $\cal{F}$ is
the gauge-invariant two-form field strength on the D5-brane.  The integral in (\ref{cs}) picks out the six-forms $C_6$, $C_4 \wedge {\cal{F}}$, $C_2 \wedge {\cal{F}}\wedge {\cal{F}}$, and $C_0 \wedge {\cal{F}}\wedge {\cal{F}} \wedge {\cal{F}}$.

Now suppose that $\mathcal{W} = {\cal{M}}^{3,1} \times \Sigma_2$, with $\Sigma_2$ a two-cycle in the internal six-manifold $X_6$.
If the field strength ${\cal{F}}$ obeys\footnote{We
define the string length to be $\ell_s \equiv 2 \pi \sqrt{\alpha'}$.}
\begin{equation}
\frac{1}{\ell_s^2} \int_{\Sigma_2} {\cal{F}} = \n \in \mathbb{Z}\,,
\end{equation}
then the Chern-Simons coupling becomes
\begin{equation} \label{d3cscoupling}
  \mu_5 \int_{\mathcal{W}} \!C_4 \wedge {\cal{F}}  \to  \n \mu_3 \int_{{\cal{M}}^{3,1}} \!C_4.
\end{equation}
The interaction (\ref{d3cscoupling}) is precisely $\n$ times the Chern-Simons coupling of a single D3-brane to the Ramond-Ramond four-form potential $C_4$, under which the D3-brane is electrically (and also magnetically) charged.  The coupling (\ref{d3cscoupling}) should be understood as a generalization of the worldline coupling
\begin{equation} \label{emL}
{\cal{L}}_{\rm{int}} = -\frac{e}{c}\int \! A_{\mu}\, \ud x^{\mu}
\end{equation} in electromagnetism.  In particular, (\ref{d3cscoupling}) shows that a D5-brane wrapping $\Sigma_2$, with
$\n$ units of ${\cal{F}}$ flux on  $\Sigma_2$, carries $\n$ units of D3-brane charge.  Equivalently, the D5-brane can be said to contain $\n$ D3-branes dissolved in the D5-brane.
This fact, while well-known, will be crucial for our considerations.

The $\Sigma_2$-wrapping D5-brane
can fluctuate in the space orthogonal to $\mathcal{M}^{3,1} \times \Sigma_2$. We denote these
corresponding canonically-normalized fluctuations as $X^i$. Defining the dimensionless field
\begin{equation}
b \equiv \frac{1}{\ell_s^2} \int_{\Sigma_2} B_2\,,
\end{equation}
we may expand the DBI action (\ref{dbi}) to second order in these fluctuations,\footnote{As in Appendix \ref{sec:conventions}, we denote $\mathcal{M}^{3,1}$ indices with $\mu, \nu$, etc.; $\Sigma_2$ indices with $a,b$, etc.; directions orthogonal to $\mathcal{M}^{3, 1}\times \Sigma_2$ with indices $i, j$, etc.; and we parameterize $\Sigma_2$ using the coordinates $y$ and $z$, with $\ud y \wedge \ud z = \ud^2 z$; see Table \ref{tab:indexguide}.}
\begin{equation}\label{eq:reducedDBI}
	S_\lab{DBI} = -\frac{T_5}{2} \int\!\ud^4 x\, \ud^2 z\, \sqrt{-g_4} \Bigg(\underset{\circled{1}}{\vphantom{\frac{1}{2} \frac{\tilde{g}_2}{\tilde{g}_2 + \ell_s^4 b^2} \partial_m X^i \partial^m X^i}\sqrt{4\tilde{g}_2 + \ell_s^4 b^2}}  + \underset{\circled{2}}{\vphantom{\frac{1}{2} \frac{\tilde{g}_2}{\tilde{g}_2 +  b^2} \partial_m X^i \partial^m X^i}\partial_\mu X^i \partial^\mu X^i} + \underset{\circled{3}}{ \frac{4\tilde{g}_2}{4\tilde{g}_2 + \ell_s^4 b^2} \partial_a X^i \partial^a X^i} +\dots \!\Bigg),
\end{equation}
where  $\tilde{g}_2$ is the determinant of the metric on $\Sigma_2$. Upon integrating over $\Sigma_2$ and denoting its volume as $\ell^2$, \circled{1} yields a four-dimensional potential for $b$
\begin{equation}
V(b) = \frac{\varepsilon}{(2\pi)^3 \alpha'^2} \sqrt{\left(\frac{\ell}{\ell_s}\right)^4 + \frac{b^2}{4}},
\label{eq:DBIpotential}
\end{equation}

In the absence of a wrapped D5-brane, $b$  would enjoy an approximate continuous shift symmetry, $b \mapsto b + \lab{const.}$, that is broken to a residual exact discrete shift symmetry, $b \mapsto b + 1$, by instanton effects.\footnote{Moduli-stabilizing effects further break this symmetry, as explained in \cite{McAllister:2008hb}.}
However, the potential (\ref{eq:DBIpotential}) induced by the D5-brane {\emph{completely breaks}} this symmetry.  In fact, the D5-brane introduces a monodromy, in that upon traversing the axion circle, from $b \mapsto b+1$, the potential energy is increased, rather than being periodic.  For large $b$, the potential (\ref{eq:DBIpotential}) becomes linear, as claimed in (\ref{eq:simpleL}) for the related case of an NS5-brane.

The strength of this symmetry breaking is proportional to $\varepsilon$, the \emph{warp factor} $\varepsilon = e^{4 A_\labbot}$ at the location of the fivebrane. In a warped compactification, the ten-dimensional metric takes the form
\begin{equation}  \label{eq:warpedelt}
	\ud s_{10}^2 = e^{2 A(y)} g_{\mu \nu} \ud x^\mu\, \ud x^\nu + e^{-2 A(y)} \tilde{g}_{m n} \, \ud y^m \, \ud y^n.
\end{equation}
By placing the fivebranes in a warped throat, the energy of this shift symmetry breaking can be gravitationally redshifted to an energy much smaller than the natural scale of breaking due to unwarped fivebranes.

The monodromy is closely related to the induced D3-brane charge (\ref{d3cscoupling}).  Starting from an initial configuration with $b=b_0$ and moving to $b=b_0+ \n$ (for $\n>0$) corresponds to shifting
\be
{\cal{F}} \mapsto {\cal{F}} + \n \omega_2\,,
\ee with $\omega_2$ a two-form obeying $\ell_s^{-2}\int_{\Sigma_2}\omega_2=1$.  This is an increase, of $\n$ units, of the gauge-invariant field strength ${\cal{F}}$.  This change is manifest in the potential (\ref{eq:DBIpotential}), which increases linearly.  The change is also visible in the D3-brane charge carried by the D5-brane, which increases by $\n$ units.  We refer to this process as ``winding up the axion $\n$ times.''

A justifiable complaint at this stage is that in a compact space, the total D3-brane charge should be fixed: in fact it must vanish by Gauss's law.
So winding up the axion would appear to be forbidden.
However,
to cancel the D5-brane tadpole, we may
suppose that in addition to the D5-brane wrapping $\Sigma_2$, there is an anti-D5-brane also wrapping $\Sigma_2$.  The anti-D5-brane Chern-Simons coupling differs from the D5-brane Chern-Simons coupling (\ref{cs}) by an overall minus sign.  Thus, winding up the axion $\n$ times induces $\n$ units of D3-brane charge on the D5-brane, as well as $-\n$ units of D3-brane charge on the anti-D5-brane, so that no net D3-brane charge is produced, and if Gauss's law is obeyed in the initial configuration, it is also obeyed after winding.

A coincident D5-brane and anti-D5 brane will quickly annihilate.  However, if a D5-brane wraps $\Sigma_2$, and an anti-D5-brane wraps a two-cycle  $\overline{\Sigma}_2$ that is \emph{homologous} to $\Sigma_2$, but is not coincident with $\Sigma_2$, then the D5-brane/anti-D5-brane configuration can be metastable and cosmologically long-lived \cite{Aganagic:2006ex}.  Because the induced D3-brane charges are determined by the homology classes of $\Sigma_2$ and $\overline{\Sigma}_2$, if $[\Sigma_2]-[\overline{\Sigma}_2]$ is trivial in homology then no net D3-brane charge is induced, just as in the case of a strictly coincident D5-brane/anti-D5-brane pair, and Gauss's law does not preclude winding up the axion.

Let us summarize the physics of $B_2$ monodromy from a wrapped D5-brane.  The D5-brane is a source of monodromy and gives rise to the non-periodic potential (\ref{eq:DBIpotential}).  The order parameter measuring the distance from the origin in the $b$ field space is the number of windings, $\n \in \mathbb{Z}$, which also counts the D3-brane charge induced on the D5-brane.  This is the \emph{monodromy charge} in the fivebrane model.
Winding up corresponds to moving away from the origin in field space and storing energy in the form of the D3-branes dissolved in the D5-brane, and anti-D3-branes dissolved in the anti-D5-brane: that is, the energy is stored in the monodromy charge.

The potential (\ref{eq:DBIpotential}) is that of a \emph{probe} D5-brane, in the same sense that (\ref{emL}) includes the potential energy of an electron in a background electromagnetic field.  However, just as (\ref{emL}) also encodes the fact that electrons source electromagnetic fields, the couplings (\ref{dbi}) and (\ref{cs}) encode the effects that a D5-brane has on the background fields.  To determine this \emph{backreaction} of the D5-brane on the bulk field, including the metric and the $p$-form fields, we simply include the couplings (\ref{dbi}) and (\ref{cs}) when varying the ten-dimensional action with respect to these fields $\varphi$.
Schematically,
\be
0 = \frac{\delta}{\delta\varphi} S_{\lab{10d},{\rm{bulk}}} + \frac{\delta}{\delta\varphi} S_{\rm{DBI}} + \frac{\delta}{\delta\varphi} S_{\rm{CS}}\,.
\ee
Any D5-brane serves as a source for the ten-dimensional metric (it has tension), and as a source for $C_6$.  But a D5-brane with
\be
\frac{1}{\ell_s^2}\int_{\Sigma_2} {\cal{F}} = \n \neq 0
\ee
also serves as a source for $C_4$; this is just to say that such a D5-brane carries D3-brane charge. The DBI action (\ref{dbi}) may be interpreted as the product of the brane tension and its ``effective volume,'' which grows with $\n$. This growth has two principal effects. The mass of the five-brane is also, schematically, the product of its tension and this effective volume, and thus as $\n$ grows the charged D5-brane will more strongly source the ten-dimensional metric. Furthermore, there are Kaluza-Klein excitations arising from the dimensional reduction of (\ref{dbi}) whose masses \emph{decrease} as this effective volume grows; indeed, the dimensional reduction of \circled{2} and \circled{3} in (\ref{eq:reducedDBI})---which correspond to the transverse fluctuations of the five-brane---yield Kaluza-Klein modes with masses $m_\lab{b KK}$ that are smaller than the naive estimate $m_\lab{KK} \propto \ell^{-1}$ by a factor of (see Appendix \ref{sec:eframe})
\begin{equation}
	\frac{m_\lab{b KK}}{m_\lab{KK}} \sim \frac{\ell^2}{\sqrt{\ell^4 + \ell_s^4 b^2}}\,.
\end{equation}

Axions descending from
$B_2$
generically suffer an $\eta$ problem \cite{McAllister:2008hb}, meaning that in expansion around a vacuum with stabilized moduli, the actual potential for the axion, taking into account all couplings to moduli, is very different from the potential (\ref{eq:DBIpotential}) that arises from the probe D5-brane action alone.
This problem can be
ameliorated by considering an axion descending from the Ramond-Ramond two-form $C_2$ and exchanging the D5-branes in the above discussion for NS5-branes. The analogous potential is then given by
\begin{equation}
	V(c) = \frac{\varepsilon}{(2\pi)^3 g_s \alpha'^2} \sqrt{\left(\frac{\ell}{\ell_s}\right)^4 + \frac{g_s^2 c^2}{4}}.
	\label{eq:NS5potential}
\end{equation}

	As we will argue in \S\ref{sec:genrelaxmon}, for a construction of a relaxion model via fivebrane axion monodromy in string theory one needs an extremely large winding $\n \gg 1$.  There is a correspondingly large induced D3-brane charge, the effect of which must be included in the ten-dimensional field equations. Backreaction of this charge and its effect on the five-brane cannot be neglected: the potential for the axion is no longer simply given by (\ref{eq:DBIpotential}), and new light modes appear.

\subsection{Fivebrane relaxion monodromy} \label{sec:genrelaxmon}

To
understand string theoretic constraints on the relaxion mechanism, we require an embedding of the four-dimensional potential
	\begin{equation}
		V(\phi, h) = \underset{\circled{A}}{\Bigl(M^2 - g_h M \left(\phi_\lab{init} - \phi\right)\Bigr) |h|^2} + \underset{\circled{B}}{\vphantom{\Bigl(M^2 - g_h M \left(\phi_\lab{init} - \phi\right)\Bigr) |h|^2}g M^3\phi} + \underset{\circled{C}}{\vphantom{\Bigl(M^2 - g_h M \left(\phi_\lab{init} - \phi\right)\Bigr) |h|^2}V_\lab{stop}(\phi, h)}\,,
		\label{eq:relaxpotential2}
	\end{equation}
or of something functionally equivalent,
	in a well-controlled compactification of string theory. As noted in the introduction, the ratio $g_h/g$ need not be $\mathcal{O}(1)$, and so in (\ref{eq:relaxpotential2}) we distinguish between the two.
	
	\begin{figure}
		\centering
		\includegraphics[width=0.5\textwidth]{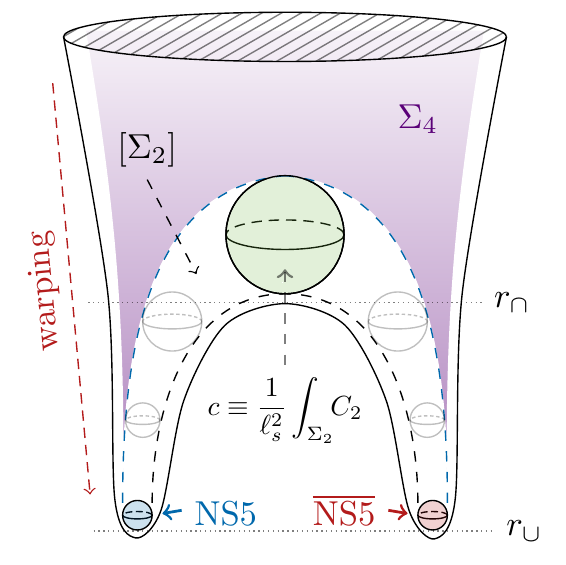}
		\caption{
			Ten-dimensional realization of \circled{B} and \circled{C} of (\ref{eq:relaxpotential2}). \circled{C} is generated by strong gauge dynamics on seven-branes wrapping a divisor $\Sigma_4$, which must necessarily intersect the minimum volume representative $[\Sigma_2]$ wrapped by the NS5-/anti-NS5-brane. \label{fig:intersect}
		}
	\end{figure}
	
	In \S\ref{sec:fivebrane}, we focused on realizing \circled{B} as the potential energy of an NS5-/anti-NS5-brane pair wrapping the minimum volume representatives of the homology class $[\Sigma_2]$ associated with the axion $c = \ell_s^{-2} \int_{\Sigma_2} C_2$, where $\phi \equiv fc$.  \circled{B} provides a potential that is self-similar (ignoring backreaction effects) over a very large distance $\Delta \phi \gg f$ in field space.
Hubble friction eventually dominates and the late-time dynamics are independent of the initial conditions for $\phi$.

Crucially, the small parameter $g$ is controlled by the warp factor at the position $r_\cup$ of the five-branes. Specifically,
	\begin{equation}
		g M^3 f \equiv \frac{2 \pi}{\ell_s^4} e^{4 A_\labbot},
		\label{eq:monodromyenergy}
	\end{equation}
	where $e^{4 A_\labbot}$ is the warp factor at $r_\labbot$, the location of the fivebranes and the bottom of the ``tooth'' in Figure \ref{fig:intersect}.  We may think of the ``roots'' of the tooth as Klebanov-Strassler or similar warped throat geometries.
Away from the tip, the warp factor is roughly $e^{4 A} \sim r^{4}/L^4$, with $L$ the characteristic size of the warped throat. A simple way to describe this warping is by the number of D3-branes it would take to form a similarly sized warped throat,
	\begin{equation}
		L^4 \sim g_s N_\lab{D3} \ell_s^4.
		\label{eq:LradiusNd3}
	\end{equation}
	As explained in \S\ref{app:dictionary}, the axion decay constant $f$ is determined by the radial position of the arch of the ``tooth,''
	\begin{equation}
		f^2 \sim g_s \frac{r_\labtop^2}{\ell_s^4},
	\end{equation}
	so the shift symmetry breaking scale is given by
	\begin{equation}
		g M^3 \sim \frac{2 \pi}{g_s^{3/2} \ell_s^3 N_\lab{D3}} \left(\frac{r_\labbot}{r_\labtop}\right) \left(\frac{r_\labbot}{\ell_s}\right)^3\!\!.
		\label{eq:sixdimmon}
	\end{equation}
	The cutoff scale $M$ depends on how the Higgs is realized and does not necessarily depend on the total D3-brane charge $N_\lab{D3}$.  However, regardless of where the Higgs is located in the
internal space,
the smallness of $g$ is necessarily tied to a large $N_\lab{D3}$. For example, if the Higgs sector is realized somewhere in the bulk geometry, then $M \propto N_\lab{D3}^0$ and $g \propto N_\lab{D3}^{-1}$, as in (\ref{eq:sixdimmon}).  If instead the Higgs sector is realized at the top of the warped throat at $r \sim L$ in Fig. \ref{fig:intersect}, then we may take $M^3 \sim L^{-3}$ and so
	\begin{equation}
		g \sim \frac{1}{(g_s N_\lab{D3})^{1/4}} \left(\frac{r_\labbot}{r_\labtop}\right) \left(\frac{r_\labbot}{\ell_s}\right)^3.
	\end{equation}
We will not consider a Higgs realized deep within the warped throat, as this would lead to an exponential suppression of $M$, corresponding to a supersymmetric resolution of the hierarchy.
	
 \begin{figure}[t]
\vspace*{-0.5cm}
\centering
\includegraphics[width=.75\textwidth]{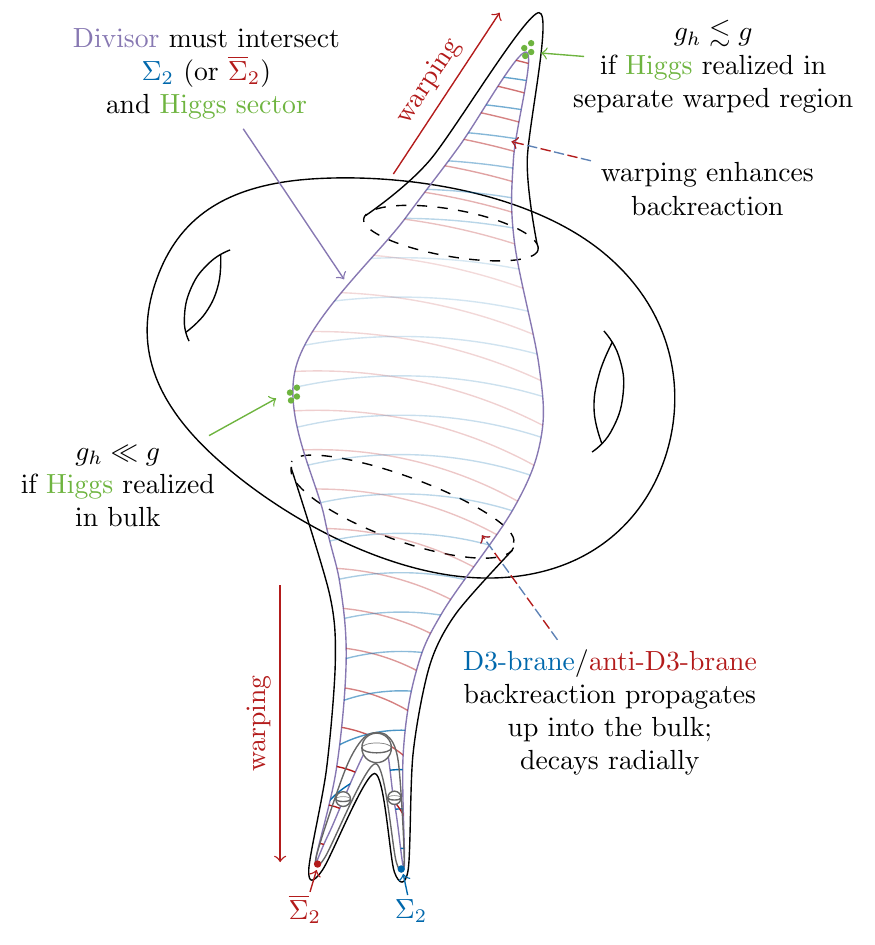}
\vspace*{0cm}
\caption{
Schematic structure of the extra dimensions, showing a string theory setup that realizes the main relaxion features and couplings. The central region is
the ``bulk'' of the extra dimensions, which does not experience position-dependent warping. The coupling $g_h$ depends on where in the bulk Calabi-Yau the Higgs sector is realized.
}
\label{fig:coreelements}
\end{figure}
		
	 We will be agnostic about the detailed origin of the Higgs coupling \circled{A}.  While its specific form would be relevant in a complete model, it is not needed to expose and quantify the issues
that concern us here, which mainly deal with the interplay between the linear potential and the stopping potential.
In the spirit of this agnosticism, we instead focus on the hierarchy generated between the string scale $M_s$ and the electroweak scale $v$.

Even so, a concrete picture of one possibility may be helpful.
The Higgs could arise from open strings stretching between stacks of D3-branes or D7-branes. The Higgs mass is then proportional to the distance between the $\lab{U}(1)_\lab{Y}$ brane and the $\lab{SU}(2)_\lab{W}$ stack. The coupling \circled{A} is generated by backreaction of the monodromy charge on the internal geometry, which changes distance between these branes and thus the Higgs mass, as in Fig.~\ref{fig:coreelements}. Because this backreaction decays as it propagates throughout the six-dimensional space, there may be an appreciable hierarchy between $g_h$ and $g$ which depends on where the Higgs sector is realized in the internal geometry.  We may mitigate this hierarchy somewhat by placing the Higgs in another warped throat---the backreaction will then be blue-shifted, leading to an increased coupling $g_h$---though, as explained above, placing the Higgs in a warped region will naturally suppress $M$.

Finally, for generic initial conditions, the relaxion traverses a distance $\Delta \phi \sim M/g_h$ in field space. This is associated with the dissipation of
		\begin{equation}
			\n \sim \frac{\Delta \phi}{ f} \sim \frac{M^4}{g_h M^3 f} \sim g_s N_\lab{D3} \left(\frac{g}{g_h}\right) \left(\frac{M}{M_s}\right)^4 \left(\frac{\ell_s}{r_\labbot}\right)^4
			\label{eq:windingreq}
		\end{equation}
units of monodromy charge.

\subsubsection*{Barriers from D7-branes}
	We will be more specific about how \circled{C} is realized. Perturbatively in $g_s$, the axion has a continuous shift symmetry $\phi \mapsto \phi + \lab{const.}$ which is broken by non-perturbative effects (in $g_s$) to the discrete shift symmetry $\phi \mapsto \phi + f$. If \circled{A} and \circled{B} are to be the only terms that
break this discrete shift symmetry, as is implicitly assumed in the relaxion construction, then $V_\lab{stop}$ must be generated non-perturbatively in $g_s$. As
noted above, we take
the Higgs and relaxion sectors
to be separated in the internal geometry, and so in order for the stopping potential to depend on both of these sectors, it must be generated by physics on one or more extended objects---by either Euclidean D$p$-branes or strong gauge dynamics on a stack of D$p$-branes.
	
For simplicity, we will assume that $V_\lab{stop}$ is generated by the strong dynamics of a gauge theory, with group $G$, realized on a stack of D7-branes wrapping a
holomorphic
four-cycle $\Sigma_4$, as illustrated in Figure \ref{fig:intersect}. The D7-branes couple to the $C_2$ axion through the Chern-Simons action
\begin{equation} \label{css}
	S_\lab{CS} \supset \mu_7 \int_{\mathcal{W}} \mathcal{F} \wedge C_2 \wedge \mathcal{F} \wedge \mathcal{F}\,.
\end{equation}
A key observation is that the D7-branes must enter the warped throat region (see Appendix \ref{d7pf} for a proof).
The coupling (\ref{css}) leads to a potential of the schematic form
	\begin{equation}
		V(\phi, v) = \Lambda_c^3\, v \cos \left(\frac{2\pi\phi}{f}\right),
		\label{eq:stoppingpotential}
	\end{equation}	
but can, in general, involve a more complicated polynomial of the Higgs vev $v$ and a general $f$-periodic function in $\phi$. The confinement scale $\Lambda_c$ is naturally related to the string scale and the D7-brane gauge coupling $g_\lab{YM}$, 
	\begin{equation}
		\Lambda_c^3 \propto \ell_s^{-3} \exp\left(-\frac{8\pi^2}{g_\lab{YM}^2\,c_G}\right)\,,
		\label{eq:barrierheight}
	\end{equation}
	where $c_G$ is a constant determined by the particular effects that generate (\ref{eq:stoppingpotential}). For example, $c_G$ is simply the dual Coxeter number of $G$ if the stopping potential is realized through gaugino condensation. In known examples, $c_G$ is at most $\mathcal{O}(10^{2})$, and we will take $c_G = 1$ henceforth.
	The generated hierarchy between the string and electroweak scales is then
	\begin{equation}
		\frac{M_s}{v}  \propto g_s N_\lab{D3} \exp\left(-\frac{8 \pi^2}{g_\lab{YM}^2}\right) \left(\frac{\ell_s}{r_\labbot}\right)^4\!.
		\label{eq:stringhierarchy}
	\end{equation}
	Since $r_\labbot \gtrsim \ell_s$, the hierarchy is controlled by the warp factor and at first sight appears to be proportional to $N_\lab{D3}$.  Thus, an arbitrarily large hierarchy could apparently be realized via substantially warping the source of monodromy. However, as we will discuss in \S\ref{analysis}, this is too naive.
		
\newpage
\section{Microphysical Constraints} \label{analysis}

Relaxation of a hierarchy by the relaxion mechanism occurs only in theories that meet several stringent requirements.  Arguably the most challenging requirements from the viewpoint of ultraviolet completion in string theory are both the \emph{large displacement} $\Delta \phi \sim M/g$, and the comparatively \emph{short stopping length}.  That is, the relaxion must evolve slowly over a large distance, gradually reducing the Higgs mass, but then rapidly come to rest after the Higgs acquires a vev.
These disparate distance scales in field space correspond to very different energy scales in the potential: the final Higgs vev $v$ is determined by the ratio
of the shift symmetry breaking scale $g M^3 f$ to the stopping potential scale $\Lambda_c^3$, cf.~(\ref{eq:hierarchy}).
In field theory, one can obtain a controllably large hierarchy by taking $g$ to be extremely small while holding $\Lambda_c$ fixed.
		
This limit is problematic in string theory. As we will show in \S\ref{sec:warpsupp}, the scale $\Lambda_c$ depends on $g$, and is 
exponentially suppressed as $g \to 0$. This dramatically limits the hierarchy that can be generated.

At the same time, the large field excursion on its own implies that the initial configuration carries a very large monodromy charge.  This gradually dissipating monodromy charge will serve as a changing source for the ten-dimensional equations of motion.   For $\n \gg 1$, this backreaction has profound effects on the compactification geometry and so on the four-dimensional relaxion potential (\ref{eq:relaxpotential2}).

It is tempting to argue that all of the corrections that result from backreaction must ultimately originate in the breaking of the axionic shift symmetry, and so must involve powers of the shift-symmetry breaking 
 parameter $g$.  This is \emph{not} correct.
In
the NS5-brane model, the breaking parameter $g$ is small because the DBI action of an NS5-brane is proportional to the warp factor at the NS5-brane location, cf.~(\ref{eq:monodromyenergy}).  Backreaction effects sourced directly through the DBI action are indeed proportional to powers of $g$.  However, the NS5-brane Chern-Simons action is not warped, and could not be: it is topological, and counts the (integer) D3-brane charge induced on the NS5-brane, i.e.~the monodromy charge $\n$.

Thus, backreaction effects sourced by the Chern-Simons action are proportional to $\n$, without factors of $g$.  For example, the integral of the R-R field strength $F_5$ over a Gaussian surface---say, an $\lab{S}^5$---surrounding the NS5-brane is simply given by $\n$, even as $g \to 0$.
One consequence, as we shall see, is that the monodromy charge provides a large correction to the stopping potential.\footnote{The backreaction sourced by this topological term does not need to propagate far to be ``detected,'' i.e.~to influence a significant term in the four-dimensional Lagrangian: see Appendix \ref{d7pf}.}

In this section we provide an array of calculations that reveal the concrete obstacles to achieving a large displacement and a short stopping length in the NS5-brane model.

\subsection{Overview of microphysical constraints} \label{Overview}

We first preview a number of constraints on relaxion monodromy constructions, which originate from microphysical limitations on string compactifications that provide the desiderata listed in \S\ref{sec:obstacles}.  Each of these constraints will be detailed in turn in \S\S\ref{sec:tadpole}-\ref{sec:fbackreaction}.
 
\begin{enumerate}
\item[\S\ref{sec:universal}] {\bf{Universal effect on the geometry.}} The shape of the warped throat region is dramatically altered by backreaction, leading to large changes in the effective action.

\item[\S\ref{sec:tadpole}] {\bf{Tadpole constraints}}. To accommodate $\n \gg 1$ units of monodromy charge without the loss of perturbative control, we must construct a background throat with $N_\lab{D3} \gg \n \gg 1$.  Gauss's law---i.e., the D3-brane charge tadpole---then implies that there must be a source that is equivalent to $-N_\lab{D3}$ D3-branes.
    To avoid the instabilities created by a large number of actual anti-D3-branes, this source must be supersymmetric, and arise from the topology of an elliptically-fibered fourfold: the D3-brane charge is then $-\chi/24$, where $\chi$ is the Euler number of the fourfold.
    The largest known Euler number of an elliptically-fibered fourfold is $1$,$820$,$448$.  So in this setting, $\n$ will have to be much smaller than $75$,$852$.

\item[\S\ref{sec:warpsupp}] {\bf{Barrier suppression from warping}}.  The D7-branes that generate the stopping potential must wrap a four-cycle $\Sigma_4$ that intersects the minimum volume two-cycle $\Sigma_2$.  The D7-brane gauge coupling function, which depends on the \emph{warped} four-volume of $\Sigma_4$, is then directly suppressed by the same warping responsible for the miniscule monodromy energy scale (\ref{eq:monodromyenergy}).
    A weakly broken shift symmetry therefore leads to extremely small barriers.

\item[\S\ref{sec:D7backreaction}] {\bf{Barrier suppression from backreaction}}. The induced D3-brane charge and tension backreact on the warped four-volume of $\Sigma_4$ and therefore perturb the D7-brane gauge coupling function. This perturbation introduces an exponential dependence of the gauge coupling on the monodromy charge $\N$, with \emph{no} powers of $g \sim N_\lab{D3}^{-1}$. This contradicts naive applications of technical naturalness: the dangerous term that arises is not negligible in the limit $g \to 0$ where the shift symmetry breaking is weak.

\item[\S\ref{sec:vmod}] {\bf{Effects on the moduli potential}}. The sources  responsible for K\"ahler moduli stabilization are exponentially sensitive to perturbations of the warp factor.  So the moduli potential depends on the relaxion field, i.e.~there are new terms in the relaxion potential not captured by (\ref{eq:relaxpotential2}).  This was extensively studied in \cite{Flauger:2009ab}.

\item[\S\ref{sec:fbackreaction}] {\bf{Effects on the axion decay constant}}. Large backreaction will also affect the axion decay constant, which depends on the volume of the cycle the axion threads as well as on the overall volume of the internal manifold.

\item[\S\ref{sec:classicalann}] {\bf{Classical annihilation of the dipole}}. The compactification detailed in \S\ref{sec:fivebrane} is metastable. The NS5-brane and anti-NS5-brane attract one another because of the induced D3-brane charge that each carries, but the fivebranes must stretch over a large-volume representative of $[\Sigma_2]$ in order to meet one another.  This costs energy, because the fivebranes have tension.  For modest windings $\N$, the tension energy can be much larger than the Coulomb energy from the D3-branes and anti-D3-branes, and the system is controllably metastable. However, for $\N \gg 1$, the Coulomb energy can overpower the tension energy, and the fivebrane/anti-fivebrane dipole can classically annihilate.

\item[\S\ref{sec:kpvinstability}] {\bf{Constraints from anti-D3-brane annihilation}}. An anti-D3-brane at the tip of a large Klebanov-Strassler throat is a metastable and cosmologically long-lived configuration. However, the barrier that ensures metastability depends on the number of anti-D3-branes in the throat. For some number $\n_\lab{KPV}$ of anti-D3-branes---and thus for windings $\n \geq N_\lab{KPV}$---the barrier disappears and the anti-D3-branes can classically annihilate against the flux of the throat.
    Thus, the accumulation of anti-D3-branes on the anti-NS5-brane creates a risk of instability.

\item[\S\ref{sec:sequester}] {\bf{Tunneling via light brane KK modes}}.  The accumulation of D3-branes in the NS5-brane pair leads to a reduction in the tension of the NS5-branes, and correspondingly a reduction in the mass of Kaluza-Klein excitations of the NS5-branes.  This Kaluza-Klein spectrum has spacing proportional to $m_0/\n$ when the axion is wound up by $\n$ cycles, with $m_0$ associated to the IR scale of the warped throat.  These light brane KK modes provide another pathway for classical annihilation of the dipole. If the throat is put at some temperature, say from a source of supersymmetry breaking elsewhere in the internal space, thermal fluctuations of the light brane KK modes could enable the NS5-branes to reach up towards one another, allowing for a quantum mechanical tunneling event.

\end{enumerate}

\subsection{Consequences of D3-brane backreaction}\label{sec:NwND3}

D3-branes and anti-D3-branes source warping, and so the D3-brane dipole that develops when the axion is wound up leads to a change in the local warp factor. The warped throat region is itself produced by some number $N_\lab{D3}$ of D3-branes that have dissolved into flux, and when the number of windings $\n$ becomes comparable to $N_\lab{D3}$, the D3-brane dipole is a large correction to the background in which it is sitting. The probe approximation is not valid for such a configuration, and the backreaction of the D3-brane dipole affects many couplings in the four-dimensional theory.

\subsubsection{Universal effect on geometry} \label{sec:universal}

The backreaction of the tension and charge of $\n$ induced D3-branes will be a small perturbation to the overall configuration as long as the ratio $g_s  \ell_s^4\n/L^4$ is small, where $L$ is the radius of the warped throat: see Appendix \ref{sec:internalbr}.
Using $N_\lab{D3}$ to denote the effective D3-brane charge of the warped throat (\ref{eq:LradiusNd3}) we must require that
\begin{equation}
	\n \ll N_\lab{D3}.
	\label{eq:nlld3}
\end{equation}
To intuitively motivate (\ref{eq:nlld3}), we may replace the $\n$ D3-branes with an $\lab{AdS}_5$ warped throat with radius
\begin{equation}
	R_{\n}^4 \sim g_s \ell_s^4 \n
\end{equation}
via a geometric transition. The perturbed geometry will be drastically different unless size of this extra throat is much smaller than the original warped throat, $R_{\n}^4 \ll L^4$. So, we require that $\n \ll N_\lab{D3}$ in order to maintain perturbative control.

The volume of the warped throat is necessarily bounded by the total volume of the internal space, $L^6 \lesssim \ell_s^6 \mathcal{V}_\lab{E}$.\footnote{We denote the total volume of the internal space $X_6$, measured in Einstein frame, as $\ell_s^6 \mathcal{V}_\lab{E}$,} which determines the four-dimensional Planck mass via $M_\lab{pl}^2 \ell_s^2 = 4 \pi \mathcal{V}_\lab{E}$. From (\ref{eq:nlld3}) we find the constraint
\begin{equation}
	\n \ll \frac{1}{g_s} \frac{M_\lab{pl}}{M_\lab{KK}}\, ,
\end{equation}
where $M_\lab{KK}=1/(\ell_s {\cal V}_E^{1/6})$.
This imposes a constraint on the number of windings for reasonable hierarchies between the compactification and Planck scales, and for reasonable values of $g_s$.

\subsubsection{Tadpole constraint}\label{sec:tadpole}

The higher-dimensional equations of motion must be satisfied in a consistent string compactification. In particular, the higher-dimensional analog of Gauss's law for the five-form flux $F_5$ becomes a powerful constraint on the ten-dimensional configuration. In a non-compact manifold, flux lines are allowed to extend to infinity and Gauss's law places no constraint on the amount of charge allowed in a given configuration. However, in a compact manifold a flux line must end on a charge and Gauss's law provides a \emph{tadpole constraint}: the total amount of D3-brane charge in the compactification must vanish. As discussed above, the warped throats pictured in Figure \ref{fig:fdoublethroat} are supported by a total of $N_\lab{D3}$ units of D3-brane charge. The tadpole constraint requires that this charge be canceled elsewhere in the Calabi-Yau geometry.

This cancellation could occur by including anti-D3-branes elsewhere in the internal space, or by forming another, oppositely charged, warped throat elsewhere with a large amount of negative D3-brane charge. In both cases, the D3-branes supporting the relaxion's warped throat and these additional anti-D3-branes will attract and the entire model will generically be unstable.

Fortunately, there exist well-known sources of supersymmetric negative D3-brane charge, and thus one may satisfy the tadpole constraint while maintaining stability. Seven-branes wrapping non-trivial cycles in the internal space provide curvature-induced negative D3-brane charge. F-theory compactified on elliptically-fibered Calabi-Yau fourfolds provides a framework for analyzing type IIB compactifications at arbitrary coupling, and the negative charge is related to the fourfold's Euler number $\chi(\lab{CY}_4)$ via
\begin{equation}
	N^\lab{CY_4}_\lab{D3} = -\frac{\chi(\lab{CY}_4)}{24}.
\end{equation}
The largest known Euler number of an elliptic-fibered Calabi-Yau fourfold is $\chi(\lab{CY}_4) = 1$,$820$,$448$ \cite{Klemm:1996ts}, which imposes the constraint
\begin{equation}
  N_\lab{D3} \leq \text{75,852}.
 \label{eq:chibound1}
\end{equation}
Requiring $\n \ll N_\lab{D3}$ to maintain control over the configuration, we then have the constraint
\begin{equation}
   \N \ll \text{75,852}.
 \label{eq:chibound}
\end{equation}
The bound (\ref{eq:chibound1}) on the Euler number of known fourfolds thus translates to a strong upper limit on the number of windings, and so constrains the maximum possible field excursion undergone by the relaxion.

The bound (\ref{eq:chibound}) applies only in the present case in which the monodromy charge is D3-brane charge.  However, in alternative axion monodromy scenarios, it would still be necessary to arrange that the background solution at zero winding carries a large background monodromy charge analogous to $N_\lab{D3}$.  In such a setting we expect topological upper bounds analogous to (\ref{eq:chibound1}) on the amount of monodromy charge that can be included without creating rapid instabilities.

\subsubsection{Suppression from warping} \label{sec:warpsupp}
		
		Relaxation of a large hierarchy requires that the shift symmetry is very weakly broken, with $g \ll 1$. In the ten-dimensional model of \S\ref{sec:genrelaxmon}, the breaking is made small by placing the source of monodromy---NS5-branes wrapping the minimum-volume two-cycles $\Sigma_2$ and $\overline{\Sigma}_2$---in a heavily warped region. However, we prove in Appendix \ref{d7pf} that supersymmetric D7-branes can generate a relaxion stopping potential only if the four-cycle $\Sigma_4$ they wrap intersects $\Sigma_2$ or $\overline{\Sigma}_2$.  So the D7-brane stack responsible for the stopping potential
necessarily descends into the warped region. As we will now see, elementary locality arguments show that the small parameter associated with this warping, $g$, in \circled{A} and \circled{B} of (\ref{eq:relaxpotential2}) then generically infects the stopping potential \circled{C} realized on the D7-brane stack, leading to an \emph{exponential suppression} of the stopping potential barriers.
		
		The gauge coupling $g_\lab{YM}$ on a spacetime-filling D7-brane wrapping a four-cycle  $\Sigma_4$ is proportional to the \emph{warped} four-volume of $\Sigma_4$ in string units,
		\begin{equation}
			\frac{1}{g_\lab{YM}^2} = \frac{1}{2 \pi \ell_s^4}\int_{\Sigma_4} \!\ud^4 \xi\, \sqrt{\tilde{g}_4}\, e^{-4A},
			\label{eq:d7g}
		\end{equation}
		with $\tilde{g}_4$ the induced, unwarped metric on $\Sigma_4$.  Defining a reference warp factor profile $\exp(4 \bar{A}) = r^4/L^4$, cf.~(\ref{eq:LradiusNd3}), we may express (\ref{eq:d7g}) as
		\begin{equation}
			g_\lab{YM}^{-2} = \alpha^{-1} g_s N_\lab{D3},
			\label{eq:g7N}
		\end{equation}
		where
		\begin{equation}
			\alpha^{-1} \propto \int_{\Sigma_4} \!\ud^4 \xi\, \sqrt{\tilde{g}_4} \,r^{-4} e^{-4(A - \bar{A})}
			\label{eq:geomfactor}
		\end{equation}
		is a dimensionless coefficient capturing the geometry of the embedding of $\Sigma_4$ in the warped throat.
		
		We may estimate $\alpha$ as follows. We have shown that $\Sigma_4$ must reach down the warped throat to intersect $\Sigma_2$ at $r_\labbot$. Assuming that $\Sigma_4$ roughly factorizes into a radial part and an angular part with volume $\breve{v}$, that it extends up into the bulk geometry as in Figure \ref{fig:coreelements}, and that the integral is dominated in the region where $A \sim \bar{A}$, we find
		\begin{equation}
			\alpha^{-1} \gtrsim \breve{v} \, \log \left(\frac{L}{r_\labbot}\right) \sim \breve{v}\,\log \left(\frac{g_s N_\lab{D3} \ell_s^4}{r_\labbot}\right).
		\end{equation}
 Importantly, $\alpha^{-1}$ is not naturally $\mathcal{O}(N_\lab{D3}^{-1})$, and in fact grows with the size of the throat, $L^4 \propto N_\lab{D3}$. So, $g_\lab{YM}^{-2} \sim \mathcal{O}(N_\lab{D3})$ unless the angular volume $\breve{v}$ is finely tuned to be exceptionally small, to one part in $g^{-1}$, which is of order the desired hierarchy.  In other words, fine-tuning the angular volume $\breve{v}$  to eliminate the effects of this warping amounts to constructing the entire hierarchy by this fine-tuning. This suppression
therefore renders the relaxation mechanism ineffectual.

From (\ref{eq:barrierheight}) and (\ref{eq:g7N}), the stopping potential
is \emph{exponentially suppressed} in $N_\lab{D3}$,
		\begin{equation}\label{eq:lambdags}
			\Lambda_c^3 \propto \ell_s^{-3} \exp\left(-\cbg N_\lab{D3}\right),
		\end{equation}
		with $\cbg \sim 8 \pi^2 /(g_s \alpha c_G)$. The hierarchy generated including this suppression is then
		\begin{equation}
			\frac{M_s}{v} \sim g_s \cbg^{-1} \left(\cbg N_\lab{D3} e^{-\cbg N_\lab{D3}}\right)\left(\frac{\ell_s}{r_\labbot}\right)^4,
			\label{eq:newhierarchy}
		\end{equation}
		and since $x e^{-x} \le e^{-1}$, the maximum resolvable hierarchy is simply
		\begin{equation}
			\frac{M_s}{v} \sim g_s \cbg^{-1} \sim \alpha\,c_G
			\label{eq:expsupphierarchy}
		\end{equation}
		which is, crucially, not $\mathcal{O}(g^{-1})$ unless $\alpha$ is severely fine tuned.

		Generically, the warping responsible for the suppression of the shift symmetry breaking energy scale
also suppresses the scale of the stopping potential. This suppression drives a runaway relaxion, and precludes the dynamical generation of a large hierarchy in the absence of an acute fine tuning.				
		
We expect this suppression to be very general.  We argued in \S\ref{sec:genrelaxmon} that the stopping potential must be generated by non-perturbative effects on a $(p+1)$-dimensional extended object, and Lorentz invariance requires this extended object to either fill spacetime and wrap an internal cycle ($p>3$) or else be instantonic. For a D$p$-brane wrapping a $p$-cycle $\Sigma_p$, the gauge coupling is given by
		\begin{equation}
			\frac{1}{g_\lab{YM, p}^2} = \frac{1}{2 \pi \ell_s^{p+1}} \int_{\Sigma_p}\!\!\ud^{p-3} \xi\, \sqrt{\tilde{g}_{p-3}} \, e^{(7-p)\Phi/4 - (p-3) A}.
		\end{equation}
		Similarly, for a Euclidean D$p$-brane wrapping the same cycle, the action is
		\begin{equation}
			S_\lab{EDp} = \frac{2 \pi}{\ell_s^{p+1}} \int_{\Sigma_p}\!\!\ud^{p+1} \xi\, \sqrt{\tilde{g}_{p+1}} \, e^{-(p-3)\Phi/4 -(p+1)A}.
		\end{equation}
		Both depend on powers of $e^{-A}$ and thus \emph{positive} powers of $N_\lab{D3}$. So, any potential barrier generated by these effects will suffer from the same exponential suppression, albeit with different powers of $N_\lab{D3}$.

		\subsubsection{Suppression from backreaction} \label{sec:D7backreaction}
 	
	The backreaction of D3-brane charge is out of  control unless the induced D3-brane charge $\n$ is a small fraction of the total D3-brane charge forming the throat, $\n/N_\lab{D3} \ll 1$, so that we may perform a perturbative expansion of the ten-dimensional field configuration in this ratio. We should thus expect corrections to (\ref{eq:relaxpotential2}) to involve powers of $\n/N_\lab{D3}$, which is consistent with the expectation that, because the monodromy charge is related to the shift symmetry breaking, any corrections due to  backreaction will come dressed with powers of $g$. Crucially, however, it is \emph{fractional} corrections to the field configurations---i.e.~$\delta\varphi/\varphi$ for some field $\varphi$---that involve powers of $\n/N_\lab{D3}$. If some quantity---say, a D7-brane gauge coupling function---also scaled with $N_\lab{D3} \propto g^{-1}$, the the absolute (additive) correction correction to this quantity is \emph{not} necessarily small when $\n/N_\lab{D3} \ll 1$.

Indeed, the monodromy charge induces a perturbation to (\ref{eq:d7g}),
		\begin{equation}
			\delta\left(\frac{8 \pi^2}{g_\lab{YM}^2}\right) \sim g_s N_\lab{D3} \int_{\Sigma_4}\!\!\ud^4 \xi\, \sqrt{\tilde{g}_4}\, e^{-4(A - A_0)} \,r^{-4} \underbrace{\left(\frac{1}{2} \frac{\delta \tilde{g}_4}{\tilde{g}_4} - \frac{\delta e^{4 A}}{e^{4A}}\right)}_{\mathcal{O}\left(\n/N_\lab{D3}\right)} \propto g_s \n \equiv \frac{\cbr \phi}{f}.
			\label{eq:g7br}
		\end{equation}
		 As discussed in detail in Appendix \ref{sec:internalbr}, the fractional perturbations are $\mathcal{O}(\n/N_\lab{D3})$ and thus the entire perturbation to the gauge coupling is $\mathcal{O}(g_s \n)$. We have again grouped specific geometric details into a coefficient $\cbr$.
		
		 	In the introduction, we gave an interpretation of this backreaction  in terms of new light states entering the spectrum of the theory upon a monodromy $\phi \mapsto \phi + f$.  Open/closed-string duality dictates that the supergravity (closed-string channel) correction (\ref{eq:g7br}) must match the one-loop correction to the gauge coupling $g_\lab{YM}$ calculated in the open-string channel. In the open-string picture of the configuration pictured in Figure \ref{fig:intersect}, we are interested in the one-loop correction to the $\lab{SU}({N_c})$ gauge theory living on the D7-brane stack wrapping $\Sigma_4$, in the presence of $\n$ D3-branes dissolved in the NS5-brane on $\Sigma_2$ and $\n$ anti-D3-branes dissolved in the anti-NS5-brane on $\bar{\Sigma}_2$. Crucially, the $\n$ D3-branes introduce $\n$ light 3-7 strings transforming in the fundamental of $\lab{SU}(N_c)$, which provide a contribution to the one-loop $\beta$-function (\ref{eq:1loopbeta}).
		
		Accounting for this backreaction
changes the structure of the potential (\ref{eq:relaxpotential2}).  In particular, from (\ref{eq:g7br}) the monodromy charge induces further relaxion-dependence of the height of the stopping potential barriers,
		\begin{equation}
			\Lambda^4(v) \to \Lambda^4(\phi) e^{-\cbr \phi/f}.
		\end{equation}
		A priori, it is not obvious that $\cbr$ is either always positive or always negative, so we will consider $\cbr>0$ and $\cbr<0$ separately.	
		Assuming that the Higgs quartic coupling takes the form
		\begin{equation}
			\mathcal{L}_h \supset -\frac{\lambda}{2} |h|^4,
		\end{equation}
		$v$ is given by
		\begin{equation}
			v(\phi) = \sqrt{\frac{g M }{\lambda}\left(\phi_\lab{h} - \phi \right)},
		\end{equation}
		where $\phi_\lab{h} = \phi_\lab{init} - M/g$ is generically $\mathcal{O}(M/g)$, and thus the corresponding induced monodromy charge when the Higgs develops a vev is very large, $\n_\lab{h} \equiv \phi_\lab{h}/ f \gg 1$. Ignoring the backreaction effect (\ref{eq:g7br}) and assuming that $f \ll \lambda v^2 / g M$, the relaxion will stop rolling when
		\begin{equation}
			\frac{\Lambda_c^3}{f} \sqrt{\frac{g M}{\lambda} \left(\phi_\lab{h} - \phi\right)} \sim g M^3
			\label{eq:stopcondnobr}
		\end{equation}
		and it will be stabilized at
		\begin{equation}
			\phi_h - \phi \sim \frac{\lambda}{g M} \left(\frac{g M^3 f}{\Lambda_c^3}\right)^2.
		\end{equation}
		
		If we now include the backreaction (\ref{eq:g7br}), (\ref{eq:stopcondnobr}) becomes
		\begin{equation}
			\frac{2 \cbr}{f} \left(\phi_\lab{h} - \phi\right) e^{2 \cbr(\phi_\lab{h} - \phi)/f} \sim \frac{2 \lambda}{g M f} \frac{\cbr}{(1 + \cbr)^2}\left(\frac{g M^3 f}{\Lambda_c^3}\right)^{2} e^{2 \cbr \phi_\lab{h}/f}.
			\label{eq:stopcondbr}
		\end{equation}
		Because $\phi_\lab{h}/f \gg 1$, the asymptotic behavior of solutions to (\ref{eq:stopcondbr}) is determined solely by the sign of $\cbr$. For $\cbr > 0$, the stopping potential barriers are exponentially suppressed by the backreaction and (\ref{eq:stopcondbr}) predicts that the relaxion stops at
		\begin{equation}
			\phi \sim -\frac{f}{2 \cbr} \log \left(\frac{2 \lambda \cbr}{(1 + \cbr)^2} \frac{g M^3 f}{\Lambda_c^4} \frac{M^2}{\Lambda_c^2}\right) < 0.
			\label{eq:suppressedstop}
		\end{equation}
		However, the linear potential $g M^3 \phi$ in (\ref{eq:relaxpotential2}) is only an approximation for a potential of the form (\ref{eq:NS5potential}) and cannot be used for arbitrarily small values of $\phi/f$. From (\ref{eq:suppressedstop}) we see that this approximation breaks down.  We should therefore understand (\ref{eq:suppressedstop}) as an indication that the relaxion stops roughly when it has dissipated all of its charge, near $\phi = 0$.  The electroweak scale is then fixed at
		\begin{equation}
			v \sim \frac{g M}{\lambda} \phi_\lab{h} \sim \frac{M^2}{\lambda},
		\end{equation}
		leaving the hierarchy unresolved.
		
		For $\beta < 0$, the barriers are exponentially \emph{enhanced}, and the relaxion stops at
		\begin{equation}
			\phi \sim \phi_\lab{h} - \frac{\lambda}{g M} \left(\frac{g M^3 f}{\Lambda_c^3}\right)^2 \frac{1}{(1 + \cbr)^2} e^{2 \cbr \phi_\lab{h}/f},
		\end{equation}
		and the electroweak scale
		\begin{equation}
			v \sim \frac{g M^3 f}{\Lambda_c^3} \frac{1}{|1 + \cbr|} e^{-|\cbr|\phi_\lab{h}/f}
			\label{eq:enhancedew}
		\end{equation}
		is suppressed by the backreaction.
		
		Can one use this barrier enhancement to save the relaxion from the exponential suppression discussed in \S\ref{sec:warpsupp}? Unfortunately, this backreaction enhancement is not enough to overcome the suppression from warping. We may combine (\ref{eq:enhancedew}) with (\ref{eq:lambdags}) to find
		\begin{equation}
			\frac{M_s}{v} \sim \frac{g_s |1 + \beta|}{(\cbg - |\cbr| \n_\lab{h}/N_\lab{D3})} \left(\frac{\ell_s}{r_\labbot}\right)^4 \left[ \left(\cbg N_\lab{D3} - |\cbr| \n_\lab{h}\right) e^{-\cbg N_\lab{D3} + |\cbr| \n_\lab{h}}\right].
			\label{eq:enhancedewstring}
		\end{equation}
		Since we require that $\n_\lab{h}/N_\lab{D3} \ll 1$  for control and we expect the geometric factors to be on the same order $\cbg \sim |\cbr|$, (\ref{eq:enhancedewstring}) implies that the necessary fine-tuning is still of the same order as the hierarchy one wishes to generate.

\subsubsection{Effects on the moduli potential}\label{sec:vmod}

In \S\ref{sec:D7backreaction} we considered the backreaction of D3-brane charge on the gauge coupling of the D7-branes that generate the stopping potential.
As shown in Appendix \ref{d7pf}, this particular D7-brane stack must enter the strongly warped region, and so the backreaction does not need to propagate far to impact them.  The result is a very large change in the gauge coupling of the D7-brane worldvolume theory, leading to exponential suppression of the stopping potential.

Let us now ask about the impact of backreaction on the moduli potential.  In the NS5-brane scenario, the K\"ahler moduli of the compactification are stabilized by nonperturbative effects on a collection of four-cycles, either Euclidean D3-branes or gaugino condensation on D7-branes.
The moduli potential also involves exponentials of the warped volumes of these cycles.
Backreaction of D3-brane charge will change the warped volumes of these cycles, and so the moduli potential will typically be a rapidly varying function of the relaxion $\phi$.

The argument of Appendix \ref{d7pf} does not imply that the four-cycles supporting the K\"ahler moduli potential enter the warped region, so in contrast to \S\ref{sec:D7backreaction}, the backreaction has to propagate across the internal geometry to influence the moduli potential.
It is tempting to argue that backreaction has a negligible effect on a sufficiently distant four-cycle.  This is \emph{not} correct.  We will give a heuristic explanation here, and refer the reader to \cite{Baumann:2014nda} for a complete quantitative treatment.

To understand whether backreaction of D3-brane charge can decouple from D7-branes on a particular four-cycle $\Sigma_4$, we work in the open string picture, where the effect of backreaction is translated into the open string one-loop threshold correction to the D7-brane gauge coupling.
On very general grounds, this effect gives non-negligible contributions to the relaxion potential unless the masses $M_{3-7}$ of the stretched open strings obey
\begin{equation}
	M_{3-7} \gtrsim M_{\rm{pl}}\,,
	\end{equation}
 for then the non-renormalizable operators coupling the relaxion to the moduli are suppressed by more than the Planck mass.
In a compact space, the diameter of the space determines an upper bound on the mass $M_{3-7}$, and one finds that at weak coupling and large volume, $M_{3-7} \ll M_{\rm{pl}}$ \cite{Baumann:2014nda} (cf. also \cite{Berg:2010ha}).  This is easily checked in simple geometries, but holds more generally.\footnote{This fact is responsible for the well-known problem that brane-antibrane potentials are generically too steep to support inflation \cite{Burgess:2001fx}.}

The upshot is that the four-cycles supporting the moduli potential cannot be taken far enough away from the source of monodromy to avoid significant backreaction: the moduli potential depends strongly on $\phi$.  One consequence is that the relaxion potential is \emph{not} simply given by the probe DBI action (\ref{eq:NS5potential}), but instead has important contributions from couplings to moduli.  This is an incarnation of the eta problem, which hinders the construction of natural models of inflation.

If all the other obstacles enumerated here could be overcome in some manner, leaving only the problem of relaxion couplings to moduli induced by backreaction, then one could 
attempt to
fine-tune the orientation of the source of monodromy with respect to the  configuration of four-cycles in the bulk of the compactification.
The idea is that if the leading multipoles of the backreaction can be made to vanish on the four-cycle ``receiver'' by fine-tuning the relative orientation, then the residual effect of the subleading multipoles might be negligible.
This approach was proposed and analyzed in \cite{Flauger:2009ab}, where it was shown that  given a suitable geometry, and for a modest winding number $N \sim 100$, the most dangerous couplings can be removed.  It is not clear that this method is applicable for the extremely large windings $N$ that arise in relaxion constructions.

\subsubsection{Effects on axion decay constants}\label{sec:fbackreaction}

As discussed in Appendix \ref{app:dictionary}, the relaxion decay constant $f$ only depends on the six-dimensional metric $\tilde{g}_{mn}$, both through the explicit factors of $\tilde{g}_{mn}$ in its definition (\ref{eq:decconstapp}) and implicitly via the defining equation of the harmonic form $\Delta \Omega = 0$. Additional D3-brane charge will not perturb $f$:  $\tilde{g}_{mn}$ is Ricci-flat in supersymmetric compactifications and additional D3-brane charge will preserve the same supercharges as the three-branes forming the warped throat. However, anti-D3-branes break the remaining supersymmetry.  The anti-D3-brane charge backreacts on the six-dimensional metric and perturbs the axion decay constant,
\begin{equation}
	\frac{\delta f^2}{M_\lab{pl}^2} = \frac{g_s}{2 \mathcal{V}_{E} \ell_s^6} \int\!\ud^6 y \sqrt{\tilde{g}} \tilde{g}^{mp} \tilde{g}^{n q} \left( 2 \Omega_{mn} \delta \Omega_{pq}+ \frac{1}{2} \tilde{g}^{r s} \delta \tilde{g}_{rs} \, \Omega_{mn} \Omega_{p q}- 2 \Omega_{mn} \Omega_{ps} \delta \tilde{g}_{qr} \tilde{g}^{rs} \right).
	\label{eq:fpert}
\end{equation}
The anti-D3-brane charge does not substantially  perturb the four-dimensional Planck mass $M_\lab{pl}^2$, as $\mathcal{V}_\lab{E}$ is dominated by the volume of the bulk Calabi-Yau. In what follows, we estimate the size of each of these terms.

The first term in (\ref{eq:fpert}) vanishes at first order, as the perturbation $\delta \Omega$ is orthogonal to the unperturbed 
$\Omega$. To analyze the contribution from the second and third terms in (\ref{eq:fpert}), we must backreact the anti-D3-brane charge on the metric $\tilde{g}_{mn}$. As detailed in Appendix \ref{sec:internalbr}, the dominant metric perturbations are
\begin{equation}
	\tilde{g}_{mn} \ud y^m \,\ud y^n \sim \left(1 + \frac{\alpha \n}{N_\lab{D3}} \left(\frac{r'}{r}\right)^8\right) \ud r^2 + r^2 \left(1 + \frac{\beta \n}{N_\lab{D3}} \left(\frac{r'}{r}\right)^{19/2}\!\!\!\mathcal{Y}^{\frac{1}{2},\frac{1}{2}, 1}(\Psi)\right) \breve{g}_{\theta \phi} \,\ud \xi^\theta \, \ud \xi^\phi
\end{equation}
where the coefficients $\alpha$ and $\beta$ and the angular function $\mathcal{Y}^{\frac{1}{2}, \frac{1}{2}, 1}(\Psi)$ depend on the details of the compactification. The perturbation to the decay constant (\ref{eq:fpert}) is then
	\begin{equation}
		\frac{\delta f^2}{M_\lab{pl}^2} \propto \frac{g_s}{2 \mathcal{V}_E \ell_s^6} \frac{\n}{N_\lab{D3}} \int_{r_\labbot}^{r_\labtop}\!\!\!\ud r \,r\, \left(\frac{r_\labbot}{r}\right)^{19/2}  \propto \frac{f^2}{M_\lab{pl}^2} \frac{\n}{N_\lab{D3}} \left(\frac{r_\labbot}{r_\labtop}\right)^{19/2}\!\!\!\!\!.
	\end{equation}
Because of the large exponent this is a comparatively weak constraint.

\subsubsection{Classical annihilation of the dipole} \label{sec:classicalann}
		
		The fivebrane configuration detailed in \S\ref{sec:fivebrane} is metastable. If the fivebranes were tensionless, the induced D3-brane charge on the NS5-brane would attract the induced anti-D3-brane charge on the anti-NS5-brane, and these branes would classically annihilate.  However, this Coulomb attraction is balanced by the fivebrane tension---in order for the fivebranes to meet, they must stretch over the large two-cycle $\Sigma_\labtop$ at the junction of the two warped throats, $r_\labtop$
		in Figure \ref{fig:fdoublethroat}, which costs an energy
		\begin{equation}
			V_t \sim \frac{2 \pi}{\ell_s^4} \frac{e^{4 A_\labtop}}{\sqrt{g_s}}\sqrt{4 \left(\frac{\vol\,\Sigma_\labtop}{\ell_s^2}\right)^2 + \N^2}.
			\label{eq:tensionenergy}
		\end{equation}
This potential energy barrier ensures the configuration is metastable, and can be exponentially long-lived. 		
However, for large enough winding, we expect the Coulomb force---which scales as $\N^2$---to overpower this ``tension force,'' allowing the fivebranes to
		classically annihilate.
		
		The potential energy density of a probe D3-brane is proportional to $\Phi_-$,
		\begin{equation}
			V = \frac{2 \pi}{\ell_s^4} \left(e^{4 A} - \alpha\right) = \frac{2 \pi}{\ell_s^4} \Phi_-.
		\end{equation}
		For $\n$ D3-branes and $\n$ anti-D3-branes, the Coulomb potential energy density is then
		\begin{equation}
			V_\lab{c} \sim \frac{2 \pi \n}{\ell_s^4} \delta \Phi_-,
		\end{equation}
		where $\delta \Phi_-$ (cf.~Appendix \ref{sec:internalbr}) is the perturbation to $\Phi_-$ due to $\N$ anti-D3-branes, measured at the location of the $\N$ D3-branes. Because the D3-branes live in a separate warped throat, $\delta \Phi_-$ must first propagate up the antibrane throat from the anti-NS5-brane location $r = r_\labbot$ to the surface $r = r_\labtop$,
		\begin{equation}
			\delta \Phi_{-,\overline{\lab{D3}}}^{I_s} \sim -\frac{\N}{N_\lab{D3}} e^{4 A_\labbot} \left(\frac{r_\labtop}{r_\labbot}\right)^{2 + \Delta_s},
			\label{eq:antid3pert}
		\end{equation}
		which then propagates down the D3-brane warped throat via the homogeneous modes
		\begin{equation}
			\delta \Phi^{I_s}_{-,\lab{D3}} = c_1 \left(\frac{r_\labtop}{r}\right)^{\Delta_s + 2} + c_2 \left(\frac{r}{r_\labtop}\right)^{\Delta_s -2}.
			\label{eq:d3pert}
		\end{equation}
		We may think of the perturbation (\ref{eq:antid3pert}) as specifying a boundary value for the perturbation (\ref{eq:d3pert}). Generically, we have
		\begin{equation}
			c_1, c_2 \sim \frac{\N}{N_\lab{D3}} e^{4 A_\labbot} \left(\frac{r_\labtop}{r_\labbot}\right)^{2 + \Delta_s}
		\end{equation}
		so that, at the position of the D3-brane charge,
		\begin{equation}
			\delta \Phi_{-, \lab{D3}}^{I_s} \sim -\frac{N}{N_\lab{D3}} e^{4 A_\labtop}+ \frac{N}{N_\lab{D3}} e^{4 A_\labtop} \left(\frac{r_\labtop}{r_\labbot}\right)^{2 \Delta_s}.
		\end{equation}
		We then find that the Coulomb energy is roughly
		\begin{equation}
			V_c \sim -\frac{2 \pi}{\ell_s^4}\frac{\n^2}{N_\lab{D3}} e^{4 A_\labtop}.
		\end{equation}
		Requiring that this be much less than the potential energy barrier (\ref{eq:tensionenergy}) yields the constraint
		\begin{equation}
			\N \ll \frac{N_\lab{D3}}{\sqrt{g_s}} + \frac{2 \,\vol\, \Sigma_\labtop}{\ell_s^2} + \mathcal{O}\left(\frac{1}{N_\lab{D3}^2}\left(\frac{\vol\,\Sigma_\labtop}{\ell_s^2}\right)^4 \right)\label{eq:coulombbound}
		\end{equation}
		
		We should also account for the interaction energy between the pair of fivebranes. By performing an open string computation in an unwarped toroidal orbifold, \cite{Conlon:2011qp} found a potential contribution that grows \emph{logarithmically} with the fivebrane separation, and argued that this would apply to warped geometries, with energy scale set by $r_\labtop$.
While it is not entirely clear that this logarithmic behavior arises in the actual NS5-brane configuration described in \S\ref{sec:axmon},
the corresponding potential energy contribution would take the schematic form
		\begin{equation}
			V_{5 \bar{5}} \sim \frac{2 \pi}{\ell_s^4}e^{4 A_\labtop} N_{\lab{NS5}}^2\log \left(\frac{L}{r_\labtop}\right) \sim \frac{2 \pi }{\ell_s^4}e^{4 A_\labtop}\, .
			\label{eq:fivebranepoten}
		\end{equation}
	We can ensure that this energy is much smaller than the uncharged tension energy (\ref{eq:tensionenergy}) by imposing
		\begin{equation}
			\vol \, \Sigma_\labtop \gg \sqrt{g_s} \ell_s^2\,  ,
			\label{eq:sigtopbound}
		\end{equation}
	which is necessary in any case to ensure the validity of the supergravity approximation.
	
	Throughout this work we have taken the homology class $[\Sigma_2]$ wrapped by the NS5-brane to be localized in the warped throat, as in Figures \ref{fig:fdoublethroat} and \ref{fig:intersect}.  That is, we assumed that the harmonic two-form dual to $[\Sigma_2]$ is principally supported in the warped region, and every holomorphic representative of $[\Sigma_2]$ is in the warped region.  This localization is automatic in the particular construction given in \cite{Flauger:2009ab}, but should also be required in alternative constructions.
A key reason is the fivebrane potential energy (\ref{eq:fivebranepoten}): if the lines of three-form flux stretching from the NS5-brane to the anti-NS5-brane passed through an unwarped region, the overall scale of supersymmetry breaking would exceed the string scale, by (\ref{eq:fivebranepoten}), and immediately destabilize the moduli.

	\subsubsection{Antibrane tunneling and annihilation} \label{sec:kpvinstability}
	
Consider a Klebanov-Strassler throat that arises from $N_\lab{D3}$ D3-branes probing a conifold with
$M_\lab{KS}$ D5-branes wrapping the shrinking two-cycle.  We take $N_\lab{D3}=M_\lab{KS}K_\lab{KS}$; then $K_\lab{KS}$ is the number of units of $H_3$ flux on the $B$-cycle.

If $\N$ anti-D3-branes are placed at the tip of  this throat, they create a metastable, exponentially long-lived state provided that $\n \lesssim 0.08 M_\lab{KS}$  \cite{Kachru:2002gs}.
With more anti-D3-branes, $\n \gtrsim 0.08 M_\lab{KS}$, the anti-D3-branes rapidly annihilate \cite{Kachru:2002gs} against the flux supporting the warped throat, decaying to the state with $K'_\lab{KS} = K_\lab{KS} - 1$.  The D3-brane charge carried by flux is then $N'_\lab{D3} = N_\lab{D3} - M_\lab{KS}$, and $M_\lab{KS}-\n$ D3-branes appear, but no anti-D3-branes remain.

While $N_\lab{D3}$ sets the overall scale of warping in the throat, $M_\lab{KS}$ sets the warp factor at the tip,
	\begin{equation}
		\left.e^{A}\right|_\labbot \sim \exp\left(-\frac{2 \pi N_\lab{D3}}{3 g_s M_\lab{KS}^2}\right)
	\end{equation}
	and
	\begin{equation}
		g_s M^2_\lab{KS} \lesssim N_\lab{D3}
	\end{equation}
	if the warping is non-negligible.
To avoid the KPV instability \cite{Kachru:2002gs}, the number of windings cannot exceed
	\begin{equation}
		N \ll 0.08 \,g_s^{-1/2} N_\lab{D3}^{1/2}.
	\end{equation}

\subsubsection{Light NS5-brane modes}\label{sec:sequester}
	As discussed in \S\ref{sec:fivebrane}, dimensional reduction of the transverse fluctuations in the NS5-brane's position yields Kaluza-Klein excitations whose mass \emph{decreases} as the relaxion is wound up. Intuitively, we may interpret the presence of two-form flux as increasing the effective volume of the NS5-brane. Since Kaluza-Klein masses will inversely scale with this effective volume, we should expect some modes to become light at large windings. As shown in Appendix \ref{sec:eframe}, to second  order the canonically normalized fluctuations are described by the action
	\begin{align}
		\pertTensor{S}{2}{\lab{NS5}} = \int\!\ud^4 x\, \sqrt{-g_4} \Bigg(& \!- V(c) - \frac{1}{2} g^{\mu \nu} \partial_\mu Y^{\hat{\imath}}_I \partial^\mu Y_{\hat{\imath}}^I -  \frac{1}{2} m^2_I(c) Y^{\hat{\imath}}_I Y_{\hat{\imath}}^{I} \nonumber \\
				&+ g(c) \left(c \partial_\mu c\right)\left( Y_I^{\hat{\imath}} \partial^\mu Y^I_{\hat{\imath}}\right) - \frac{1}{2} g(c)^2 \left(\partial c\right)^2 Y_I^{\hat{\imath}} Y^I_{\hat{\imath}} \Bigg)
		\label{eq:flucLag}
	\end{align}
	with
	\begin{equation}
		m_I^2 \sim \frac{4}{g_s \ell_\lab{E}^2} \frac{e^{2 A_\labbot}}{\n^2} \left(\frac{\ell_\lab{E}}{\ell_s}\right)^4 \quad \text{and} \quad g(\n) \sim \frac{1}{2 \n^2}
	\end{equation}
	at large winding $\n^2 \gg 4(\ell_\lab{E}/\ell_s)^4/g_s$,
	where $\ell_\lab{E}^2$ is the Einstein-frame volume of the two-cycle $\Sigma_2$ and the $\lambda_I$ are eigenvalues of $\Sigma_2$'s Laplacian, labeled by the multi-index $I$. As discussed in the introduction, the appearance of light states is generic in realizations of monodromy in string theory, and one must ensure that these do not drastically affect the phenomenology.
	
The presence of ${\cal O}(\n)$ light states in the spectrum, including 3-7 strings and KK excitations of the NS5-branes, can have a range of consequences.
For example, modes with mass $m<3H/2$ can fluctuate during inflation, storing energy and potentially impacting the late-time perturbations.
Here we will examine just one effect of the KK modes, which is an enhanced probability of NS5-brane annihilation.

%\subsubsection*{Tunneling and annihilation of the dipole}

The masses of the canonically normalized fluctuations of the NS5-brane embedding
(\ref{eq:flucLag}) decrease with $\n$, and one might worry that these light modes facilitate an additional instability. For example, if these modes are thermally excited by some source of supersymmetry breaking elsewhere in the compact space, then the NS5-branes can more readily reach each other and either classically or quantum-mechanically annihilate.
	
A complete analysis of this process is beyond the scope of this work. We will instead use an approximate criterion for the onset of instability.
The dominant instanton in the four-dimensional field theory responsible for the transition between the metastable and stable states (i.e., the states with and without the NS5/anti-NS5-brane dipole, respectively) will be $\lab{SO}(4)$-symmetric, with radius determined by
	\begin{equation}
		R_* \sim \frac{T_\lab{D}}{\Delta V},
	\end{equation}
	where, in the thin-wall approximation, $T_\lab{D}$ is the tension of the domain-wall interpolating between the two vacua and $\Delta V$ is their difference in energy. We then assume a loss of control when the typical thermal fluctuations of a spatial region of size $R_*$ are comparable to the distance between the two fivebranes, $r_\lab{RMS} \sim r_\labtop$.
	
	The difference in energies, in the probe approximation, is the potential energy contribution from the NS5-branes,
		\begin{equation}
			\Delta V = \frac{2 \pi}{\ell_s^4}\n \,e^{4 A_\labbot}
		\end{equation}
		while the tension of the domain wall is determined by an NS5-brane
winding $\n$ times around the minimum-volume three-cycle whose endpoints are $\Sigma_2$ and $\bar{\Sigma}_2$. As described in Appendix \ref{sec:eframe}, the tension of the domain wall follows from the NS5-brane action and is roughly
		\begin{equation}
			T_\lab{D}  \sim \frac{1}{\ell_s^3} N_\lab{D3}^{3/4} \left(e^{4 A_\labtop} - e^{4 A_\labbot}\right)
		\end{equation}	
		and so
		\begin{equation}
			R_* \sim \ell_s \frac{N_\lab{D3}^{3/4}}{\n} \left(\frac{r_\labtop}{r_\labbot}\right)^4
		\end{equation}
		If the NS5-brane is in thermal equilibrium at temperature $T$, then a smooth excitation of size $R_*$ in the canonically normalized fluctuations $Y^{\hat{\imath}}_I$ gains a thermal expectation value
		\begin{equation}
				\langle Y^{\hat{\imath}}_{I} Y^{\hat{\jmath}}_{J}\rangle \sim \frac{T}{m_I^2} \frac{1}{R_*^3} \delta_{IJ} \delta^{\hat{\imath} \hat{\jmath}}.
		\end{equation}
		The thermal fluctuation in the radial direction, averaged over $\Sigma_2$, is roughly
		\begin{equation}
			\langle \delta r^2 \rangle \sim g_s \ell_s^2 \left( \ell_s T\right) \frac{\n^4}{N_\lab{D3}^{3/4}}\left(\frac{\ell_s}{\ell}\right)^2 \left(\frac{r_\labbot}{r_\labtop}\right)^{12}
		\end{equation}
		where $\ell^2$ is the unwarped volume of the two-cycle $\Sigma_2$.
		The requirement that this is much smaller than the size of the dipole $\langle \delta r^2 \rangle \ll r_\labtop^2$ imposes the weak constraint		
		\begin{equation}
			\n^4 \ll e^{4 A_\labtop}\frac{N_\lab{D3}^{5/4}}{g_s^{1/2} (\ell_s T)} \left(\frac{\ell_\lab{E}}{\ell_s}\right)^2 \left(\frac{r_\labtop}{r_\labbot}\right)^{10}\!\!.
		\end{equation}

\section{Discussion and Outlook} \label{sec:summary}

We have identified many obstacles to realizing the relaxion mechanism in string theory.  Some of these obstacles are extremely general,
while others apply only to NS5-brane monodromy, the particular example we studied in detail.  We will now step back and give some perspective on our results, explaining their scope of validity.

Our first observation was that axion monodromy in string theory proceeds by the accumulation of monodromy charge,
and the backreaction of this charge substantially changes the couplings of the axion.  This applies to any realization of axion monodromy in string theory.  Thus, any ultraviolet completion in string theory of a relaxation mechanism that involves axion displacements $\Delta \phi > f$ will be vulnerable to the backreaction of monodromy charge.

The effects of this backreaction will vary from one model to another.
We focused on NS5-brane monodromy because this is, to our knowledge, the only scenario where the smallness of the shift symmetry breaking parameter $g$ is natural---in this case, because of warping---while in alternative constructions in string theory, one must fine-tune discrete data to achieve small $g$.
In the NS5-brane model, we
found that the barriers in the stopping potential are exponentially small in the winding number $N \equiv \phi/f$, leading to a runaway relaxion.
We expect this barrier suppression phenomenon to be rather general, but not universal.  However, the particular effects of backreaction on the axion decay constants detailed in \S\ref{sec:fbackreaction}, and the constraints from annihilation in \S\ref{sec:classicalann} and \S\ref{sec:kpvinstability}, could be very different in other models.

Some of the challenges that we have identified might be milder in non-supersymmetric compactifications of string theory.
In particular, in compactifications that break all supersymmetry at the Kaluza-Klein scale as in e.g.~\cite{Saltman:2004jh,McAllister:2014mpa}, tadpole constraints on the total charge need not be a serious limitation.  On the other hand, ensuring metastability of such a configuration can be very challenging.
Moreover, for an embedding of the relaxion in a non-supersymmetric compactification, the absence of spacetime supersymmetry below the KK scale might require either the KK scale or even the string scale to arise as the regulator of the relaxion setup at the relaxion cutoff scale $M$.

We assumed that the periodic stopping potential arises from non-perturbative effects that 
couple locally to the axion. This local coupling then exposes the stopping potential to an exponential suppression from warping.
However, the stopping mechanism could instead arise from other effects, for example from heavy states~\cite{Flauger:2016idt} coupled to the relaxion, or from the exponential production of massive particles~\cite{Hook:2016mqo}, which are not necessarily susceptible to the same failure modes.\footnote{We thank E. Silverstein for illuminating discussions of these points.}

\subsection{Exact discrete shift symmetries for relaxions} \label{sec:discretesymm}

Throughout this work we have considered axion monodromy, in which a source of monodromy completely breaks the shift symmetry of an axion.
An important alternative is \emph{alignment} of multiple periodic contributions to the axion potential, leaving an unbroken discrete shift symmetry.
We now briefly outline this possibility and mention some of the obstacles to embedding this scenario in string theory.

The essential feature of the relaxion potential (\ref{eq:relaxpotential}) is the combination of a slowly-varying term \circled{B}, the linear $g M^3 \phi$, and a quickly-varying term \circled{C}, the oscillatory $\Lambda_c^3 v \cos(2 \pi \phi/f)$.
As written, \circled{B} explicitly and completely breaks the discrete shift symmetry $\phi \mapsto \phi + f$.
Alternatively, \circled{B} could represent the leading term in the expansion of a function that is invariant under a much larger discrete shift $\phi \mapsto \phi + k f$.
For example, we could have
\begin{equation}
V_{\scalebox{0.75}{\circled{B}}}= g f M^3 \sin\left(\frac{2\pi \phi}{k f}\right),
\end{equation}
where $\phi$'s field space diameter is actually $k$ times larger than would be naively inferred by only considering small displacements.
We refer to these two cases as having an \emph{explicitly broken symmetry} or an \emph{exact discrete symmetry}, respectively.
Thus far, we have only concentrated on the former. The explicit breaking is induced by a source of monodromy, an NS5-brane, and we have shown that the accumulation of monodromy charge leads to backreaction effects that spoil the relaxation mechanism.
Given this difficulty, one might ask whether the relaxion mechanism could be more readily realized in a solution of string theory with an exact discrete shift symmetry.

As in the models with explicit breaking, the main difficulty in realizing a discrete shift-symmetric relaxion lies in ensuring that the potential has structure over two---and only two---disparate scales. That is, the potential must roughly be
the
sum of two terms---a slowly varying term with periodicity $f$
that apes the linear term \circled{B} in (\ref{eq:relaxpotential}), and a quickly varying stopping potential with periodicity $f_s \ll f$. One might take, as a toy model, a potential that is generated only by the instantons with winding number 1 and $k$, so that the potential takes the schematic form
\begin{equation}
	V = M^2 e^{-S_1} \cos \left(\frac{2 \pi\phi}{f}\right) |h|^2 + M^4 e^{-S_1} \cos\left(\frac{ 2 \pi \phi}{f}\right) +  M^3 e^{-S_k} v \cos \left(\frac{2 \pi k \phi}{f}\right)
\end{equation}
For $\phi \ll f$, the potential is approximately a ``monomial with modulations'' with $f_s = f/k$, and has the \circled{A} \circled{B} \circled{C} structure of (\ref{eq:relaxpotential}), with $g \propto e^{-S_1}$ and $\Lambda_c^3 \sim M^3 e^{-S_k}$. The analogue of (\ref{eq:hierarchy}) in this two period model is then
\begin{equation}
	\frac{v}{M} \gtrsim \frac{1}{k} \left(\frac{M}{\Lambda_c}\right)^3 \sim \frac{1}{k} e^{S_k - S_1}.
\end{equation}
Naively, the generated hierarchy grows with $k$.

However, there are many problems with this toy model. First and foremost, the action for a $k$-instanton is typically $S_k \geq k S_1$, and we require that $S_1 \gg 1$ in order to trust the instanton expansion. The stopping potential barriers will then shrink with $k$,
 \begin{equation}
 	|V_\lab{stop}| \propto e^{-k S_1} \propto g^k
 \end{equation}
and, reminiscent of the suppression due to warping discussed in \S\ref{sec:warpsupp}, the maximum achievable hierarchy actually shrinks with $k$. The stopping potential is too small to stop the evolution near the point where the Higgs is massless. If some mechanism were able to enhance the $k$-instanton contribution, one must
still explain the absence of $j$-instanton effects, with $1 < j < k$, and we find it implausible that all such effects could be negligible for $k \gg 1$.\footnote{Such a situation appears to conflict with the lattice form of the Weak Gravity Conjecture, but is already implausible regardless.} Furthermore, one would have to explain why the Higgs couples to instantons of winding 1 and $k$ differently, and why the $1$-instanton and $k$-instanton contributions do not \emph{both} vanish when $v = 0$.

Many of these problems may be mitigated in models with multiple axions, as in the Kim-Nilles-Peloso mechanism~\cite{Kim:2004rp} and kinetic alignment setups~\cite{Bachlechner:2014hsa, Bachlechner:2014gfa} in
the inflationary context.  Scenarios for aligned relaxions have been presented in~\cite{Choi:2015fiu, Kaplan:2015fuy, Fonseca:2016eoo}.\footnote{See also~\cite{Flacke:2016szy} for a recent discussion of naturalness constraints on such scenarios.} A general multi-axion Lagrangian can be written (cf.~e.g.~\cite{Bachlechner:2014gfa}) as
\begin{equation}
	\mathcal{L} = -K^{ij} \partial \phi_i \partial \phi_j - \sum_{a} \Lambda_a^4 \exp\left(-Q\indices{_a^i} S_i\right) \left[1 - \cos\left(2\pi Q\indices{_a^i} \phi_i\right)\right],
\end{equation}
where $K^{ij}$ is a positive definite kinetic matrix of real numbers, while $Q\indices{_a^i}$ is a charge matrix containing integers. We imagine that there are  ``slow'' and ``fast'' linear combinations of canonically normalized axions with effective decay constants $f$ and $f_s \ll f$, respectively. The addition of another direction in field space solves several of the problems mentioned previously, at the cost of introducing much more complicated dynamics.

Foremost among the advantages is that the stopping potential is no longer necessarily suppressed. In the single axion model, the hierarchy between $f_s$ and $f$---and so between $v$ and $M$---was generated by a hierarchy in the charge matrix $Q\indices{_a^i}$, and a high charge contribution is exponentially suppressed relative to a low charge contribution. In
a multi-axion model, $f_s \ll f$ may instead be realized in the kinetic matrix $K^{ij}$, and this does not impose an exponential hierarchy in the associated barrier heights. Of course, the hierarchy in the kinetic matrix must then be explained, but it is much easier to realize a hierarchy in the eigenvalues of a matrix of real numbers than in a matrix of
bounded integers, and one does not need to explain why instantons with winding $j$, $1 < j < k$, do not contribute.

A very mild degree of alignment has been demonstrated in explicit examples \cite{Long:2016jvd}, but whether alignment can yield large effective axion decay constants in string theory is an important open question, even for the ${\cal O}(100)$ enhancements that could suffice for inflation.  It is not obvious to us that the vastly larger enhancements needed for a relaxion scenario are possible in known compactifications.  For example, the ``clockwork'' mechanism~\cite{Choi:2014rja,Kaplan:2015fuy} requires a specific  matrix of axion charges of instantons, and it remains to be seen whether this particular pattern of charges can arise in string theory.
However, it is very plausible that linearly independent combinations of axions couple differently to the Higgs.

In summary, relaxion scenarios with exact discrete symmetries, built on the alignment of multiple instanton effects for one or more axions, are qualitatively different from the axion monodromy scenarios, with explicitly broken symmetry, considered in this work.
However, \emph{both} classes of models are vulnerable to ultraviolet physics.  Axion monodromy scenarios suffer from the backreaction of monodromy charge, as we have explained.  Aligned scenarios could avoid this problem, but require extremely special axion charges.  These charges are ultimately topological data dictated by the ultraviolet theory, and it is not clear that string theory allows strong enough alignment to permit relaxation of a large hierarchy. Furthermore, these multi-axion models have much more complicated dynamics, and it is not clear that the dynamical generation of a large hierarchy can proceed in a robust way.

\subsection{Constraints from the Weak Gravity Conjecture}

The Weak Gravity Conjecture (WGC), a class of conjectures asserting that gravity must be the weakest force \cite{ArkaniHamed:2006dz,Cheung:2014vva,delaFuente:2014aca,Tom,Madrid,Madison,Bachlechner:2015qja,Brown:2015lia,Heidenreich:2015wga,Heidenreich:2015nta,Saraswat:2016eaz}, leads to (still conjectural) constraints on axion theories.  One could therefore ask whether the WGC constrains relaxion monodromy scenarios.
It does \cite{Ibanez:2015fcv,Hebecker:2015zss}, as we will briefly explain, but the known WGC constraints are far weaker than the limitations we have exposed in this work, which are independent of WGC considerations.

The WGC constrains monodromy scenarios by placing upper limits on the tension of domain walls.
In the four-dimensional description of axion monodromy due to Kaloper, Lawrence, and Sorbo~\cite{KaloperSorbo1,KaloperSorbo2,Kaloper:2016fbr}, Brown-Teitelboim domain walls connect different branches of the scalar potential.  At the same time, when instanton effects lead to modulations of the axion potential, distinct critical points are connected across four-dimensional field theory domain walls, via Coleman-de Luccia tunneling.
It turns out that the electric form of the WGC places constraints \cite{Ibanez:2015fcv} on the domain walls of the
Kaloper-Lawrence-Sorbo model, while the magnetic WGC places constraints on the field theory domain walls associated with instanton modulations~\cite{Hebecker:2015zss}.
In both cases one finds a bound on the domain wall tension~\cite{Ibanez:2015fcv,Hebecker:2015zss}
\begin{equation}
T < m f M_\lab{pl},
\end{equation}
where $m$ is the mass of the axion.
For a relaxion model
this implies a bound on the relaxion cutoff scale $M$ of roughly the same order as the constraints already given in~\cite{Graham:2015cka}.

We conclude that the constraints arising from very general four-dimensional quantum gravity considerations, such as the WGC, do not automatically
capture all of the effects of actual embeddings in quantum gravity.  Examining such embeddings is therefore crucial for assessing the viability of
the relaxion mechanism in string theory.\footnote{We note that this view concurs with
the
results of~\cite{Kaloper:2016fbr}, where some leading effects of backreaction on the axion Lagrangian are captured by a series of higher powers of gauge-invariant field strengths, whose coefficients must necessarily be determined in the UV theory, and in which strong backreaction effects can drive one far from the ``natural'' bottom-up estimates.}

\section{Conclusions} \label{conc}

Could a portion of the observed hierarchy between the weak scale and the Planck scale be a consequence of dynamical relaxation of the Higgs mass during cosmological evolution?  This striking idea is the core of the relaxion mechanism~\cite{Graham:2015cka}.
In this scenario, the relaxation of the Higgs mass is driven by the slow evolution of an axion field, the relaxion, whose shift symmetry is very weakly broken by a potential term that introduces monodromy.  After relaxation over many cycles of monodromy, the Higgs mass passes through zero, causing barriers to appear in the axion potential, and so halt the evolution.  In effective field theory, the hierarchy that is generated is determined by the weak breaking parameter, and so is technically natural.

In this work we asked whether the relaxion mechanism survives ultraviolet completion in string theory.  Do the essential components for the scenario exist in a well-controlled compactification, and do these components work in concert in string theory as they do in effective field theory?

We found that the key components of the scenario can indeed be realized in string theory.  The mechanism of axion monodromy, first developed in the context of large-field inflation in string theory, can produce---in the probe approximation---the secular relaxion potential needed for slow relaxation over many fundamental axion periods.  Moreover, the extremely low scale of the secular potential required for the relaxion mechanism can be explained by situating the source of monodromy in a strongly warped region.  This is possible in one known scenario for axion monodromy in string theory, the NS5-brane model of \cite{McAllister:2008hb}, in which two-form axions acquire their potential from NS5-branes wrapping curves in a warped region.

However, our main result is that the structures required for monodromy in string theory present formidable and very general obstacles to a successful relaxion scenario in string theory.  Monodromy proceeds by the accumulation of monodromy charge on a source of monodromy.
As the relaxion rolls over $N$ fundamental axion periods, it necessarily accumulates or discharges $N$ units of monodromy charge.
This large quantity of monodromy charge sources backreaction in the internal space, completely invalidating the probe approximation, and changing the couplings in the effective theory.  The impact of monodromy charge is visible in a dual description as the appearance of $N$ new light states.

We argued that the backreaction of monodromy charge can lead to disastrously large changes to the secular potential in {\it{any}} realization of the relaxion scenario
via axion monodromy
in string theory.  In the specific case of the NS5-brane model, we computed the detailed form of these changes.
The accumulation of monodromy charge suppresses the gauge coupling of the D7-brane gauge theory that generates the stopping potential.  In the dual description, the $N$ light states are charged under the D7-brane gauge group, and give a large threshold correction to the gauge coupling.  The result is that the stopping potential is suppressed by a factor $\exp(-\cbr N)$, where $\cbr$ is a constant determined by the geometry.  The stopping potential is therefore completely negligible, and cannot halt the evolution when the Higgs mass passes through zero.  The Higgs mass indeed relaxes to smaller values, but this process continues far into the tachyonic regime, in a relaxion runaway.

In summary, we have shown that the physics of ultraviolet completion in string theory does not decouple from the dynamics of the relaxion mechanism.
Our results do not exclude the dynamical relaxation of hierarchies in string theory, but in our view they do exclude technically natural dynamical relaxation driven by axion monodromy.  It would be valuable to understand whether some of the difficulties we have uncovered result from limitations in existing constructions, or if instead they are consequences of general structures in quantum gravity.

\section*{Acknowledgments}
We thank Nima Afkhami-Jeddi, Tom Hartman, Nemanja Kaloper, David E. Kaplan, Eric Kuflik, Cody Long, Miguel Montero, Surjeet Rajendran, Michael Stillman, Amir Tajdini, Irene Valenzuela, and Timo Weigand for valuable discussions. The work of L.M.~is supported in part by an NSF CAREER Award.
The work of G.S.~is supported in part by the Helmholtz Association.  The work of A.W.~is supported by the ERC Consolidator Grant STRINGFLATION under the HORIZON 2020 contract no. 647995.

\appendix
\addtocontents{toc}{\protect\setcounter{tocdepth}{1}}

\section{Axions in String Theory}
\label{app:dictionary}

There is an extensive literature on axions in string theory, but for the reader's convenience we now gather a few salient facts.  We begin
with the example of the Neveu-Schwarz two-form gauge potential $B_2$.
	
	Integrating a ten-dimensional $p$-form gauge potential $C_p$ over a non-trivial $p$-cycle in the compactification manifold will give rise to an axion in four dimensions. The number of independent, non-trivial $p$-cycles then determines the maximal number of axions arising from $C_p$. The two-form $B_2$ has an associated field strength $H_3 \equiv \ud B_2$, and appears in the ten-dimensional type II and heterotic supergravity actions as\footnote{Normalization conventions appear in Appendix \ref{sec:conventions}.}
	\begin{align}
		S_\lab{SUGRA} &\supset -\frac{1}{4 \kappa_{10}^2}\int\!\ud^{10} X\, \sqrt{\minus G^S}\, e^{- 2\Phi} \,|H_3|^2 \label{eq:axionkinsugra}
	\end{align}
	Reducing this action along a six-dimensional compact space $X_6$, each non-trivial two-cycle $\Sigma_2^I$ with its associated harmonic form $\omega_2^I$, $\ell_s^{-2} \int_{\Sigma_2^I} \omega_2^J = \delta^J_{~I}$, gives rise to a four-dimensional axion $b_I(x)$,
	\begin{equation}
		b_I(x) \equiv \frac{1}{\ell_s^2} \int_{\Sigma_2^I} \!B_2,
	\end{equation}
	with $B_2 = \sum_{I} b_I(x) \omega_2^I$. Upon dimensional reduction, the first term of  (\ref{eq:axionkinsugra}) yields kinetic terms for the $b_I$ axions,
	\begin{equation}
		S_\lab{kin} = -\frac{1}{2} \int\!\ud^4 x\, \sqrt{-g}\,\gamma^{IJ} \partial_\mu b_I \partial^\mu b_J \qquad \qquad \frac{\gamma^{IJ}}{M_\lab{pl}^2} = \frac{g_s}{2 \mathcal{V}_\lab{E} \ell_s^6} \int_{X_6} \!\!\hodge_6 \,\omega_2^I \!\wedge \omega_2^J.
		\label{eq:decconstapp}
	\end{equation}
	If a basis of harmonic forms $\omega_2^I$ is chosen such that $\gamma^{IJ}$ is diagonal, then $\phi_I = f_I b_I$ (no sum) are the canonically normalized axion fields, whose decay constants are the eigenvalues of $\gamma$,  $f_I = \lab{eig}_I \, \gamma^{JK}$. For example, if the compact space is a product of two-spheres, $X_6 = \lab{S}^2 \times \lab{S}^2 \times \lab{S}^2$, each with volume $L^2 \ell_s^2$, then we simply have $f_I^2 = M_\lab{pl}^2 L^{-4}/2$.

	If $X_6$ is Calabi-Yau, then the axion decay constants for two-form axions $b_I$ and $c_I$---arising from the two-form potentials $B_2$ and $C_2$, respectively, in type IIB string theory---may be simply computed from the
intersection numbers $\kappa_{IJK}$, the volumes $\ell_s^2 v^I$ of the two-cycles $\Sigma_2^I$, and the overall total volume $\mathcal{V}_\lab{E}$ of $X_6$. For example, for an axion $c_- = \ell_s^{-2} \int_{\Sigma_2^-} \!C_2$ associated to an orientifold-odd cycle $\Sigma_2^-$ the axion decay constant is
	\begin{equation}
		\frac{f^2}{M_\lab{pl}^2} \sim g_s \frac{\kappa_{I--} v^I}{\mathcal{V}_\lab{E}}.
		\label{eq:fn1}
	\end{equation}

When fivebranes are introduced to create monodromy,
the axion that experiences this monodromy will in general be a linear combination of the $\omega^I_2$, which we call $\Omega$.
For example, in a variant of the axion monodromy construction detailed in \S\ref{sec:axmon}, the (rel)axion $c(x)$ arises from a two-form $\Omega$ dual to the blowup cycle of an orbifold whose fixed point locus is $\Sigma_\lab{o}$, with $\lab{dim}_{\mathbbm{C}}\, \Sigma_\lab{o} = 1$. As shown in \cite{Flauger:2009ab}, the support of $\hodge \Omega \wedge \Omega$ is localized about $\Sigma_\lab{o}$.
The six-dimensional metric is approximately a cone,
\begin{equation}
	\tilde{g}_{mn} \ud y^m \, \ud y^n \approx \ud r^2 + r^2 \breve{g}_{\theta \phi} \,\ud \Psi^\theta \, \ud \Psi^\phi
\end{equation}
and $\Omega_{mn} \sim \Omega_{\theta \phi}$ to good approximation has its legs along the angular directions, so
\begin{equation}
	\frac{f^2}{M_\lab{pl}^2} \sim \frac{g_s}{\mathcal{V}_\lab{E}} \frac{1}{\ell_s^6} \int_{r_\labbot}^{r_\labtop}\!\!\ud r\, r \int\!\ud^5 \Psi \sqrt{\breve{g}}\, \breve{g}^{\theta \phi} \breve{g}^{\psi \chi} \Omega_{\theta \psi} \Omega_{\phi \chi} \approx \frac{g_s}{\mathcal{V}_\lab{E}} \frac{r_\labtop^2}{\ell_s^2}.
	\label{eq:relaxdecconst}
\end{equation}
Locally, we may think of the blow-up cycle as an Eguchi-Hanson space fibered over $\Sigma_\lab{o}$.
Since the integrand is highly localized about $\Sigma_\lab{o}$, we have $\int\!\hodge \Omega\wedge \Omega \approx \vol(\Sigma_\lab{o})$, and because of the conical nature of the six-dimensional metric, $\vol(\Sigma_\lab{o})$ is dominated by the contribution at $r_\labtop$, so $\int \!\hodge_6 \,\Omega \wedge \Omega \propto r_\labtop^2$.

The axion enjoys a continuous shift symmetry to all orders in perturbation theory in both $g_s$ and $\alpha'$. However, this continuous shift symmetry does not survive at the nonperturbative level, and is broken to a discrete shift symmetry by instantons carrying axion charge.
In particular, fundamental strings are charged under $B_2$, via the coupling
\begin{equation} \label{eq:swsb}
S(\mathcal{W}) = \ldots + \frac{i}{2\pi\alpha'}\int_{\mathcal{W}} \!\ud^2\sigma \,\sqrt{-h}\, \epsilon^{mn} B_{mn}+\ldots,
\end{equation}
where $h$ is the metric on the string worldsheet $\mathcal{W}$, and $m,n$ are two-dimensional indices tangent to $\mathcal{W}$. The string path integral receives a contribution from a Euclidean string whose worldsheet wraps $\Sigma_2^I$, termed a worldsheet instanton. This contribution will be proportional to $e^{-S_I}$, where $S_I = S(\Sigma_2^I)$. Because of the coupling (\ref{eq:swsb}),
	\begin{equation}
		S_I \supset 2 \pi i b_I,
	\end{equation}
	and so the potential generated by these nonperturbative effects is still invariant under discrete shifts $b_I \mapsto b_I + \n$, $\n \in \mathbbm{Z}$, as $e^{-S_I} \mapsto e^{-S_I + 2 \pi i \n} = e^{-S_I}$. Thus, worldsheet instantons break the perturbative, continuous shift symmetry of $b_I$ to the discrete shift $b_I \mapsto b_I + 1$.

The real part of the action (\ref{eq:swsb}) is proportional to the volume of $\Sigma_2^I$ in string units, and worldsheet instanton contributions become more important as the volume shrinks. These contributions are difficult to compute, so requiring computational control of the effective action constrains the sizes of cycles, $\Sigma_2^I$, and thus the sizes of the axion decay constants. A standard
requirement for control is
that the sizes of all cycles are much larger than the string length, $v^\alpha \gg 1$.

However, the two-cycle $\Sigma_2^I$ may sit in a warped region, with warp factor $e^{A}$. For two-form axions, (\ref{eq:decconstapp}) is unchanged---there is no explicit dependence on the warp factor. However, a ten-dimensional string will see a warped volume, and in particular the real part of the worldsheet instanton action is enhanced by factor $e^{-2 A}$. This allows the two-cycle volumes $v^\alpha$ to be smaller by a factor of $e^{-2A}$ without loss of control, and so the axion decay constant can be very small in a highly warped throat.

If we take $v^I$ to measure the warped volume of $\Sigma_2^I$ in string units, i.e.~the volume a ten-dimensional string would measure, then we may write
\begin{equation}
	\frac{f^2}{M_\lab{pl}^2} \sim g_s \frac{\kappa_{I--} v^{I}}{\mathcal{V}_\lab{E}} \left.e^{2 A}\right|_{\Sigma_2^{I}} \ll g_s \frac{\kappa_{I--} v^{I}}{\mathcal{V}_\lab{E}},
\end{equation}
keeping the constraint that $v^{I} \gg 1$.

\section{Necessity of the Intersection} \label{d7pf}

In NS5-brane axion monodromy, D3-brane charge accumulates on an NS5-brane that wraps a two-cycle $\Sigma_{\lab{NS5}}$ (denoted $\Sigma_2$ elsewhere in the text).
Taking $c(x) \equiv \int_{\Sigma_{\lab{NS5}}} \!C_2$ to be the relaxion field, a stopping potential can be generated by strong gauge dynamics in a gauge theory $G$ to which the relaxion has a nonvanishing axionic coupling $\lambda\, c(x) F\wedge F$, with $\lambda$ a constant.
We will take $G$ to be realized on a stack of D7-branes wrapping a
holomorphic
four-cycle $D$ (denoted $\Sigma_4$ elsewhere in the text).
The backreaction of the D3-brane charge changes the supergravity background, with the strongest effects occurring near $\Sigma_{\lab{NS5}}$. In this appendix we show that any $D$ for which $\lambda\neq 0$ \emph{necessarily intersects} $\Sigma_\lab{NS5}$.  Thus, one cannot mitigate the backreaction by arranging
that $D$ is outside of the warped region.

For our purposes, it suffices to show that $D$ and $\Sigma_\lab{NS5}$ have at least one point in common, even though the intersection number $[D] \cap [\Sigma_\lab{NS5}]$ of the corresponding homology classes may vanish.
We will use $\cap_s$ to denote intersection as point sets, as distinct from the topological intersection $[\Sigma_1] \cap [\Sigma_2]$,\footnote{Two submanifolds $M$, $N$, of $X$ have $M \cap_s N \neq \emptyset$ if and only if $M$ and $N$ have at least one point in common, without regard to the orientation of $M$ and $N$.}
and we will show that $\lambda \neq 0$ implies that $D \cap_s \Sigma_\lab{NS5}$.

Consider a D7-brane that fills spacetime and wraps a smooth four-cycle $D \subset X$ in the internal space $X$.
The D7-brane couples to $C_2$ axions via the Chern-Simons action
					\begin{equation}
						S_\lab{CS} = \mu_7 \int_{\mathcal{W}} \sum_{p} \iota^* C_p \wedge e^{\mathcal{F}} \supset \frac{\mu_7}{3!} \int_{\mathcal{W}}\!\iota^* C_2 \wedge \mathcal{F} \wedge \mathcal{F} \wedge \mathcal{F},
					\end{equation}
					where $\mathcal{W} = \mathcal{M}^{3,1} \times D$ is the D7-brane worldvolume,
$\iota: D \to X$ is the inclusion map of $D$ into $X$, $\iota^*$ denotes the pullback onto $D$, $F_2$ is the field strength of the worldvolume gauge theory, and $\mathcal{F} = \iota^* B_2 + 2\pi \alpha' F_2$.
The axionic coupling to the gauge theory $G$ on a stack of D7-branes wrapping $D$ is therefore
					\begin{equation}
						S_\lab{CS} \supset \frac{\mu_7}{2!} \left(\int_D \iota^*C_2 \wedge \mathcal{F}\right) \left(\int_{\mathcal{M}^{3,1}}\!\!\lab{tr}\,\mathcal{F} \wedge \mathcal{F}\right)\,.
					\end{equation}
Poincar\'{e} duality in $D$ relates the flux $\mathcal{F}$ to a two-cycle $S_\mathcal{F} \subset D$ which may further be viewed as a two-cycle $\iota_* S_\mathcal{F}$ in $X$, so that
\begin{equation} \label{istar}	
					\vartheta \equiv \int_{D} \iota^* C_2 \wedge \mathcal{F} = \int_{S_\mathcal{F}} \!\iota^* C_2 = \int_{\iota_* S_\mathcal{F}} \!\!C_2.
					\end{equation}
The holomorphic representative $\iota_* S_\mathcal{F}$ of the class
$[\iota_* S_\mathcal{F}]$ is contained, as a point set, in $D$.
So establishing the condition
\begin{equation}
		\iota_* S_\mathcal{F} \cap_s \Sigma_\lab{NS5} \neq \emptyset
		\label{eq:isc}
	\end{equation}
will imply our desired result $D \cap_s \Sigma_\lab{NS5}$.

Now we choose a basis of nontrivial two-cycle classes, $\{[\Sigma_i]\}$, $I = 1, \dots, p\equiv h^{1,1}$,  to span $H_2(X, \mathbbm{Z})$.
Without loss of generality, we may take the NS5-brane class $[\Sigma_\lab{NS5}]$ to be an element of this basis, say $[\Sigma_1] \equiv [\Sigma_\lab{NS5}]$. There exists a dual basis of harmonic two-forms $\omega^J$ such that $\int_{\Sigma_I}\omega^J = \delta^{J}_{I}$.
Expanding $\ell_s^{-2} C_2(x) = \sum_{i=1}^{p} c_I(x) \omega^I$, the relaxion field is $c_1(x) \equiv c(x)$.
We may also expand
\begin{equation} \label{iotaexp}
\iota_* S_\mathcal{F} = a_1 [\Sigma_1] + \dots + a_{p} [\Sigma_p] + (\lab{boundary})\,,
\end{equation}
for some integers $a_I$.
Comparing to (\ref{istar}), we see that $\lambda \neq 0$
if and only if $a_1 \neq 0$.

The relation (\ref{iotaexp}) with $a_1 \neq 0$ does not, on its own, imply (\ref{eq:isc}).
For example, consider a basis of homology $\{[\Sigma_1], [\Sigma_2]\}$, with minimum volume representatives $\{\Sigma_1, \Sigma_2\}$ that obey $\Sigma_I \cap_s \Sigma_J = \delta_{IJ}$. If $[S] = a_1 [\Sigma_1] + a_2 [\Sigma_2]$, then $S \cap_s \Sigma_1 \neq \emptyset \iff a_1 \neq 0$, regardless of the value of $a_2$.
But working in the basis $\{[\Sigma_1], [\Sigma_2'] = [\Sigma_2] - [\Sigma_1]\}$, for $a_1 \neq 0$ and $a_2=0$ we again have $S \cap_s \Sigma_1 \neq \emptyset$, while if $a_1=a_2$ we have instead $S \cap_s \Sigma_1 = \emptyset$.  Thus, the condition (\ref{eq:isc}) depends on the relation between $[\Sigma_1]$ and $[\Sigma_2],\ldots [\Sigma_p]$, which we have not yet specified.

We may view this issue in a dual picture. The coupling (\ref{istar}) can be written as the triple intersection of three divisors in $X$,
\begin{equation}
\vartheta = [D] \cap [D_{\cal{F}}] \cap [\lab{D}_{NS5}]\,,
\end{equation}
where $D_{\cal{F}}=\lab{PD}_{X}(\iota_*\cal{F})$ and $D_\lab{NS5}=\lab{PD}_{X}(\omega^1)$, with $\lab{PD}_{X}$ denoting the Poincar\'{e} dual in $X$.
The divisor $D_\lab{NS5}$ is dual to the curve $\Sigma_\lab{NS5}$, in that $D_\lab{NS5}$ is Poincar\'{e} dual to the two-form $\omega^1$ that is the dual vector to $\Sigma_\lab{NS5}$ with respect to the pairing $\int_{\Sigma_I}\omega^J = \delta^{J}_{I}$.
It follows that $[\Sigma_\lab{NS5}] \cap [D_\lab{NS5}]=1$.
Moreover, the requirement of a nonvanishing axionic coupling, $\vartheta \neq 0$, implies that $[D_{NS5}]\cap [D] \neq 0$.
Now although $[\Sigma_\lab{NS5}] \cap [D_\lab{NS5}] \neq 0$ and $[D_\lab{NS5}] \cap [D] \neq 0$,
it appears that $D_\lab{NS5}$ could stretch between $D$ and $\Sigma_\lab{NS5}$, intersecting each, even though $D$ and $\Sigma_\lab{NS5}$ remain widely separated.

To exclude this possibility, we use further facts about $\Sigma_\lab{NS5}$ and $\iota_* S_\mathcal{F}$.
Preserving supersymmetry in the D7-brane worldvolume requires that $\mathcal{F} \in H^{2}(D, \mathbbm{Z})$ be of type $(1,1)$, and so
its Poincar\'{e} dual $\iota_* S_\mathcal{F}$ is a
holomorphic curve.  Heuristically, $\iota_* S_\mathcal{F}$ can be viewed as  the curve wrapped by a D5-brane dissolved in $D$: if the D7-brane were annihilated by introducing an anti-D7-brane, a D5-brane on $\iota_* S_\mathcal{F}$ would remain.
Moreover, $\Sigma_\lab{NS5}$ is itself an irreducible holomorphic curve.  (In a construction in which $\Sigma_\lab{NS5}$ is a sum of irreducible holomorphic curves, this argument can be applied to each component.)
We can therefore express $\iota_* S_\mathcal{F}$ uniquely as a finite sum of distinct irreducible holomorphic curves $\{\sigma_A\}$, $A=1,\ldots K$ (with $\Sigma_\lab{NS5}\equiv \sigma_1$):
\begin{equation} \label{semif}
						\iota_* S_\mathcal{F} = a_1 \Sigma_\lab{NS5} + a_2 \sigma_2 + \dots + a_K \sigma_K, \qquad a_A \in \mathbbm{Z} \geq 0\,,
					\end{equation}
and we have shown above that a relaxionic coupling, $\lambda \neq 0$, requires $a_1 \neq 0$.
Because the $\sigma_A$ are distinct irreducible holomorphic curves, they intersect each other at most at points.
So $\iota_* S_\mathcal{F}$ contains all but finitely many of the points of $\Sigma_\lab{NS5}$.  The condition (\ref{eq:isc}) then follows, and so $D$ must intersect $\Sigma_\lab{NS5}$, which is what we set out to prove.

Note that if the $\sigma_A$ were simply a set of distinct, irreducible simplicial complexes, the relation (\ref{eq:isc}) would not be automatic.  If $\sigma_1$ intersected some of $\sigma_2,\ldots \sigma_K$ along suitable two-simplices, then $\sum_i a_A \sigma_A$ might have no points in common with $\sigma_1$, because adding $a_2\sigma_2 + \ldots a_K\sigma_K$ could subtract all the points of $\sigma_1$.  For curves intersecting at most at points, this is not possible.
			
	\section{Type IIB Supergravity with Fivebranes} \label{sec:conventions}

			\subsection{Conventions for type IIB supergravity}
			The bosonic part of the type IIB supergravity action in Einstein frame is
			\begin{equation}
				S_\lab{IIB} = \frac{1}{2 \kappa_{10}^2} \int\!\ud^{10} X\, \sqrt{-G_\lab{E}} \left(R_\lab{E} - \frac{|\partial \tau|^2}{2 (\lab{Im}\, \tau)^2} - \frac{|G_3|^2}{2\, \lab{Im}\, \tau} - \frac{1}{4} |\tilde{F}_5|^2\right) - \frac{i}{8 \kappa_{10}^2} \int\!\frac{C_4 \wedge G_3 \wedge \bar{G}_3}{\lab{Im}\, \tau}
			\end{equation}
			with $2\kappa_{10}^2 = \ell_s^{8}/2\pi$, $G_3 \equiv F_3 - \tau H_3$, $\tau \equiv C_0 + i e^{-\Phi}$, $F_{p+1} = \ud C_{p}$, $H_3 = \ud B_2$, $\tilde{F}_5 = F_5 - \tfrac{1}{2} C_2 \wedge H_3 + \tfrac{1}{2} B_2 \wedge F_3$, and $\tilde{F}_5 = \hodge_{10} \tilde{F}_5$ is imposed at the level of the equations of motion.
			
			We define the string length
			\begin{equation}
				\ell_s^2 = (2 \pi)^2 \alpha'.
			\end{equation}
			
			The actions for extremal D$p$-branes and NS5-branes are given by
			\begin{equation}
				S_{\lab{D}p} = -\mu_p \int\!\ud^{p+1} \xi\, e^{-\Phi} \sqrt{-\det\left(G_{ab} + B_{ab} + 2 \pi \alpha' F_{ab}\right)} + S_\lab{CS}
				\label{eq:dbi}
			\end{equation}
			and
			\begin{equation}
				S_\lab{NS5} = -\mu_5 \int\!\ud^6 \xi \, e^{-2 \Phi} \sqrt{-\det \left(G_{ab} - e^{\Phi} \left(C_{ab} + 2 \pi \alpha' F_{ab}\right)\right)} + S_\lab{CS},
			\end{equation}
			respectively, where $\mu_p = 2\pi/\ell_s^{p+1}$, and $F_{ab}$ is the gauge field strength on the brane worldvolume. The Chern-Simons term for a D$p$-brane reads
			\begin{equation}
				S_\lab{CS} = \mu_p \int \!\sum_{n} C_n \wedge e^{\mathcal{F}},
			\end{equation}
			where we have introduced the notation $\mathcal{F} = B_2 + 2 \pi \alpha' F$.
			The Chern-Simons piece sets the flux quantization condition
\begin{equation}
				\frac{1}{\ell_s^{p+1}} \int_{\Sigma^{p}} \!\! F_{p+1} \in \mathbbm{Z}.
			\end{equation}
We expand in a basis of $H^{2}(X_6)$,
denoted $\omega^{I}(y)$, with normalization
			\begin{equation}
				\int_{\Sigma^J} \!\omega^{I} = \ell_s^2 \delta^{I}_{J}.
			\end{equation}
We list our index conventions in Table \ref{tab:indexguide}.
			\begin{table}[h]
\begin{center}
\begin{tabulary}{\textwidth}{C C}
	\toprule
		\textbf{Directions} & \textbf{Indices} \\ \midrule
		(3+1)-dim spacetime & $\mu, \nu, \rho, \dots$  \\
		6-dim internal space & $m, n, \dots$ \\
		5-dim angular space & $\theta, \phi, \dots$ \\
		brane worldvolume & $a, b, \dots$ \\
		transverse to worldvolume & $i, j, k,  \dots $\\
		transverse vielbein & $\hat{\imath}, \hat{\jmath}, \hat{k}, \dots$ \\
		along cycle $\Sigma_2$ & $\alpha, \beta, \dots$ \\
\bottomrule
\end{tabulary}
\end{center}
\caption{Guide to index conventions.
\label{tab:indexguide}}
\end{table}

		\subsection{Einstein-frame potentials for fivebranes} \label{sec:eframe}
			The DBI action for a D5-brane is
			\begin{equation}
				S_\lab{DBI} = -\frac{2 \pi}{\ell_s^6} \int\!\ud^6 \xi\, e^{-\Phi} \sqrt{-\det\left(G^S_{ab} + \mathcal{F}_{ab}\right)}
			\end{equation}
			where
$G^S_{ab}$ and $B_{ab}$ are the pull-backs onto the brane worldvolume of the ten-dimensional string-frame metric and NS-NS two-form $B_2$, respectively.
 		
 			The ten-dimensional string-frame metric is related to the Einstein-frame metric via
			\begin{equation}
				G_{MN}^{S} = e^{\Phi/2} G^{E}_{MN},
			\end{equation}
			which we assume takes a warped product form
			\begin{equation}
				G^{E}_{MN}\, \ud X^M\, \ud X^N = e^{2 A(y)} g_{\mu \nu} \, \ud x^\mu \, \ud x^\nu + e^{-2 A(y)} \tilde{g}_{mn} \, \ud y^m \, \ud y^n,
			\end{equation}
			where $g_{\mu \nu}$ and $\tilde{g}_{m n}$ are metrics on the
four-dimensional spacetime and the six-dimensional internal space, respectively.
			
			Defining
			\begin{equation}
				b(x) = \frac{1}{\ell_s^2} \int_{\Sigma_2} \!B_2 = \frac{1}{\ell_s^2} \int_{\Sigma_2} \! \ell_s^2\, b(x) \,\ud y \wedge \ud z,
			\end{equation}
			choosing coordinates on $\Sigma_2$ such that $\ud y  \wedge \ud z$ is harmonic, and setting $F_{ab} = 0$, we may write
			\begin{equation}
				e^{\Phi/2} G_{ab}^{E} + B_{ab} = \begin{pmatrix} e^{\Phi/2} e^{2 A(y)} g_{\mu \nu} & 0 \\ 0 & \mathbf{m} \end{pmatrix}
			\end{equation}
			with
			\begin{equation}
				\mathbf{m} = \begin{pmatrix} e^{\Phi/2} e^{-2 A} \tilde{g}_{yy} & e^{\Phi/2} e^{-2 A} \tilde{g}_{yz} + \ell_s^2 b/2 \\ e^{\Phi/2} e^{-2 A} \tilde{g}_{yz} - \ell_s^2 b/2 & e^{\Phi/2} e^{-2 A} \tilde{g}_{zz} \end{pmatrix}.
			\end{equation}
			With $\tilde{g}_2 \equiv \tilde{g}_{yy} \tilde{g}_{zz} - \tilde{g}_{yz}^2$, we have
			\begin{equation}
				\sqrt{-\det\left(G_{ab} + B_{ab}\right)} = e^{\Phi + 4 A} \sqrt{-\det g } \sqrt{e^{\Phi} e^{-4 A} \tilde{g}_2 + \ell_s^4 b^2/4}.
			\end{equation}
			Upon integration over $\Sigma_2$, we may
take $e^{-4 A} \tilde{g}_2 \to \ell_\lab{E}^4$, where $\ell_\lab{E}^2$ is the characteristic size of $\Sigma_2$ in the ten-dimensional \emph{Einstein frame}. The DBI action, upon dimensional reduction, then yields a four-dimensional potential for the $b$ axion,
			\begin{equation}
				S_\lab{DBI} = \int\!\ud^4 x\, \sqrt{-g} \left(-\frac{\pi e^{4 A}}{\ell_s^4} \sqrt{\frac{g_s}{4} \left(\frac{\ell_\lab{E}}{\ell_s}\right)^4 + b^2}\right).
			\end{equation}
The dimensional reduction of the NS5-brane action follows similarly.

We will also be interested in the spectrum of Kaluza-Klein excitations, which we now compute.  Setting $F_{ab} = 0$, the NS5-brane action is
			\begin{equation}
				S_\lab{NS5} = -\frac{2 \pi}{\ell_s^6} \int\!\ud^6 \xi\, e^{-2 \Phi} \sqrt{-\det \mathbf{M}}
			\end{equation}
			with $\mathbf{M}_{ab} = G_{ab} - e^{\Phi} C_{ab}$. Expanding in fluctuations $\delta \mathbf{M}_{ab}$, we have
			\begin{equation}
				\sqrt{-\det\left(\overline{\mathbf{M}} + \delta \mathbf{M}\right)} = \sqrt{-\det \overline{\mathbf{M}}} \left(1 + \frac{1}{2} \,\lab{tr}\,\overline{\mathbf{M}}^{-1} \delta \mathbf{M}\right),
			\end{equation}
			with
			\begin{equation}
				\overline{\mathbf{M}} = \begin{pmatrix}
					e^{\Phi/2} e^{2 A} g_{\mu \nu} & 0 \\
					0 & \mathbf{m}
				\end{pmatrix}
			\end{equation}
			and
			\begin{equation}
				\mathbf{m} = \begin{pmatrix}
					e^{\Phi/2} e^{-2 A} \tilde{g}_{yy} & e^{\Phi/2} e^{-2 A}\tilde{g}_{yz} - e^{\Phi} \ell_s^2 c/2 \\
					e^{\Phi/2} e^{-2 A} \tilde{g}_{yz} + e^{\Phi} \ell_s^2 c/2 & e^{\Phi/2} e^{-2 A} \tilde{g}_{zz}
				\end{pmatrix}.
			\end{equation}
			As above,
			\begin{equation}
				\sqrt{-\det \overline{\mathbf{M}}} = e^{\Phi+4 A} \sqrt{-g} \sqrt{e^{\Phi} e^{-4 A} \tilde{g}_2 + e^{2 \Phi} \ell_s^4 c^2/4}.
			\end{equation}
			The fluctuation $\delta \mathbf{M}$ arises from allowing the embedding of the NS5-brane to fluctuate. We may explicitly write the pull-back as
			\begin{equation}
				G_{ab} - e^{\Phi} C_{ab} = \Pi^{MN}_{ab} \left(G_{MN} - e^{\Phi} C_{MN}\right).
			\end{equation}
			If we take the embedding of the fivebrane to be specified by $X^{M}(\xi^{a})$ and allow the brane to fluctuate in the transverse directions  $X^M(\xi^a) = \delta^{M}_{a}\xi^{a} + \delta^{M}_{j} X^j(\xi^b)$, the projection operator is then
			\begin{equation}
				\Pi_{ab}^{MN} \equiv \frac{\partial X^M}{\partial \xi^a} \frac{\partial X^N}{\partial \xi^b} = \delta_{a}^M \delta_{b}^N + \delta_a^M \delta^N_j \frac{\partial X^j}{\partial \xi^b} + \delta^{N}_b \delta^M_i \frac{\partial X^i}{\partial \xi^b} + \delta^M_i \delta^N_j \frac{\partial X^i}{\partial \xi^a} \frac{\partial X^j}{\partial \xi^b}.
			\end{equation}
			Assuming a product metric $G^{E}_{a i} = 0$, we have
			\begin{equation}
				\delta \mathbf{M}_{ab} = e^{\Phi/2} e^{-2 A} \tilde{g}_{ij} \frac{\partial X^i}{\partial \xi^a} \frac{\partial X^j}{\partial \xi^b}.
			\end{equation}
		 and the NS5-brane action may be written
			\begin{align}
				S_\lab{NS5} = -\frac{2 \pi}{\ell_s^6} \int\!\ud^6 \xi\, e^{-\Phi} e^{4 A} \sqrt{-g} &\sqrt{e^{\Phi} e^{-4 A} \tilde{g}_2 + e^{2 \Phi} \ell_s^4 c^2 /4} \Bigg( 1 + \frac{1}{2} e^{-4 A} g^{\mu \nu} \tilde{g}_{ij} \partial_\mu X^i \partial_\nu X^j \nonumber \\
				&+ \frac{1}{2} \frac{e^{\Phi} e^{-4 A} \tilde{g}_2}{e^{\Phi} e^{-4 A} \tilde{g}_2 + e^{2 \Phi} \ell_s^4 c^2/4} \tilde{g}_{ij} \tilde{g}^{\alpha \beta} \tilde{\nabla}_\alpha X^i \tilde{\nabla}_\beta X^j\Bigg).
			\end{align}
			We define the canonically normalized fields as
			\begin{equation}
				Y^{\hat{\imath}} = F E\indices{^{\hat{\imath}}_j} X^j
			\end{equation}
			with
			\begin{equation}
				F^2(x^\mu, y, z) = \frac{2 \pi}{\ell_s^6} \frac{e^{-\Phi}}{\sqrt{\tilde{g}_2}} \sqrt{e^{\Phi} e^{-4 A} \tilde{g}_2 + e^{2 \Phi} \ell_s^4 c^2/4} \qquad\text{and}\qquad \tilde{g}_{ij} = \delta_{\hat{\imath} \hat{\jmath}} E\indices{^{\hat{\imath}}_{i}} E\indices{^{\hat{\jmath}}_{j}}.
			\end{equation}
			We have assumed that $\tilde{g}_{i \alpha} = 0$, and thus $\tilde{\nabla}_{\alpha} E\indices{^{\hat{\jmath}}_j} = 0$.
Decomposing in real $\tilde{g}_2$ harmonics gives
			\begin{equation}
				\tilde{\nabla}^2 \mathcal{Y}^{I} = -\frac{e^{-2 A}}{\ell_\lab{E}^2}\lambda^I \mathcal{Y}^{I} \qquad Y^{\hat{\imath}} = \sum_{I} Y^{\hat{\imath}}_{I} \mathcal{Y}^{I} \qquad \int\!\ud^2 \sigma \sqrt{\tilde{g}_2}\, \mathcal{Y}^I\, \mathcal{Y}^J  = \delta^{IJ}.
			\end{equation}
			The action is
			\begin{align}
				S_\lab{NS5} = \int\!\ud^4 x\, \sqrt{-g} \Bigg(& - V(c) - \frac{1}{2} g^{\mu \nu} \partial_\mu Y^{\hat{\imath}}_I \partial^\mu Y_{\hat{\imath}}^I -  \frac{1}{2} m^2_I(c) Y^{\hat{\imath}}_I Y_{\hat{\imath}}^{I} \nonumber \\
				&+ g(c) \left(c \partial_\mu c\right)\left( Y_I^{\hat{\imath}} \partial^\mu Y^I_{\hat{\imath}}\right) - \frac{1}{2} g(c)^2 \left(\partial c\right)^2 Y_I^{\hat{\imath}} Y^I_{\hat{\imath}} \Bigg)
			\end{align}
			with
			\begin{subequations}
			\begin{align}
				V(c) &\equiv \frac{\pi}{\ell_s^4} \frac{e^{4 A}}{\sqrt{g_s}}\sqrt{4\left(\frac{\ell_\lab{E}}{\ell_s}\right)^4 + g_s c^2}\\
				g(c) &\equiv \frac{g_s}{2} \left(4 \left(\frac{\ell_\lab{E}}{\ell_s}\right)^4 + g_s c^2\right)^{-1}\\
				m_I^2(c) &\equiv \frac{\mu_5^2 e^{-\Phi}}{F^4} \frac{e^{-2 A}}{\ell_\lab{E}^2} \lambda_I = 4\lambda_I \frac{e^{2 A}}{\ell_\lab{E}^2} \left(\frac{\ell_\lab{E}}{\ell_s}\right)^4 \left(4\left(\frac{\ell_\lab{E}}{\ell_s}\right)^4 + g_s c^2\right)^{-1}.
			\end{align}
			\end{subequations}

		Finally, we will be interested in the tension of the domain wall interpolating between the metastable and stable states of the NS5-brane axion monodromy scenario. In the thin-wall approximation,
the domain wall corresponds to an NS5-brane winding $n$ times around the minimum volume three-cycle $\Sigma_3$ whose endpoints are $\Sigma_2$ and $\bar{\Sigma}_2$, the two-cycles wrapped by the NS5-brane and anti-NS5-brane, respectively. The tension can then be read off by reducing the action
		\begin{equation}
			S_\lab{NS5}= -\frac{2 \pi}{\ell_s^6} \int_{\mathcal{D}} \!\ud^6 \xi \,e^{-2 \Phi} \sqrt{-\det\left(G^S_{ab} - e^{\Phi} C_{ab}\right)} \to -T_D \int_{\mathcal{D}^{2,1}} \!\!\!\ud^3 x\, \sqrt{-\det \mathcal{P}(g_4)}.
		\end{equation}
		where $\mathcal{P}(g_4)$ denotes the pullback of the spacetime metric $g_{\mu\nu}$ onto the world-volume of the domain wall $\mathcal{D}^{2,1}$.
		
		We can gain intuition for this tension by modeling the three-cycle $\Sigma_3$ as
		\begin{equation}
			\ud s_{\Sigma_3}^2 = \ud r^2 + r^2 \left(\ud y^2 + \ud z^2\right),
		\end{equation}
where the two-torus volume form is $\ud y \wedge \ud z$, and $r \in [r_\labbot, r_\labtop]$. Then
		\begin{equation}
			G^S_{ab} - e^{\Phi} C_{ab} = \left(\begin{array}{c c c}
				e^{\Phi/2} e^{2 A} \mathcal{P}(g_4)_{\mu \nu} & 0 & 0 \\
				0 & e^{-2 A} e^{\Phi/2} & 0 \\
				0 & 0 & \mathbf{m} 				
			\end{array}\right)
		\end{equation}
		where
		\begin{equation}
			\mathbf{m} = \begin{pmatrix} e^{-2 A} e^{\Phi/2} r^2  & -e^{\Phi} \ell_s^2 c/2 \\
			-e^{\Phi} \ell_s^2 c/2 & e^{-2 A} e^{\Phi/2} r^2 \end{pmatrix},
		\end{equation}	
		and so
		\begin{equation}
			S_\lab{NS5} = -\left(\frac{2 \pi}{\ell_s^6} \int_{r_\labbot}^{r_\labtop}\!\ud r\, e^{-3 \Phi/4} e^{3 A} \sqrt{e^{-4 A} r^4 + e^{\Phi} \ell_s^4 c^2/4}\right) \int_{\mathcal{D}^{2,1}}\!\!\!\ud^3 x\, \sqrt{-\det \mathcal{P}(g_4)}.
		\end{equation}
		The tension then takes the form
		\begin{equation}
			T_\lab{D} = \frac{2 \pi}{\ell_s^3} \frac{L^3}{g_s^{3/4} \ell_s^3} \frac{1}{4} \left(e^{4 A_\labtop} - e^{4 A_\labbot}\right) \sqrt{1 + \frac{g_s \ell_s^4 c^2}{4 L^4} } \sim \frac{1}{\ell_s^3} N_\lab{D3}^{3/4} \left(e^{4 A_\labtop} - e^{4 A_\labbot}\right),
		\end{equation}
		since we must have $g_s \ell_s^4 c^2/4 L^4 \sim \n^2/N_\lab{D3} \ll 1$ to avoid the KPV instability.
		
		\section{Backreaction on the Internal Space} \label{sec:internalbr}

When one introduces a source of monodromy in a compactification, and explicitly breaks supersymmetry, the corresponding stress-energy will backreact on the metric, affecting the parameters in the low-energy effective theory.
In the NS5-brane model detailed in \S\ref{sec:fivebrane}, a key source of stress-energy is anti-D3-brane charge induced on the anti-NS5-brane. Because D3-brane charge preserves the same supersymmetry as the background, it will not backreact on the internal metric at leading order.\footnote{In the presence of anti-brane charge, the D3-brane charge will backreact on the internal metric at second order. Similarly, if the D3-brane charge is large enough a better description becomes available in which the
D3-branes
are dissolved into flux and a new warped throat is formed, corresponding to the analysis of \S\ref{sec:universal}. In what follows, we will take $\n \ll N_\lab{D3}$ and assume that the induced anti-D3-brane charge may be thought of as a small perturbation to the geometry.} However, the anti-D3-brane charge will break the remaining supersymmetry of the background and perturb the internal metric. Furthermore, both D3-brane and anti-D3-brane charge will perturb the warp factor $e^{4A}$.  In this appendix we calculate the perturbations to
the internal metric and the warp factor.

			At the level of the supergravity equations of motion, we may approximate the $\n$ units of induced anti-D3-brane charge as $\n$ anti-D3-branes smeared about the anti-NS5-brane. These anti-D3-branes do not source the metric directly, but do so through a combination of the warp factor $e^{4A}$ and the $\tilde{F}_5$ field which we denote $\Phi_-$, where $\Phi_\pm \equiv e^{4 A} \pm \alpha$, and
\begin{equation}
	F_5 = \left(1 + \hodge_{10}\right) \ud \alpha(y) \wedge \ud \vol_{\mathbbm{R}^{1,3}}.
\end{equation}
The equations of motion for $\Phi_{\pm}$ and the internal Einstein equation read
\begin{subequations}
	\begin{align}
		\tilde{\nabla}^2 \Phi_+ &= \frac{2}{\Phi_+ + \Phi_-} (\tilde{\nabla} \Phi_+)^2 + \frac{1}{2} g_s \ell_s^4 \left(\Phi_+ + \Phi_-\right)^2 \sum_{i} \delta(\lab{D3}_i) \label{eq:phipeom}\\
		\tilde{\nabla}^2 \Phi_- &= \frac{2}{\Phi_+ + \Phi_-} (\tilde{\nabla} \Phi_-)^2 + \frac{1}{2} g_s \ell_s^4 \left(\Phi_+ + \Phi_-\right)^2 \sum_{i} \delta(\overline{\lab{D3}}_i) \label{eq:phimeom}\\
		\tilde{R}_{mn} &= \frac{2}{(\Phi_+ + \Phi_-)^2} \tilde{\nabla}_{(m} \Phi_+ \tilde{\nabla}_{n)} \Phi_-
	\end{align}
	\label{eq:sugraeom}
\end{subequations}
where we use $\tilde{\nabla}$, etc., to denote quantities related to the unwarped, internal metric $\tilde{g}_{mn}$. We treat the anti-D3-branes as a perturbation to an imaginary self-dual background, in which
	\begin{equation}
		\Phi_+ \approx \frac{2 r^4}{L^4} \qquad \text{and}\qquad \Phi_- = 0,
		\label{eq:isdbg}
	\end{equation}
	$L^4 \propto g_s N_\lab{D3} \ell_s^4$, and the internal metric is taken to be the conifold, a cone over $\lab{T}^{1,1}$,
	\begin{equation}
		\tilde{g}_{mn} \ud y^m \, \ud y^n = \ud r^2 + r^2 \ud s_{\lab{T}^{1,1}}^2 = \ud r^2 + r^2 \breve{g}_{ij} \ud \Psi^i\, \ud \Psi^j.
	\end{equation}
	We linearize the system of equations (\ref{eq:sugraeom}) using the expansions\footnote{We use the conventions of \cite{Gandhi:2011id}, except that angular indices are denoted $\theta, \phi, \dots$, $\tilde{g}^\lab{here}_{mn} = g^\lab{there}_{mn}$, and $\breve{g}_{ij}^\lab{here} = \tilde{g}^\lab{there}_{\theta \phi}$.}
	\begingroup
	\allowdisplaybreaks
	\begin{subequations}
		\begin{align}
			\tilde{g}_{mn} &= \pertTensor{\tilde{g}}{0}{mn} + \delta \tilde{g}_{mn}\\
			\delta \tilde{g}_{rr} &= \sum_{I_s} \tau^{I_s}(r) \mathcal{Y}^{I_s}(\Psi) \\
			\delta \tilde{g}_{r \theta} &= \sum_{I_v} b^{I_v}(r) \mathcal{Y}^{I_v}_\theta (\Psi) \\
			\delta \tilde{g}_{\theta \phi} &= \sum_{I_s} \frac{1}{5} \pi^{I_s}(r) \breve{g}_{\theta \phi} \mathcal{Y}^{I_s}(\Psi) + \sum_{I_t} \phi^{I_t}(r) \mathcal{Y}^{I_t}_{\theta \phi}(\Psi) \\
			\Phi_- &= \pertTensor{\Phi}{0}{-} + \pertTensor{\delta\Phi}{1}{-}(r, \Psi) = \sum_{I_s} \delta\Phi_-^{I_s}(r) \mathcal{Y}^{I_s}(\Psi)
		\end{align}
	\end{subequations}
	\endgroup
	where $\mathcal{Y}^{I_s}(\Psi)$, $\mathcal{Y}^{I_v}_\theta (\Psi)$ and $\mathcal{Y}^{I_t}_{\theta \phi}(\Psi)$ are the scalar, transverse vector, and transverse traceless tensor harmonics on $\lab{T}^{1,1}$, with appropriate Laplacian eigenvalues $\lambda({I_s})$, $\lambda({I_v})$, and $\lambda({I_t})$, respectively.
	
	The Einstein metric on $\lab{T}^{1,1}$ is
		\begin{equation}
			\ud s_{\lab{T}^{1,1}}^2 = \frac{1}{6} \left(\ud \theta_1^2 + \sin^2 \theta_1 \,\ud \varphi_1^2\right) + \frac{1}{6} \left(\ud \theta_2^2 + \sin^2 \theta_2 \, \ud \varphi_2^2 \right) + \frac{1}{9} \left(\cos \theta_1\, \ud \varphi_1 + \cos \theta_2 \, \ud \varphi_2 + \ud \psi\right)^2,
		\end{equation}
		and in these coordinates a basis of scalar harmonics is given by
		\begin{equation}
			\mathcal{Y}^{I_s}(\varphi_1, \theta_1, \varphi_2, \theta_2, \psi) = e^{i \frac{R}{2} \psi} e^{i m_1 \varphi_1} e^{i m_2 \varphi_2} d_{m_1 \frac{R}{2}}^{(j_1)} (\theta_1) d^{(j_2)}_{m_2 \frac{R}{2}} (\theta_2)
		\end{equation}
		where $I_s \equiv \{j_1, m_1, j_2, m_2, R\}$ is a multi-index, $d^{(j)}_{m_1 m_2}(\theta)$ is the Wigner (small) d-matrix, and $\theta_i \in [0, \pi)$, $\phi_i \in [0, 2\pi)$, and $\psi \in [0, 4 \pi)$. We will only need the scalar harmonics in the following, with eigenvalues $\lambda_s(j_1, j_2, R) = 6(j_1(j_1+1)+j_2(j_2+1)-R^2/8)$.
	
	We will only focus on the radial scaling of the dominant metric perturbation. The presence of the anti-D3-brane charge on the NS5-brane may be interpreted as $N$ anti-D3-branes smeared over the two-cycle wrapped by the NS5-brane. The backreaction is heavily dependent on the geometric details of this smearing, so the reported radial scalings may be reduced by suitable geometric tuning. However, we expect a generic smearing to source all possible angular modes and any order-of-magnitude estimates to be set by the dominant mode.
	
	When placed in the background (\ref{eq:isdbg}), the anti-D3-branes will feel a force towards small $r$.  In the actual configuration, interactions with the anti-NS5-brane provide a stabilizing force that keeps the three-brane charge localized around the two-cycle, but the system (\ref{eq:sugraeom}) does not account for this force.  The effects of the stabilizing force could be included by sourcing appropriate perturbations in the warped throat, but doing so would leave the radial scaling of the dominant perturbation unchanged, and so our analysis applies in any case.
	
	For an anti-D3-brane at $(r', \Psi')$, we find that
	\begin{equation}
		\delta\Phi_-^{I_s}(r; r', \Psi') = -\frac{1}{\Delta_s}\frac{g_s \ell_s^4}{L^4}\frac{r'^4}{L^4}\left(\left(\frac{r'}{r}\right)^{2 + \Delta_s} \!\!\!\!\theta(r - r') + \left(\frac{r}{r'}\vphantom{\frac{r'}{r}}\right)^{\Delta_s-2}\!\!\!\!\theta(r'-r)\right) \mathcal{Y}^{I_s}(\Psi'),
	\end{equation}
	so in the area of interest,
	\begin{equation}
		\delta\Phi_-^{I_s}(r; r', \Psi') \propto -\frac{1}{\Delta_s}\frac{\n}{N_\lab{D3}} \frac{r'^4}{L^4} \left(\frac{r'}{r}\right)^{2 + \Delta_s}\!\!\!, \label{eq:dphipert}
	\end{equation}
	where $\Delta_s \equiv \sqrt{4+\lambda({I_s})}$.  This $\Phi_-$ profile induces a metric perturbation
	\begin{subequations}
	\begin{align}
		\pi^{0}(r) = 0 \qquad & \qquad \pi^{I_s}(r) \propto r^2 \frac{\n}{N_\lab{D3}} \left(\frac{r'}{r}\right)^{6 +\Delta_s} , \\
		\tau^{0}(r) \propto  \frac{\n}{N_\lab{D3}} \left(\frac{r'}{r}\right)^8 \quad & \qquad \tau^{I_s}(r) \propto -\frac{\n}{N_\lab{D3}} \left(\frac{r'}{r}\right)^{6 + \Delta_s}.
	\end{align}
		\label{eq:metperts}
	\end{subequations}
	From the spectroscopy of $\lab{T}^{1,1}$, the lowest scalar mode has quantum numbers $(\frac{1}{2}, \frac{1}{2}, \pm1)$ and $\Delta_s = 7/2$, so the dominant metric perturbation is
	\begin{equation}
		\delta \tilde{g}_{mn} \ud y^m \, \ud y^n \propto \frac{\n}{N_\lab{D3}} \left(\frac{r'}{r}\right)^{19/2} \mathcal{Y}^{\frac{1}{2}, \frac{1}{2}, 1}(\Psi) \,r^2 \breve{g}_{\theta \phi} \,\ud \Psi^\theta \,\ud \Psi^\phi
		\label{eq:dommetpert}
	\end{equation}
	where $\mathcal{Y}^{\frac{1}{2}, \frac{1}{2}, 1}(\Psi)$ is some real superposition of angular harmonics with $(j_1, j_2, R) = (\tfrac{1}{2}, \tfrac{1}{2}, \pm 1)$.

		The perturbations in both the warp factor and internal metric will alter the gauge coupling function,
		\begin{equation}\label{oneg}
			\frac{8 \pi^2}{g_\lab{YM}^2} = \frac{4 \pi}{\ell_s^4} \int_{\Sigma_4}\!\ud^4 \xi \, \sqrt{\tilde{g}_4}\, e^{-4 A},
		\end{equation}
		on a stack of D7-branes wrapping a divisor $\Sigma_4$, such that
		\begin{equation}\label{eq:gcfpert}
			\delta\left(\frac{8 \pi^2}{g_\lab{YM}^2}\right) = \frac{4 \pi}{\ell_s^4} \int_{\Sigma_4}\!\ud^4 \xi \, \sqrt{\tilde{g}_4} \Bigl(-2 \Phi_+^{-2} \left(\pertTensor{\delta \Phi}{1}{+} + \pertTensor{\delta \Phi}{1}{-}\right) + \Phi_+^{-1} \tilde{g}^{ab} \delta \tilde{g}_{ab}\Bigr).
		\end{equation}
		We proved in Appendix \ref{d7pf} that, in order for these D7-branes to couple to $c = \ell_s^{-2} \int_{\Sigma_2} C_2$ supersymmetrically, $\Sigma_4$ must not only descend into the warped throat but actually intersect $\Sigma_2$. We expect the supergravity description to break down near the intersection and a local model to be more apt and, from the open-string picture discussed in \S\ref{sec:intro}, we expect this contribution to be $\mathcal{O}(N)$. Furthermore, away from the intersection, the supergravity approximation becomes accurate and we have shown above $\N$ D3-branes induce an $\mathcal{O}(N/N_\lab{D3})$ fractional perturbation. This contribution is then
		\begin{equation}
			\delta\left(\frac{8 \pi^2}{g_\lab{YM}^2}\right) \propto N_\lab{D3} \int_{\Sigma_4}\!\ud^4 \xi \, \sqrt{\tilde{g}_4}\, r^{-4}\, \Bigl(\underbrace{\vphantom{\Bigl(}-2 \Phi^{-1}_+ \left(\pertTensor{\delta \Phi}{1}{+} + \pertTensor{\delta \Phi}{1}{-}\right) + \tilde{g}^{ab} \delta \tilde{g}_{ab}}_{\mathcal{O}(N/N_\lab{D3})}\Bigr) \propto N.
		\end{equation}

\bibliographystyle{JHEP_mod}
\bibliography{runaway_relaxion_monodromy_arxiv}
\end{document}